\documentclass{svjour3}                     
\smartqed  
\usepackage{graphicx}
\usepackage{graphics}
\usepackage{mathptmx}      
\usepackage{latexsym}
\usepackage{multirow}
\usepackage{epsfig}
\usepackage[latin1]{inputenc}
\usepackage[T1]{fontenc}
\usepackage{wrapfig}
\usepackage{savesym}
\usepackage{amsmath}
\savesymbol{iint}
\usepackage{txfonts}
\restoresymbol{TXF}{iint}
\journalname{Experimental Astronomy}
\begin{document}
\title{Ground-based calibration and characterization of the \\ Fermi Gamma-Ray Burst Monitor Detectors}
\titlerunning{Fermi GBM ground-based calibrations}        
\author{E.~Bissaldi         \and
        A.~von Kienlin      \and
        G.~Lichti           \and
        H.~Steinle          \and
        P.~N.~Bhat          \and
        M.~S.~Briggs        \and
        G.~J.~Fishman       \and
        A.~S.~Hoover        \and
        R.~M.~Kippen        \and
        M.~Krumrey          \and
        M.~Gerlach          \and
        V.~Connaughton      \and
        R.~Diehl            \and        
        J.~Greiner          \and
        A.~J.~van der Horst \and
        C.~Kouveliotou      \and
        S.~McBreen          \and        
        C.~A.~Meegan        \and
        W.~S.~Paciesas      \and
        R.~D.~Preece        \and
        C.~A.~Wilson-Hodge
}
\institute{E. Bissaldi \and A. von Kienlin \and G. Lichti \and H. Steinle \and
           R.~Diehl \and J.~Greiner \and S.~McBreen  \at
              Max-Planck-Institut f\"ur extraterrestrische Physik, 
              Giessenbachstrasse 1, 85748 Garching, Germany \\
              Tel.: +49-89-30000-3816 \\
              Fax: +49-89-30000-3569  \\
              \email{ebs@mpe.mpg.de}        
         \and
         A.~S. Hoover \and R.~M. Kippen \at
         Los Alamos National Laboratory, Los Alamos, NM, USA 
         \and
         G.~J. Fishman  \and C.~Kouveliotou \and 
         A.~J.~van der Horst \and C.~A.~Meegan \and C.~A.~Wilson-Hodge \at
         Marshall Space Flight Center, Huntsville, AL, USA
         \and
         P.~N. Bhat \and M.~S. Briggs \and V.~Connaughton \and  W.~S.~Paciesas  
         \and R.~D.~Preece \at
         University of Alabama, NSSTC, Huntsville, AL, USA
         \and
         M. Krumrey \and M. Gerlach \at
         Physikalisch-Technische Bundesanstalt, Berlin, Germany
}
\date{Received: date / Accepted: date}    
\maketitle
\begin{abstract}\label{abstract}
One of the scientific objectives of NASA's Fermi Gamma-ray Space Telescope
is the study of Gamma-Ray Bursts (GRBs).
The Fermi Gamma-Ray Burst Monitor (GBM) was designed to detect 
and localize bursts for the Fermi mission. 
By means of an array of 12 NaI(Tl) (8~keV to 1~MeV) and two BGO (0.2 to 40~MeV)
scintillation detectors, GBM extends the energy range (20~MeV to >~300 GeV) 
of Fermi's main instrument, the Large Area Telescope, 
into the traditional range of current GRB databases.
The physical detector response of the GBM instrument to GRBs is determined 
with the help of Monte Carlo simulations, which are supported and verified by on-ground 
individual detector calibration measurements. We present the principal instrument properties, 
which have been determined as a function of energy and angle, including the channel-energy 
relation, the energy resolution, the effective area and the spatial homogeneity.
%
%
\end{abstract}
%
%
\section{Introduction}\label{intro}
The Fermi Gamma-ray Space Telescope (formerly known as GLAST),
which was successfully launched on June 11, 2008, is an international 
and multi-agency space observatory \cite{ATW94,MIC96} that studies 
the cosmos in the photon energy range of 8 keV to greater than 300 GeV.
The scientific motivations for the Fermi mission comprise a wide range 
of non-thermal processes and phenomena that can best be studied in 
high-energy gamma rays, from solar flares to pulsars and cosmic rays in our Galaxy,
to blazars and Gamma-Ray Bursts (GRBs) at cosmological distances \cite{GLAST01}. 
Particularly in GRB science, the detection of energy emission 
beyond 50 MeV \cite{DIN01,KAN08} still represents a puzzling topic, 
mainly because only a few observations by the Energetic Gamma-Ray 
Experiment Telescope (EGRET) \cite{THO93} on-board the Compton 
Gamma-Ray Observatory (CGRO) \cite{HUR94,GON03} and more recently 
by AGILE \cite{GIU08} are presently available above this energy . 
Fermi's detection range, extending approximately an order of magnitude 
beyond EGRET's upper energy limit of 30 GeV, will hopefully expand the catalogue of 
high-energy burst detections. A greater number of detailed observations of burst 
emission at MeV and GeV energies should provide a better understanding of bursts, thus
testing GRB high-energy emission models \cite{MES94,WAX97,DER00,SAR01}.
Fermi was specifically designed to avoid some of the limitations of EGRET,
and it incorporates new technology and advanced on-board software that will allow
it to achieve scientific goals greater than previous space experiments. 

The main instrument on board the Fermi observatory is the Large Area Telescope
(LAT), a pair conversion telescope, like EGRET, operating in the energy range between 20
MeV and 300 GeV. This detector is based on solid-state technology, obviating the need 
for consumables (as was the case for EGRET's spark chambers, whose detector gas needed to be 
periodically replenished) and greatly decreasing (< 10 $\mu$s) dead time (EGRET's high dead time 
was due to the length of time required to re-charge the HV power 
supplies after event detection).
These features, combined with the large effective area and excellent background rejection,
allow the LAT to detect both faint sources 
and transient signals in the gamma-ray sky. 
Aside from the main instrument, the Fermi Gamma-Ray Burst Monitor (GBM)
extends the Fermi energy range to lower energies (from 8 keV to 40 MeV). The GBM
helps the LAT with the discovery of transient events within a larger FoV and 
performs time-resolved spectroscopy of the measured burst emission. 
In case of very strong and hard bursts, the GRB position, which is usually communicated by the GBM 
to the LAT, allows a repointing of the main instrument, in order to search for higher energy 
prompt or delayed emission. 

\begin{figure}[t!]
\centering
\begin{tabular}{cc}
\includegraphics[height=65mm,bb=54 63 486 569,clip]{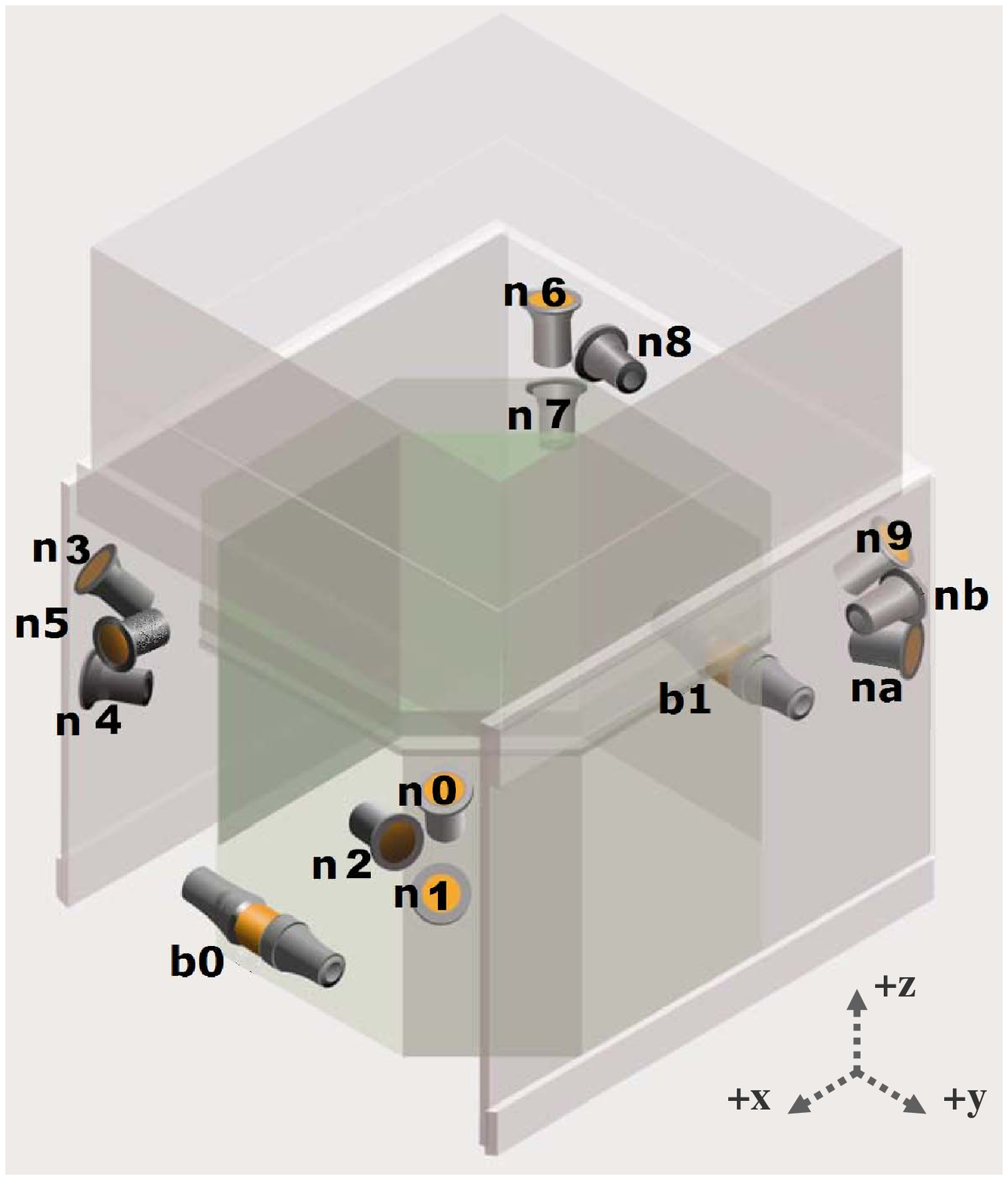} & 
\includegraphics[height=65mm,bb=9 0 603 792,clip]{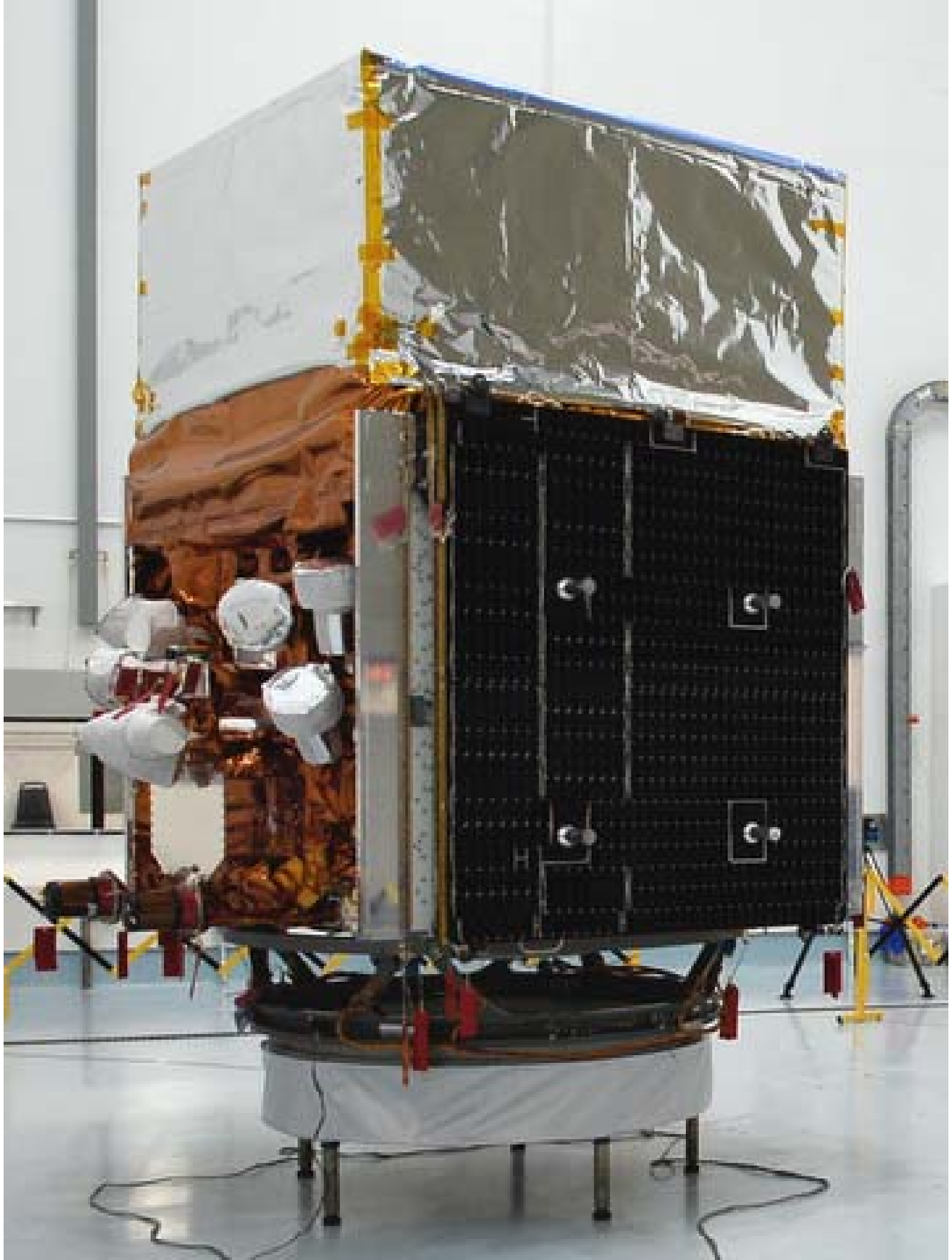}
\end{tabular}
\caption{{\it On the left}: Schematic representation of the Fermi spacecraft, showing the placement of the 
14 GBM detectors: 12 NaI detectors (from {\it n0} to {\it nb}) are located in
groups of three on the spacecraft edges, while two BGOs ({\it b0} and {\it b1})
are positioned on opposite sides of the spacecraft.
{\it On the right}: Picture of Fermi taken at Cape Canaveral few days before the launch. 
Here, six NaIs and one BGO are visible on the spacecraft's side.
Photo credit: NASA/Kim Shiflett (\texttt{http://mediaarchive.ksc.nasa.gov})}
\label{Schema_Spacecraft}
\end{figure}

The GBM is composed of unshielded and uncollimated scintillation detectors (12 NaI (Tl) and two BGO) 
which are distributed around the Fermi spacecraft with different viewing angles, 
as shown in Fig. \ref{Schema_Spacecraft}, in order to determine the direction to a burst by 
comparing the count rates of different detectors. The reconstruction of source locations and the 
determination of spectral and temporal properties from GBM data requires very detailed 
knowledge of the full GBM detectors' response. This is mainly derived from computer modeling 
and Monte Carlo simulations \cite{KIP04,HOO08b}, which are supported and verified by 
experimental calibration measurements.
%
%

In order to perform the above validations, several calibration campaigns
were carried out in the years 2005 to 2008. The calibration of each individual detector 
(or detector-level calibration) comprises three distinct campaigns: a main campaign with 
radioactive sources (from 14.4 keV to 4.4 MeV), which was performed in the laboratory 
of the Max-Planck-Institut f\"ur extraterrestrische Physik (MPE, Munich, Germany), 
and two additional campaigns focusing on the low energy calibration of the NaI 
detectors (from 10 to 60 keV) and on the high energy calibration of the BGO detectors (from 4.4 to 17.6 MeV), 
respectively. The first one was performed at the synchrotron 
radiation facility of the Berliner Elektronenspeicherring-Gesellschaft 
f$\rm{\ddot{u}}$r Synchrotronstrahlung (BESSY, Berlin, Germany), with the support and collaboration of the German
Physikalisch-Technische Bundesanstalt (PTB), while the second was carried out
at the SLAC National Accelerator Laboratory (Stanford, CA, USA).

Subsequent calibration campaigns of the GBM instrument were performed at system-level,
that comprises all flight detectors, the flight Data Processing Unit (DPU) and the 
Power Supply Box (PSB). These were carried out in the laboratories of the 
National Space Science and Technology Center (NSSTC) and of the
Marshall Space Flight Center (MSFC) at Huntsville (AL, USA)
and include measurements for the determination of the channel-energy relation 
of the flight DPU and checking of the detectors' performance before and
after environmental tests. After the integration of GBM onto the spacecraft, 
a radioactive source survey was performed in order to verify the spacecraft 
backscattering in the modeling of the instrument response. These later
measurements are summarized in internal NASA reports and will be not further 
discussed.

This paper focuses on the detector-level calibration campaigns of the GBM instrument,
and in particular on the analysis methods and results, which crucially support the development 
of a consistent GBM instrument response. It is organized as follows: Section \ref{GBM Det} outlines
the technical characteristics of the GBM detectors; Section \ref{Calib_camp} describes the various
calibration campaigns which have been done, highlighting 
simulations of the calibration in the laboratory environment performed at MPE (see Section \ref{Sec_Lab_Sim});
Section \ref{Line_Ana_Proc} discusses the analysis system for the calibration data
and shows the calibration results.
In Section \ref{Concl}, final comments about the scientific capabilities of GBM are given 
and the synergy of GBM with present space missions is outlined.
%
%
\begin{figure}[b!]
\centering
\begin{tabular}{c}
\includegraphics[width=84mm,bb=126 83 769 423,clip]{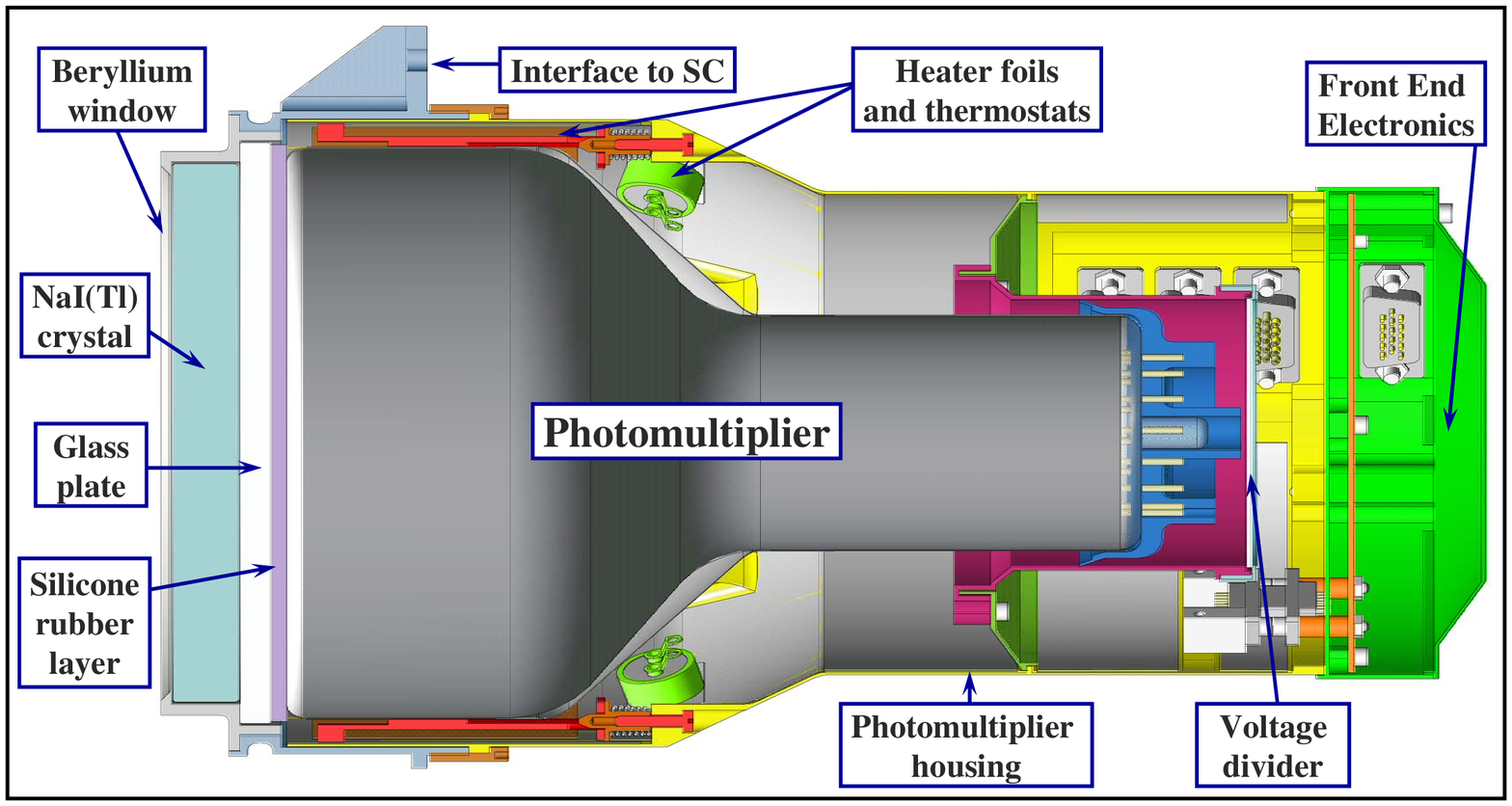}  \\
\includegraphics[width=84mm,bb=94 0 886 425,clip]{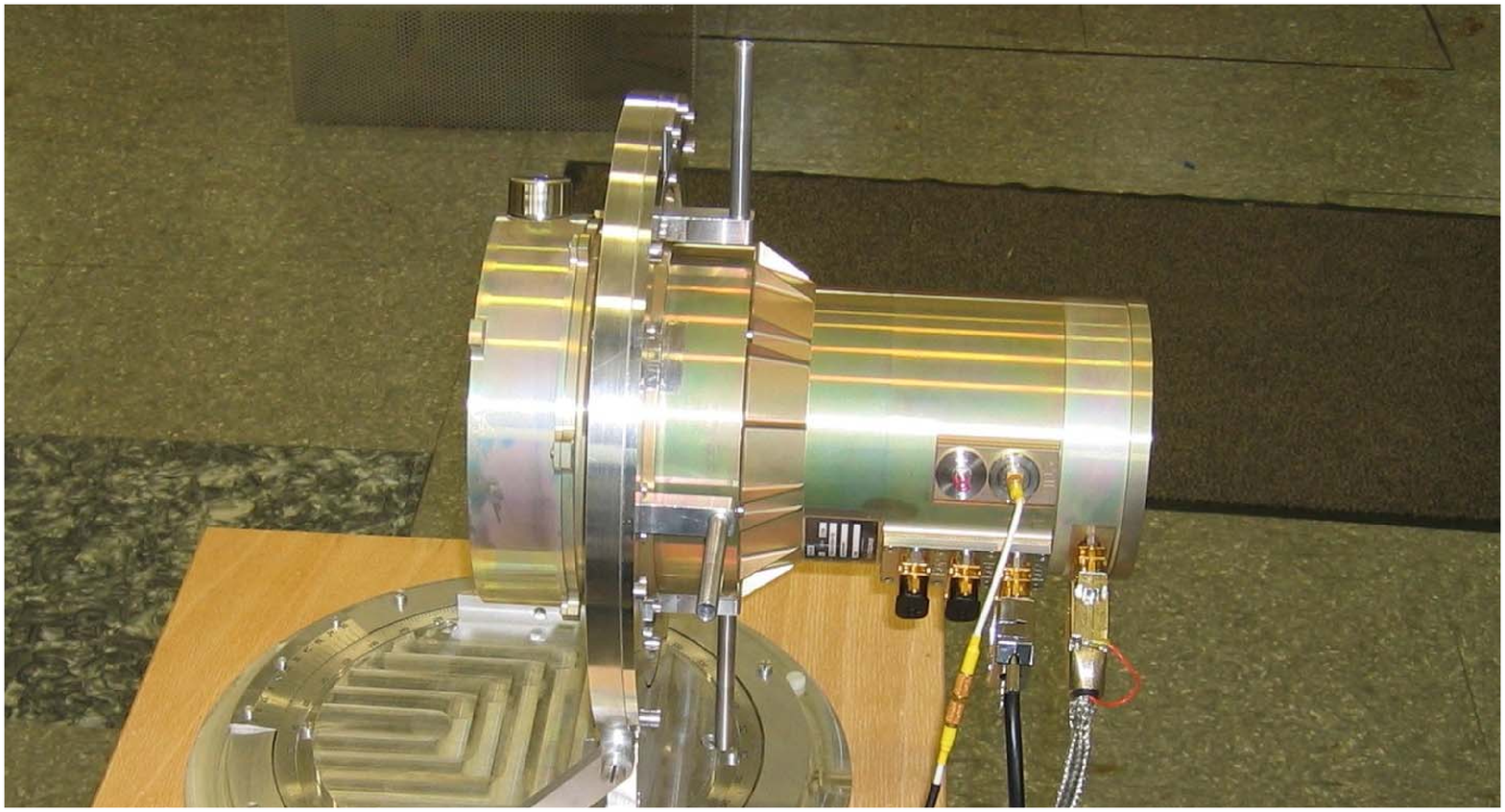}
\end{tabular}
\caption{On the {\it left}: Schematic cross-section of a GBM NaI detector showing the
main components. A picture of a detector flight unit mounted on the calibration stand
was taken in the laboratory during detector level calibration 
measurements and is shown in the {\it right panel}
}
\label{Fig_NaI_detectors}
\end{figure}
%
%
%
%
\section{The GBM Detectors}\label{GBM Det}
The GBM flight hardware comprises a set of 12 Thallium activated
Sodium Iodide crystals (NaI(Tl), hereafter NaI), two Bismuth Germanate crystals
(Bi$_4$ Ge$_3$~O$_{13}$, commonly abbreviated as BGO), a DPU, and a PSB. 
In total, 17 scintillation detectors were built: 12 flight module (FM) NaI detectors, two FM BGO detectors,
one spare NaI detector and two engineering qualification models (EQM), one for each detector type.
Since detector NaI FM 06 immediately showed low-level performances, it was decided to replace it
with the spare detector, which was consequently numbered FM 13. Note that the detector numbering scheme
used in the calibration and
adopted throughout this paper is different to the one used for in-flight analysis,
as indicated in Table~\ref{Tab_Num_Dec} (columns 2 and 3) in the appendix.

\begin{figure}[t!]
\centering
\includegraphics[width=84mm,bb=99 90 693 504,clip]{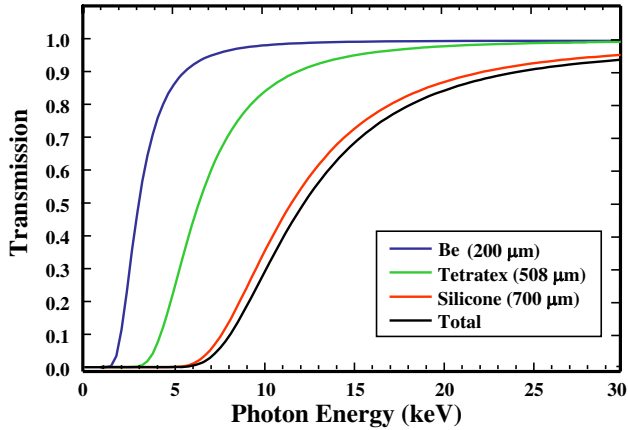}
\caption{X-ray transmission of all individual layers of a NaI detector entrance window
(the 0.2~mm thick Beryllium sheet in {\it blue}, the 0.5~mm thick Tetratex layer
in {\it green}, the 0.7~mm thick Silicone layer in {\it red}
and the sum of all components in {\it black}. The transmission at 10 keV is $\sim$~30\%
}
\label{Fig_NaI_Trans}
\end{figure}

The cylindrical NaI crystals (see Fig. \ref{Fig_NaI_detectors}) 
have a diameter of 12.7~cm (5'') and a thickness 
of 1.27~cm (0.5''). For light tightness and for sealing the crystals against atmospheric 
moisture (NaI(Tl) is very hygroscopic) each crystal is packed light-tight in a 
hermetically sealed Al-housing (with the exception of the glass window to which the
PMT is attached). In order to allow measurements of X-rays down to 5 keV
(original project goal \cite{GBM99}) the radiation entrance window
is made of a 0.2~mm thick Beryllium sheet. However, due to mechanical stability reasons, 
an additional 0.7~mm thick Silicone layer had to be mounted between the Be window and the crystal,  
causing a slight increase of the low-energy detection threshold. Moreover,
an 0.5~mm thick Tetratex layer was placed in front of the
NaI crystal, in order to improve its reflectivity.
The transmission probability as a function of energy for all components of the
detector window's system is shown in Fig. \ref{Fig_NaI_Trans}.
Consequently, NaI detectors are able to detect gamma-rays in the energy 
range between $\sim$~8~keV and $\sim$~1~MeV. The individual detectors are mounted around the spacecraft and are 
oriented as shown schematically in Fig. \ref{Schema_Spacecraft} ({\it left panel}). This arrangement 
results in an exposure of the whole sky unocculted by the earth in orbit.

The Hamamatsu R877 photomultiplier tube (PMT) is used for all the GBM detectors.
This is a 10-stage 5-inch phototube made from borosilicate glass with a bialkali (CsSb) photocathode,
which has been modified (R877RG-105) in order to fulfill the GBM mechanical load-requirements. 

\begin{figure}[t!]
\centering
\begin{tabular}{c}
\includegraphics[width=84mm,bb=36 36 559 230,clip]{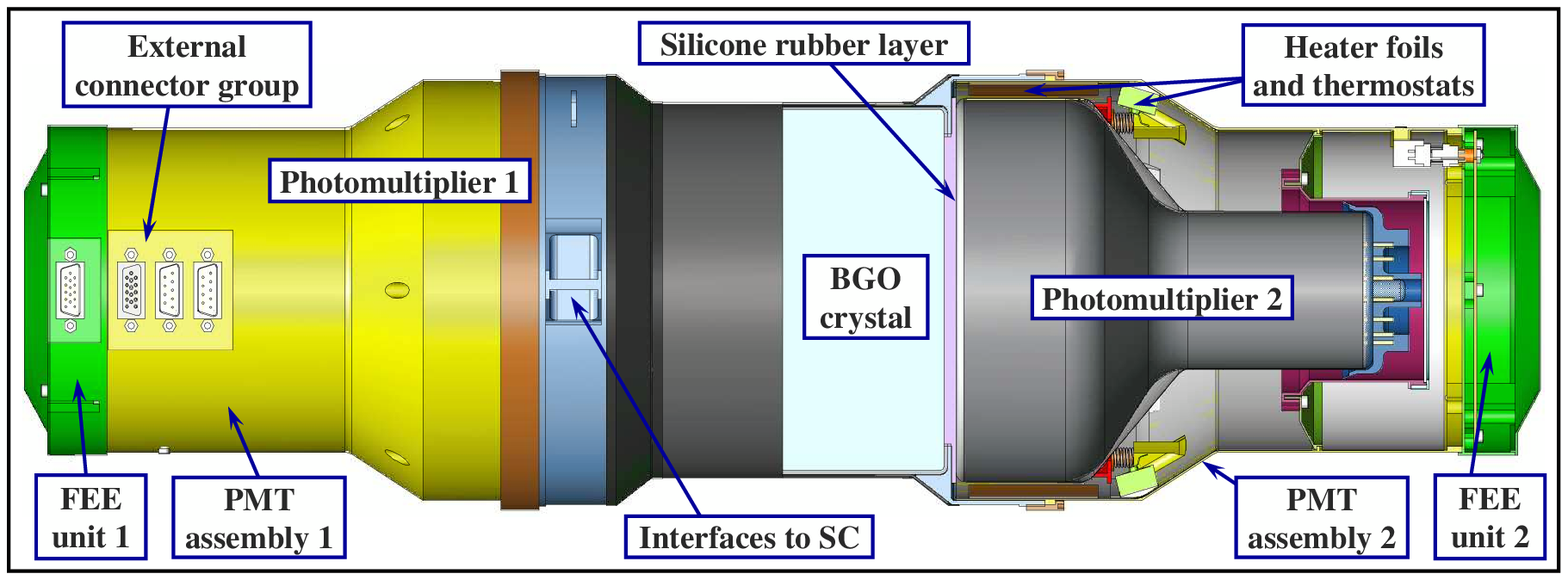} \\
\includegraphics[width=84mm,bb=53 120 558 314,clip]{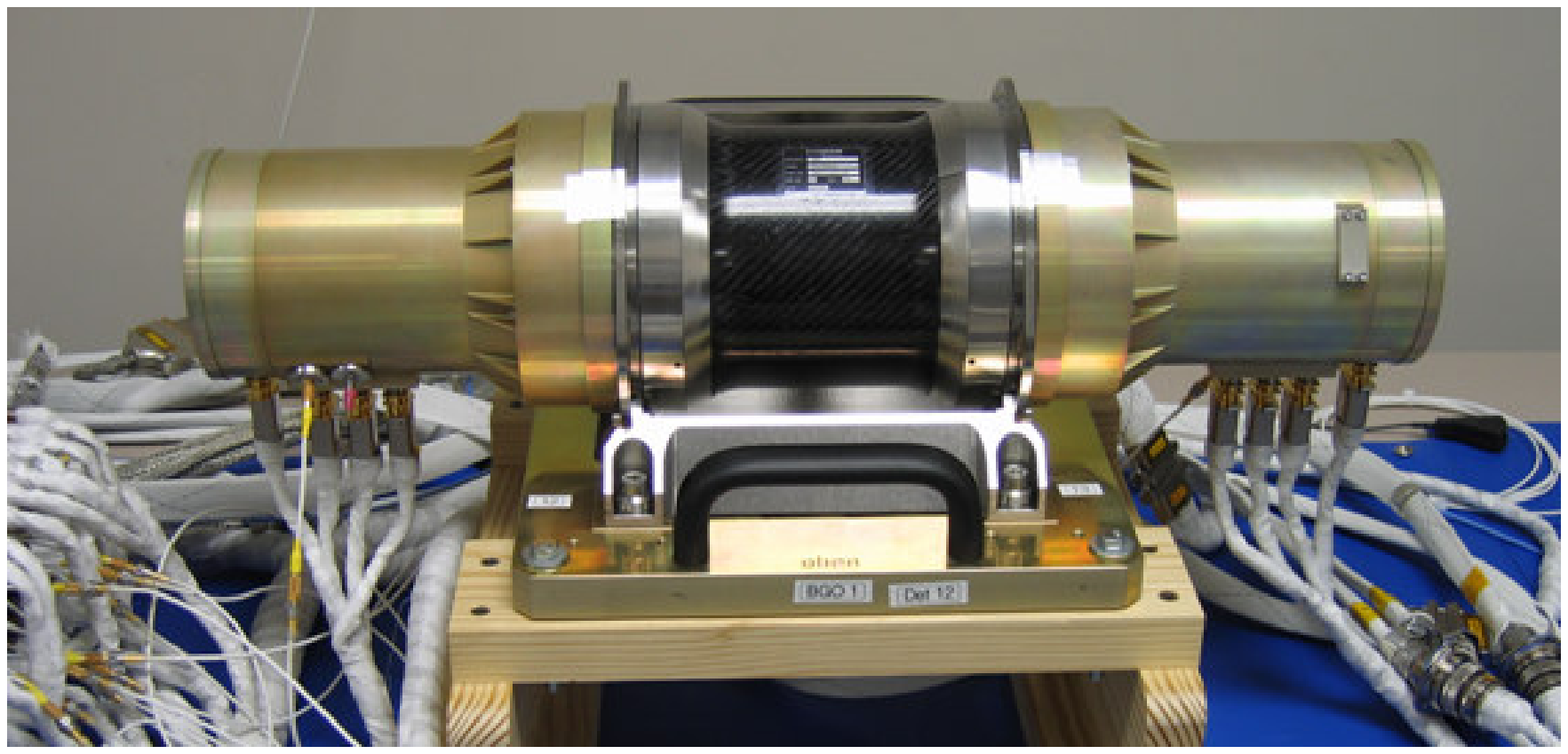}
\end{tabular}
\caption{On the {\it top}: Schematic cross-section of a GBM BGO detector.
The right hand side of the schematic is a cut away view, whereas the left hand side
is an external view. The central portion contains the BGO crystal ({\it light blue}),
which is partially covered by the outer surface of the detector's assembly.
A picture of a BGO detector flight unit taken in the laboratory
during system level calibration measurements is shown in the {\it bottom panel}
}
\label{Fig_BGO_detectors}
\end{figure}

With their energy range extending between $\sim$~0.2 and $\sim$~40~MeV, two BGO detectors
provide the overlap in energy with the LAT instrument and are crucial for in-flight inter-instrument calibration. 
The two cylindrical BGO crystals (see Fig. \ref{Fig_BGO_detectors}) have a diameter and a length of 12.7 cm (5'') and 
are mounted on opposite sides of the Fermi spacecraft (see Fig. \ref{Schema_Spacecraft}), 
providing nearly a 4 $\pi$ sr FoV.
The BGO housings are made of CFRP (Carbon Fibre Reinforced Plastic), which provides the light tightness and 
improves the mechanical stability of the BGO unit. For thermal reasons, the interface
parts are fabricated of Titanium.
On each end, the circular side windows of the crystal are polished in mirror quality 
and are viewed by a PMT (same type as used for the NaI detectors). Viewing the crystal by
two PMTs guarantees a better light collection and a higher level of redundancy. 

The output signals of all PMTs (for both NaIs and BGOs) are first amplified via linear charge-sensitive amplifiers.
The preamplifier gains and the HVs are adjusted so that they produce a $\sim$5 V signal for a 1
MeV gamma-ray incident on a NaI detector and for a 30 MeV gamma-ray
incident on a BGO detector. Due to a change of the BGO HV settings after launch, 
this value changed to 40 MeV, thus extending the original BGO energy range \cite{GBM99}.
Signals are then sent through pulse shaping stages to an output 
amplifier supplying differential signals to the input stage of the DPU,
which are combined by a unity gain operational amplifier in the DPU before 
digitizing. In the particular case of BGO detectors, outputs from the two PMTs are divided by two 
and then added at the preamplifier stage in the DPU.

In the DPU, the detector pulses are continuously digitized by a separate flash ADC 
at a speed of 0.1~$\mu$s. The pulse peak is measured by a separate software running 
in a Field Programmable Gate Array (FPGA). This scheme allows a fixed energy independent 
commendable dead-time for digitization. The signal processor digitizes the amplified 
PMT anode signals into 4096 linear channels. Due to telemetry limitations, these channels 
are mapped (pseudo-logarithmic compression) on-board into (1) 128-channel resolution 
spectra, with a nominal temporal resolution of 4.096 s (Continuous High SPECtral resolution or CSPEC data)
and (2) spectra with a poorer spectral resolution of eight channels and better temporal resolution
of 0.256 s (Continuous high TIME resolution or CTIME data)
by using uploaded look-up tables.\footnote{The temporal resolution of CTIME and 
CSPEC data is adjustable: nominal integration times are decreased when a trigger occurs.}
These were defined with the help of the on-ground
channel-energy relations (see Section \ref{CE Relation}).
Moreover, time-tagged event (TTE) data are continuously stored by the DPU.
These data consist of individually digitized pulse height events from the GBM
detectors which have the same channel boundaries as CSPEC and 2 microsecond
resolution. TTE data are transmitted only when a burst trigger occurs or by command.
More details on the GBM data type as well as a block diagram of the GBM 
flight hardware can be found in \cite{MEE07}.

Besides processing signals from the detectors, the DPU processes commands, 
formats data for transmission to the spacecraft and controls high and
low voltage (HV and LV) to the detectors. Changes in the detector gains can be due to several
effects, such as temperature changes of the detectors and of the HV power supply, 
variations in the magnetic field at the PMT, and PMT aging. GBM adopts a technique
previously employed on BATSE, that is Automatic Gain Control (AGC). In this way,
long timescale gain changes are compensated by the GBM flight
software by adjusting the PMT HV to keep the background 
511 keV line at a specified energy channel. 
\begin{table}[t!]
\centering
\caption{Properties of radioactive nuclides used for NaI and BGO calibration campaigns:
(1) Half-lives in years (y) or days (d); 
(2) Decay type producing the gamma-ray ($\gamma$) or 
X-ray (e.g. K and L) radiation $-$ For nuclides which are part of
decay chains, the daughter nuclides producing the corresponding radiation are also given;
(3) Line energies in keV;
(4) Photon-emission probabilities for the corresponding decays
}
\begin{tabular}{ccccc}
\hline\noalign{\smallskip}
\multirow{3}{15mm}{\centering \bf Nuclide}  &  \multirow{3}{15mm}{\centering \bf (1) \\ \bf Half-life}  &  \multirow{3}{15mm}{\centering \bf (2) \\ \bf Line origin}  & \multirow{3}{20mm}{\centering \bf (3) \\ \bf Line Energies \\ (keV)} & \multirow{3}{25mm}{\centering \bf (4) \\ \bf Transition \\ \bf Probability}   \\
 & & &  \\
 & & &  \\
\noalign{\smallskip}\hline\noalign{\smallskip}
\multirow{2}{*}{$\mathbf{^{22}}${\bf Na}}  &  \multirow{2}{*}{950.5(4) d}  &  Annih.     &	 511       &  1.798    \\
																					 &															 &  $\gamma$	 &	 1274.54 	 &	0.9994	 \\
\noalign{\smallskip}\hline\noalign{\smallskip}
$\mathbf{^{40}}${\bf K}	  								 &  1.277(8)E9 y			    &  $\gamma$   &  1460.83 	 &  0.1067  \\
\noalign{\smallskip}\hline\noalign{\smallskip}
$\mathbf{^{54}}${\bf Mn} 	                 &	312.15(8) d		        & 	$\gamma$   &  834.84	   &	0.999750(12)  \\ 
\noalign{\smallskip}\hline\noalign{\smallskip}
  	                                       &                        &  $\gamma$  &	 14.41     &	0.0916(15)  \\
$\mathbf{^{57}}${\bf Co}                   &	271.83(8) d						&  $\gamma$	 &  122.06		 &  0.8560(17)  \\
																					 &												&  $\gamma$  &	136.47 		 &	0.1068(8)	  \\
\noalign{\smallskip}\hline\noalign{\smallskip}
\multirow{2}{*}{$\mathbf{^{60}}${\bf Co}}  &  \multirow{2}{*}{5.2712(11) y} &  $\gamma$	 &  1173.23   &  0.9985(3)    \\
																					 &															  &	 $\gamma$	 &	1332.49   &  0.999826(6)  \\	
\noalign{\smallskip}\hline\noalign{\smallskip}
\multirow{2}{*}{$\mathbf{^{88}}${\bf Y}}   &  \multirow{2}{*}{106.630(25) d} &  $\gamma$ &	898.04 	  &  0.940(3)   \\
																					 &																 &  $\gamma$ &	1836.06   &	 0.9933(3)  \\
\noalign{\smallskip}\hline\noalign{\smallskip}		
                                           &                        &  Ag$-$SumK$\alpha$ &	22.1 		  &	 0.836(6) 	  \\
$\mathbf{^{109}}${\bf Cd}                  &  462.1(14) d		        &	 Ag$-$SumK$\beta$  &	25 				&	 0.1777(19)	  \\
																					 &												&  $\gamma$					 &  88.03     &	 0.03626(20)  \\
\noalign{\smallskip}\hline\noalign{\smallskip}		
																					 &											  &  Ba$-$SumK$\alpha$ &	32.06 	  &	 0.0553(10)   \\
$\mathbf{^{137}}${\bf Cs} 								 &  30.13(24) y 	        &	 Ba$-$SumK$\beta$  &  36.6 		  &  0.01321(27)  \\
																					 &												&	 Ba$-$137m 				 &	661.66  	&	 0.8500(20)   \\
\noalign{\smallskip}\hline\noalign{\smallskip}		
$\mathbf{^{203}}${\bf Hg} 								 &  46.604(17) d					&	 $\gamma$					 &	 279.2    &	 0.8146(13)   \\
\noalign{\smallskip}\hline\noalign{\smallskip}
$\mathbf{^{232}}${\bf Th}									 &		1.405(6)E10 y			  &	 $^{208}$Tl $\left(\gamma\right)$	&		2614.53 &  0.3564 		\\
\noalign{\smallskip}\hline\noalign{\smallskip}		
$\mathbf{^{241}}${\bf Am}	                 &  432.2(7) y 	          &  $\gamma$					&	 59.4 	    &	 0.359(4) 	\\
\noalign{\smallskip}\hline\noalign{\smallskip}
$\mathbf{^{241}}${\bf Am}/$\mathbf{^{9}}${\bf Be}	&		432.2 (7) y		&		$\gamma$				&	 4430       &	 0.00004	 \\
\noalign{\smallskip}\hline
\end{tabular}
\label{Tab_Nuclides}
\end{table}

%
%
%
\section{Calibration Campaigns}\label{Calib_camp}
To enable the location of a GRB and to derive its spectrum, a detailed knowledge of 
the GBM detector response is necessary. The information regarding
the detected energy of an infalling gamma-ray photon, which is
dependent on the direction from where it entered the detector, is stored into
a response matrix. This must be generated for each detector using computer simulations. 
The actual detector response at discrete incidence angles and energies has to be measured
to verify the validity of the simulated responses. 
The complete response matrix of the whole instrument system (including LAT and the spacecraft structure) is
finally created by simulation of a dense grid of energies and infalling photon directions 
using the verified simulation tool \cite{KIP07}.

The following subsections are dedicated to the descriptions of the three calibration campaigns
at detector level. The most complete calibration of all flight and engineering qualification models
was performed at the MPE laboratory using a set of 
calibrated radioactive sources whose type and properties are listed in Table \ref{Tab_Nuclides}. \\
Due to the lack of radioactive sources producing lines below 60 keV and in order to study spatial
homogeneity properties of NaI detectors, a dedicated calibration campaign was performed at
PTB/BESSY. Here, four NaI detectors (FM 01, FM 02, FM 03 and FM 04)\footnote{Detectors were delivered
to MPE for detector level calibration in batches of four, and shortly there after shipped to the 
US for system level calibration. Therefore, as the PTB/BESSY facility was only available
for a short time, only one batch of NaIs could be calibrated there.} were exposed to
a monochromatic X-ray beam with energy ranging from 10 to 60 keV, and the whole detector's surface was additionally
raster-scanned at different energies with a pencil beam perpendicular to the detector's surface. \\
In order to extend the BGO calibration range, another dedicated calibration campaign was carried 
out at the SLAC laboratory. Here, the BGO EQM detector\footnote{The BGO flight modules
were not available for calibration at the time of measurements, since they had already been shipped 
for system integration.} 
was exposed to three gamma-ray lines (up to 17.6 MeV) produced by the interaction of a proton beam 
of $\sim$340 keV, generated with a small Van-de-Graaff accelerator, with a LiF-target.
A checklist showing which detectors were employed at each detector-level calibration 
campaign is given in Table \ref{Tab_Num_Dec} (columns 4 to 6).
%
\begin{figure}[b!]
\centering
\begin{tabular}{cc}
\includegraphics[height=50mm,bb=0 0 792 612,clip]{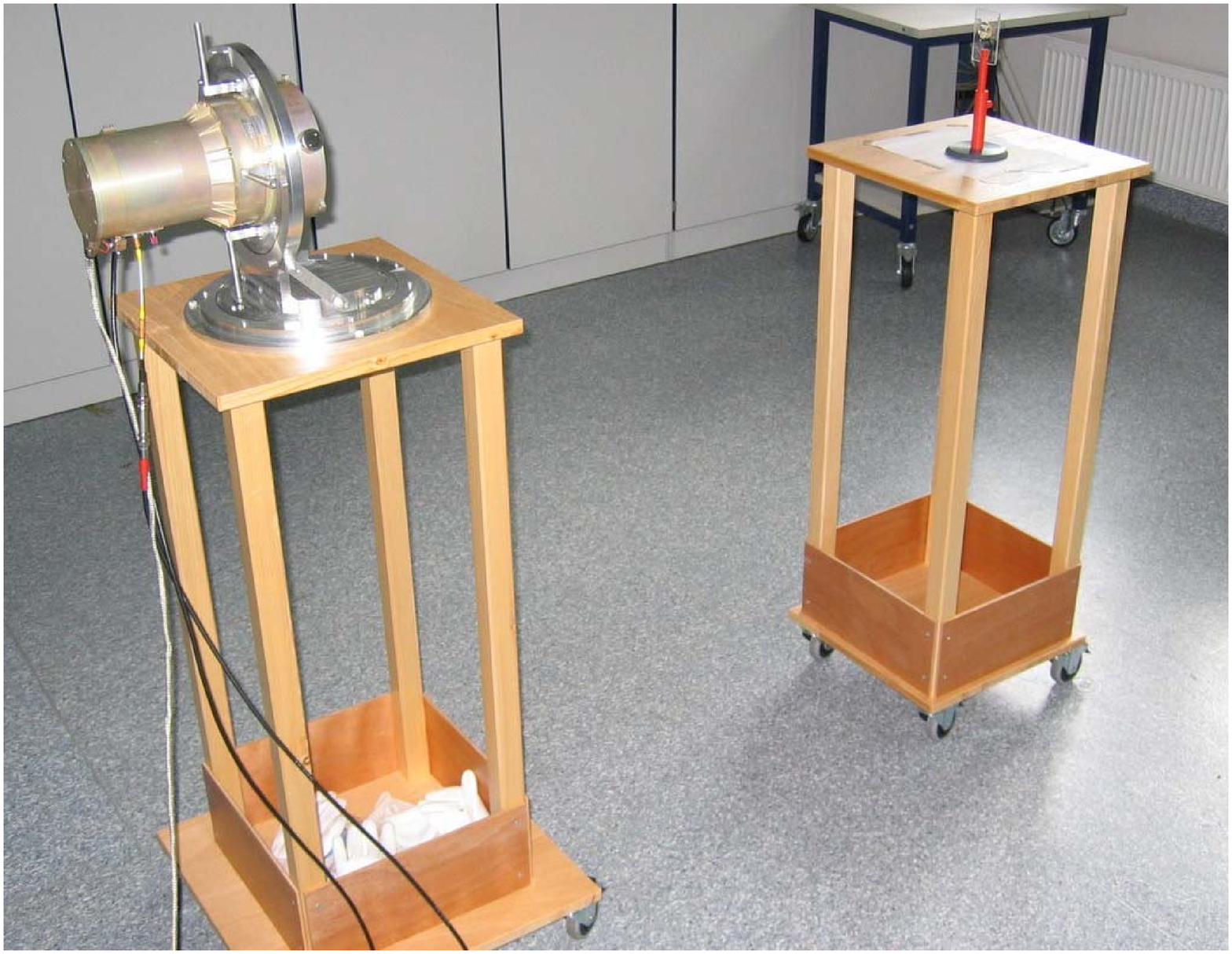} & 
\includegraphics[height=50mm,bb=0 0 612 792,clip]{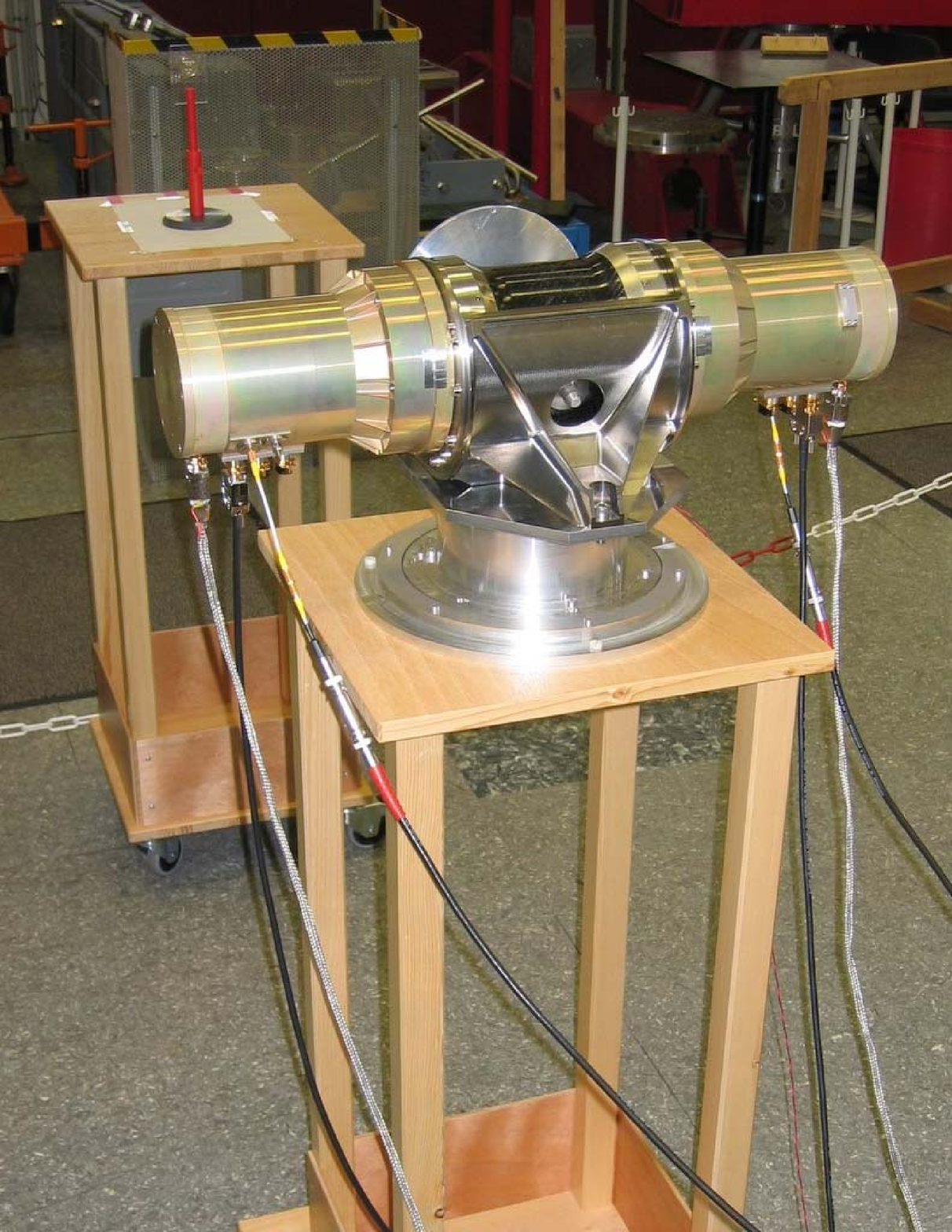}
\end{tabular}
\caption{
Detector holders with NaI ({\it left panel}) and BGO detectors ({\it right panel}) are positioned in front
of the radioactive source (on top of the red holder) on their wooden stands during the calibration at the MPE laboratory
}
\label{Fig_MPE_Lab}
\end{figure}
%
\subsection{Laboratory Setup and Calibration Instrumentation at MPE}\label{MPE_calib_setup}
The measurements performed at MPE resulted in an energy calibration with various radioactive 
sources and, in addition, a calibration of the angular response of the detectors at different 
incidence angles of the radiation. The detectors and the radioactive sources were
fixed on special holders which were placed on wooden 
stands above the laboratory floor to reduce scattering from objects close to them 
(see Fig. \ref{Fig_MPE_Lab}). The radioactive sources were placed almost always at the same 
distance ({\it d}) from the detector. The position of the detector's 
wooden stand with respect to the laboratory was never changed during measurements. 
Due to the unavailability of the flight DPU and PSB, commercial HV and LV power supplies
were used and the data were read out by a Breadboard DPU.

The determination of the angular response of the detectors was achieved in the following way.
The center of the NaI detector calibration coordinate system was chosen 
at the center of the external surface of the Be-window of the detector unit, 
with the X axis pointing toward the radioactive source, the Y axis pointing 
toward left, and Z axis pointing up (see Fig. \ref{Fig_Lab_sim}, {\it left
panel}). The detectors were mounted on a specially developed holder in such a 
way that the front of the Be-window was parallel to the Y/Z plane (if the detector 
is pointed to the source; i.e. 0$^o$ position) and so that detectors could be 
rotated around two axes in order to achieve all incidence angles of the radiation.
The detector rotation axes were the Z-axis (Azimuth) and around the X-axis (roll). 
For BGO detectors, the mounting was such that the very center of the detector (center of crystal) 
was coincident with the origin of the coordinate system and the 0$^o$ position was defined as the 
long detector axis coincident with the Y-axis.
The BGO detectors were only rotated around the Z-axis, and no roll angles were measured
in this case.
%
%
\begin{figure}[t!]
\centering
\begin{tabular}{cc}
\includegraphics[width=84mm,bb=81 0 873 450,clip]{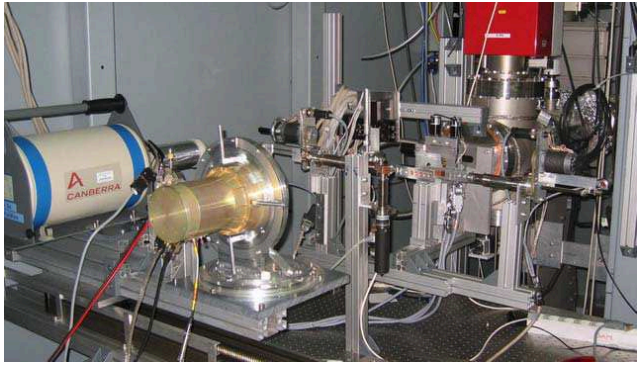}
\end{tabular}
\caption{NaI FM 04 detector photographed inside the measurement cave of the BAMline during the low-energy 
calibration campaign at the electron storage ring BESSY II in Berlin. 
The HPGe detector is located left of the GBM detector. Both are mounted on the XZ table which was moved
by step motors during the scans. The beam exit window of the BAMline is located
below the red box visible at the top right corner. 
The Cu and Al filters holders are placed horizontally between the window and the detectors
}
\label{Fig_Lab_BESSY}
\end{figure}
%
\subsection{NaI Low-Energy Calibration at PTB/BESSY}\label{Sec_BESSY}
The calibration of the NaI detectors in the low photon energy range 
down to 10 keV was performed with monochromatic synchrotron radiation 
with the support of the PTB. A pencil beam of about 
0.2 x 0.2 mm$^{\,2}$ was extracted from a wavelength-shifter beamline, 
the ``BAMline'' \cite{RIE05}, at the electron storage ring BESSY II,
which is equipped with a double-multilayer monochromator (DMM) 
and a double-crystal monochromator (DCM) \cite{GOR01}. In the photon
energy range from 10 keV to 30 keV DCM and DMM were operated in series 
to combine the high resolving power of the DCM with the high spectral 
purity of the DMM. Above 30 keV, a high spectral purity with higher order 
contributions below 10$^{-4}$ was already achieved by the DCM alone. 
The tunability of the photon energy was also used to investigate the 
detectors in the vicinity of the Iodine K-edge at 33.17 keV.

The absolute number of photons in the pencil beam was independently 
determined by two different methods: firstly by taking at each photon 
energy a spectrum with a high-purity germanium detector (HPGe) for which 
a quantum detection efficiency (QDE) of unity had been determined earlier, 
and secondly by using silicon photodiodes which in turn had been 
calibrated against PTB primary detector standards such as a cryogenic 
radiometer and a free-air ionization chamber \cite{KRU06}. As these 
photodiodes are operated in the photovoltaic mode, the photon fluxes 
had to be about four orders of magnitude higher than for the counting 
detectors. Different pairs of Cu and Al filters were designed for 
different photon energy ranges so that the transmittance of one filter 
was in the order of 1 \%\ which can easily be measured. Two identical 
filters were used in series to achieve the required reduction in flux 
by four orders of magnitude. A picture of the calibration
setup is shown in Fig. \ref{Fig_Lab_BESSY}.

The effective area of the detectors as a function of the photon energy 
was determined by scanning the detectors at discrete locations
in x- and y-direction over the 
active area while the pencil beam was fixed in space. During the scan, 
the intensity was monitored with a photodiode operated in transmission. 
The effective area is just the product of the average QDE and the active area. 
In addition, the spatial homogeneity of the QDE was determined by these 
measurements (see Section \ref{Effective_Area}).

The measurements presented in this paper were recorded at 18 different 
energies, namely from 10 to 20 kev in 2 keV steps, from 30 to 37 keV in 
1 keV steps and at 32.8, 40, 50 and 60 keV. These accurate measurements 
allowed to exactly determine the low-energy behavior of the channel-energy 
relation of the NaI detectors (see Section \ref{NaI CE}) and to fine tune the 
energy range around the Iodine K-edge at 33.17 keV (see Section \ref{Iodine_Reg}). 
Moreover, three rasterscans of the detector's surface were performed at 10, 
36 and 60 keV in order to study the detectors' spatial
homogeneity (see Section \ref{QDE_Uni} for more details).
%
\begin{figure}[t!]
\centering
\begin{tabular}{cc}
\includegraphics[height=50mm,bb=119 576 248 707,clip]{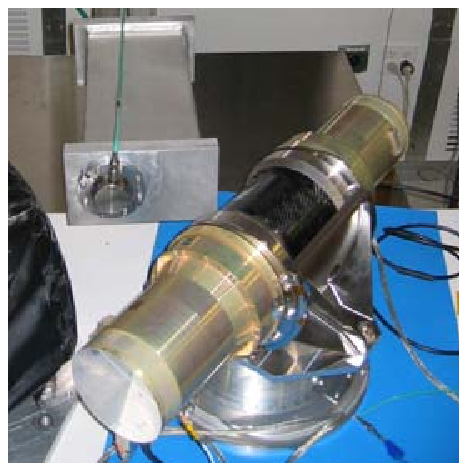} &                       
\includegraphics[height=50mm,bb=55 55 556 556,clip]{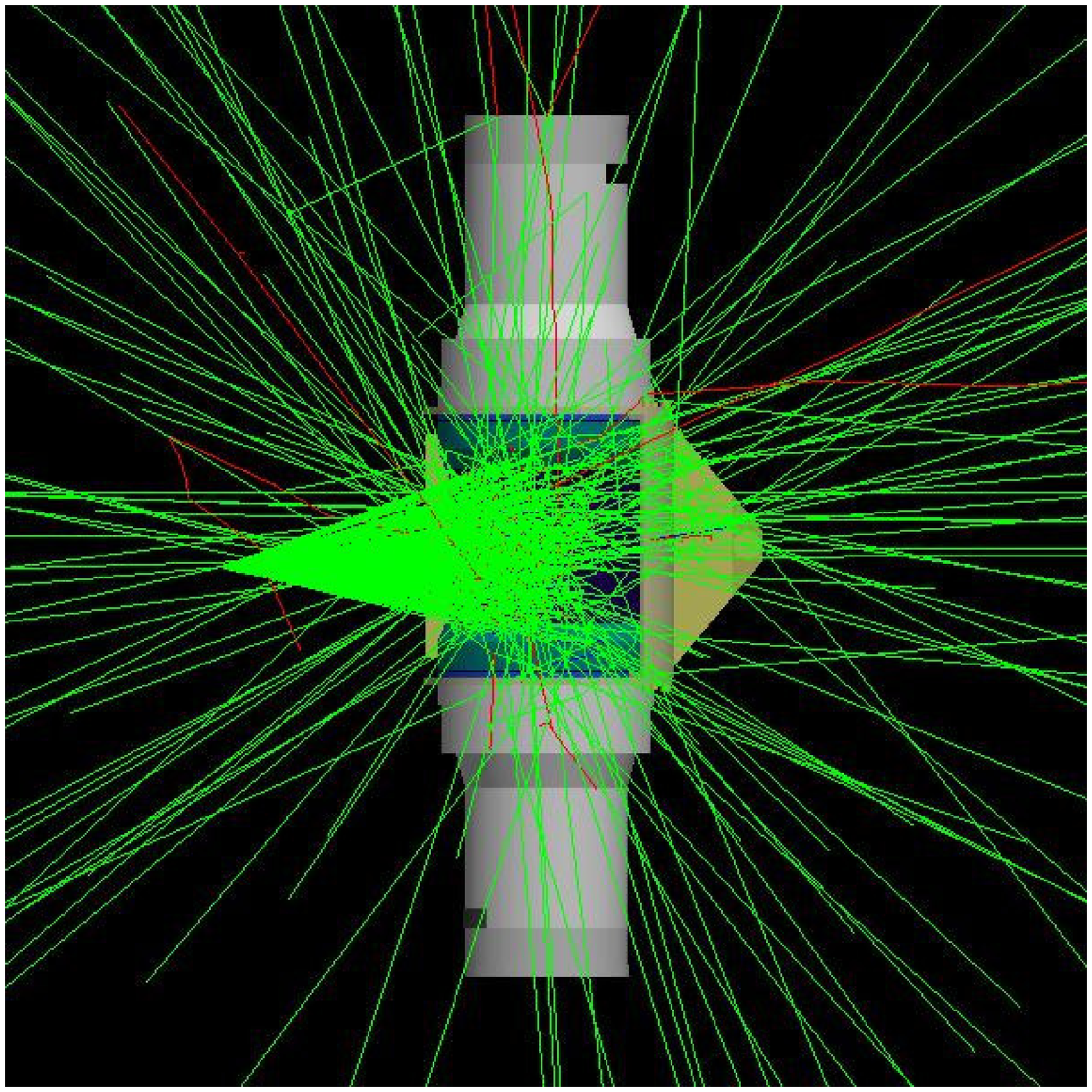}
\end{tabular}
\caption{
BGO EQM detector photographed in the SLAC laboratory during the high-energy 
calibration campaign ({\textit{left panel}}). The grey box on the left is
the end of the proton beam, inside which the LiF target was placed in order to
react and produce the desired gamma lines (see Eq. \ref{eq_SLAC_1} and \ref{eq_SLAC_2}).
The \textit{right panel} shows a simulation of the gamma-ray interaction with the detector.
Only gamma-rays whose first interaction is within the detector crystal are shown for clarity
}
\label{Fig_Lab_SLAC_sim}
\end{figure}
%
\subsection{BGO High-Energy Calibration at SLAC}\label{Sec_SLAC}
In order to better constrain the channel-energy relation and the energy resolution at energies higher than
4.4 MeV, an additional high-energy calibration of the BGO EQM detector was performed at SLAC with a small 
electrostatic Van-de-Graaff accelerator \cite{KIE07}. This produces a proton 
beam up to $\sim$350 keV and was already used to verify the 
LAT photon effective area at the low end of the Fermi energy range (20 MeV).
When the proton beam produced by the Van-de-Graaff accelerator 
strikes a LiF target, which terminates the end of the vacuum pipe (see Fig. \ref{Fig_Lab_SLAC_sim}, {\it left panel}),
gammas with energies of 6.1 MeV, 14.6 MeV, and 17.5 MeV are produced via the reactions
\begin{eqnarray}
p\,(\sim 340 keV)\;+\;^7 Li & \rightarrow \, ^8 Be\,(1+)^* \;\rightarrow & ^8 Be \; + \; \gamma\,(14.6 \, or \, 17.5 \, MeV)     \label{eq_SLAC_1} \\
p\,(\sim 340 keV) \;+\; ^{19} F  & \rightarrow \,  ^{16} O^* \,+\,\alpha  \; \rightarrow & ^{16} O \; + \; \gamma\,(6.1 \, MeV)  \label{eq_SLAC_2} \, .
\end{eqnarray}
The highly excited 17.5 MeV state of $^8$Be is created by protons in a resonance 
capture process at 340 keV on $^7$Li (see Eq. \ref{eq_SLAC_1}). At lower energies, photons 
are still produced from the Breit-Wigner tail ($\Gamma$ =12 keV) of the $^8$Be$^*$ resonance. 
The narrow gamma-ray line at 17.5 MeV is produced by the transition to the $^8$Be ground state, 
in which the quantum energy is determined by $h\nu$ = $Q$ + 7/8 $E_p$, where Q = 17.2 MeV is the
energy available from the mass change and $E_p$ = 340 keV is the proton beam energy. 
The gamma-ray line observed at 14.6 MeV, which corresponds to transitions to the 
first excited state of $^8$Be, is broadened with respect to the experimental 
resolution, because of the short lifetime of the state against decay into two 
alpha-particles. Finally, Equation \ref{eq_SLAC_2} shows that 6.1 MeV gamma-rays 
are generated when the narrow ($\Gamma$ =3.2 keV ) $^{16}$O resonance at 340 keV is hit.

For performing the measurements, the EQM detector was placed as close as possible to the LiF-target
at an angle of $\sim$45$^o$ with respect to the proton-beam line,
in order to guarantee a maximized flux of the generated gamma-rays.
Unfortunately, measurements for the determination of the detector's effective area could not
be obtained, since the gamma-ray flux was not closely monitored.
%
\begin{figure}[b!]
\centering
\begin{tabular}{cc}
\includegraphics[height=48mm,bb=121 78 439 395,clip]{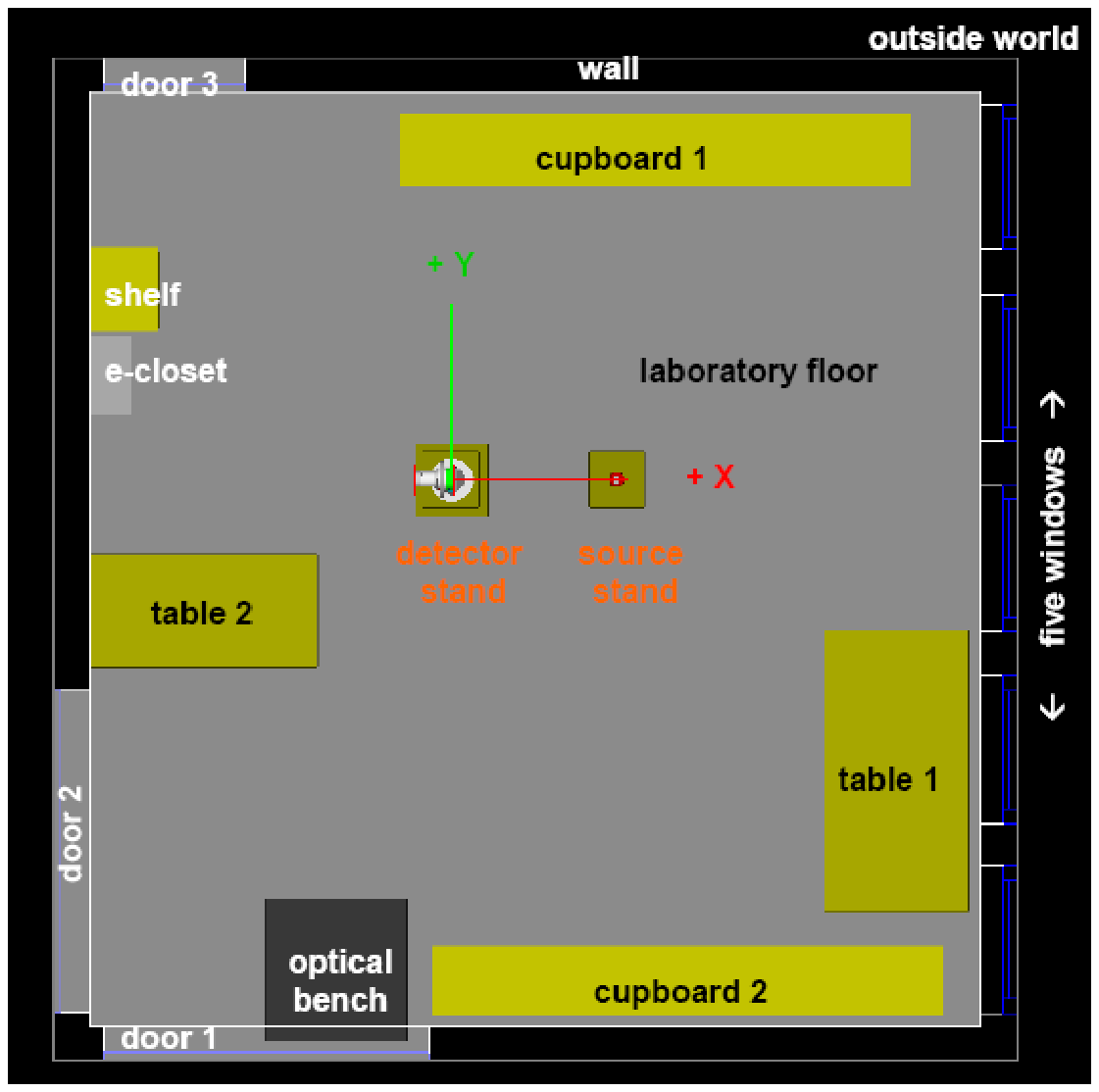} & 
\includegraphics[height=48mm,bb=1 70 713 620,clip]{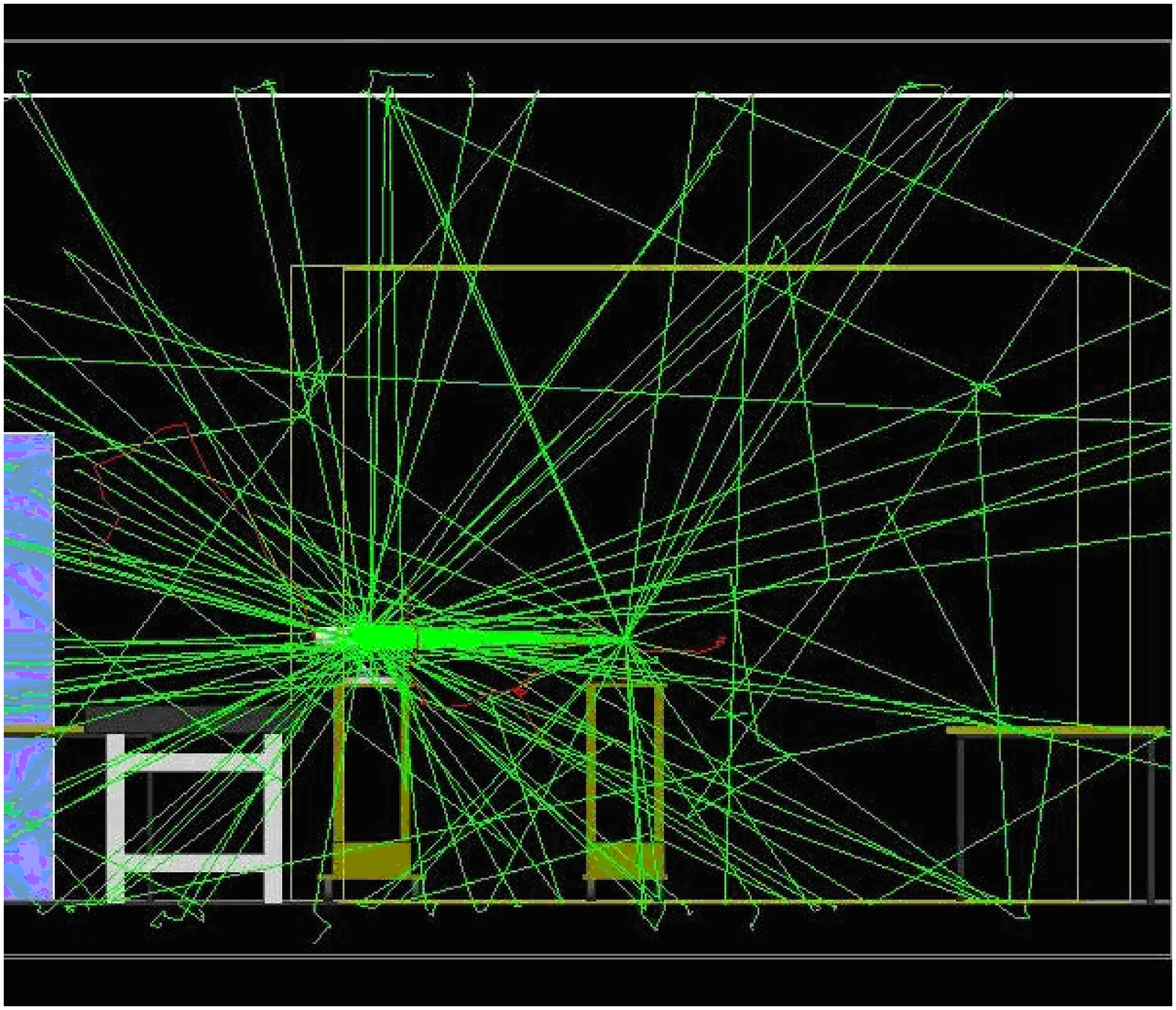}
\end{tabular}
\caption{Simulation of the laboratory environment. The {\it left panel} shows in a top view of 
the laboratory the components of the simulation model of all objects which were present
during the calibration campaign. Also shown is the coordinate system adopted (X and Y 
axis of the right-handed system; +Z axis pointing upward). The {\it right panel} shows an
example of the simulated scattering of the radiation in the laboratory. 
In a view from the -Y axis, the path of the first 100 photons interacting with the 
detector are shown. The radioactive source emitted radiation of 1.275 MeV isotropically. 
The major part of the detected photons is directly incident, but a significant 
fraction is scattered radiation by the laboratory environment (see Table \ref{Tab_sim_contr})
}
\label{Fig_Lab_sim}
\end{figure}
%
\subsection{Simulation of the laboratory and the calibration setup at MPE}\label{Sec_Lab_Sim}
In order to simulate the recorded spectra of the calibration campaign at MPE to gain confidence 
in the simulation software used, a very detailed model of the environment in which the calibration 
took place had to be created. The detailed modeling of the laboratory was necessary as all scattered 
radiation from the surrounding material near and far had to be included to realistically simulate 
all the radiation reaching the detector\footnote{An important argument driving the decision 
not to use a collimator for measurements with radioactive sources was the fact that
the simulation of the laboratory environment with all it's scattering represented a 
necessary and critical test for the simulation software, which later had to include the spacecraft \cite{WAL07}.}.
Background measurements with no radioactive sources present were taken to subtract the ever-present 
natural background radiation in the laboratory.  However, the source-induced ``background'' radiation created by 
scattered radiation of the non-collimated radioactive sources had to be included in the simulation 
to enable a detailed comparison with the measured spectra. 

\begin{table}[t!]
\centering
\caption{Contribution of simulated laboratory components to the detected photons}
\begin{tabular}{c|ccc|c}
\hline\noalign{\smallskip}
  &   \multicolumn{3}{c|}{NaI}	&		BGO \\
\noalign{\smallskip}\hline\noalign{\smallskip}
component	            &  22 keV	    &    122 keV   & 	  1.275 MeV	 &  4.43 MeV   \\
\noalign{\smallskip}\hline\noalign{\smallskip}
direct incidence      &	  94.0 \%\	&    91.0 \%\	 &     75.0 \%\	 &   70.0 \%\  \\				
scattered rad. total	&    6.0 \%\	&     9.0 \%\	 &     25.0 \%\	 &   30.0 \%\  \\
walls	                &  < 0.1 \%\	&     0.6 \%\	 &     12.0 \%\	 &   13.0 \%\  \\
source holder	        &    4.6 \%\	&     7.5 \%\	 &      3.0 \%\	 &    2.0 \%\  \\
source stand	        &    0.1 \%\  &   < 0.1 \%\	 &    < 0.1 \%\	 &  < 0.1 \%\  \\
detector stand        &	 < 0.1 \%\	&   < 0.1 \%\	 &      2.0 \%\	 &  < 0.1 \%\  \\
floor	                &  < 0.1 \%\	&     0.8 \%\	 &      8.0 \%\	 &   15.0 \%\  \\
other furniture	      &  < 0.1 \%\	&   < 0.1 \%\	 &    < 0.1 \%\	 &  < 0.1 \%\  \\
air	                  &    1.3 \%\	&   < 0.1 \%\	 &    < 0.1 \%\	 &  < 0.1 \%\  \\
\noalign{\smallskip}\hline
\end{tabular}
\label{Tab_sim_contr}
\end{table}
The detailed modeling of the calibration setup of the MPE laboratory 
was performed using the GEANT4 -based GRESS\footnote{GEANT4 is a suite 
of high energy interaction simulation tools created at CERN \cite{AGO03}. 
GRESS is the General Response Simulation System which is developed at LANL. 
It is based on a modified (extended) version of GEANT4 \cite{KIP04}.} 
simulation software provided by the collaboration team based at the Los Alamos 
National Laboratory (LANL, USA), who also provided the software model of the detectors \cite{HOO08b}.
The modeling of the whole laboratory included laboratory walls (concrete), 
windows (aluminum, glass), doors (steel), tables (wood, aluminum), cupboards (wood), 
a shelf (wood), the electricity distributor closet (steel) the optical bench (aluminum, 
granite) and the floor (PVC). Moreover, detector and source stands (wood), source holder 
(PVC, acrylic) and detector holder (aluminum) were modeled in great detail (see Fig. \ref{Fig_Lab_sim}, {\it left panel}). 
A summary of the comparison of the measurements and the simulation, with respect to the 
influence of the various components of the calibration environment is given in Table \ref{Tab_sim_contr} 
and a sample plot of the scattered radiation is given in the {\it right panel} of Fig. \ref{Fig_Lab_sim}.

Additional simulations of the other calibration campaigns, in particular for the PTB/ BESSY one, are planned. 
In the case of SLAC measurements, the simulation tools were only used to determine the ratios
between full-energy peaks and escape peaks (see Fig. \ref{Fig_Lab_SLAC_sim}, {\it right panel}): 
no further simulation of the calibration setup is foreseen.
%
%
%
\section{Calibration Data Analysis and Results}\label{Line_Ana_Proc}
\subsection{Processing of Calibration Runs}\label{Proc_Runs}
During each calibration campaign, all spectra measured by the GBM detectors were recor-ded together with the 
information necessary for the analysis. Shortly before or after the collection of data runs, 
additional background measurements were recorded for longer periods. 
Every run was then normalized to an exposure time of 1 hour, and the background was subsequently subtracted from the data. 
In the case of measurements performed at PTB/BESSY,
natural background contribution could be neglected due to the very high beam intensities
and to the short measurement times.
%
%
\begin{figure}[t!]
\centering
\begin{tabular}{cc}
\includegraphics[height=42mm,bb=0 10 590 460,clip]{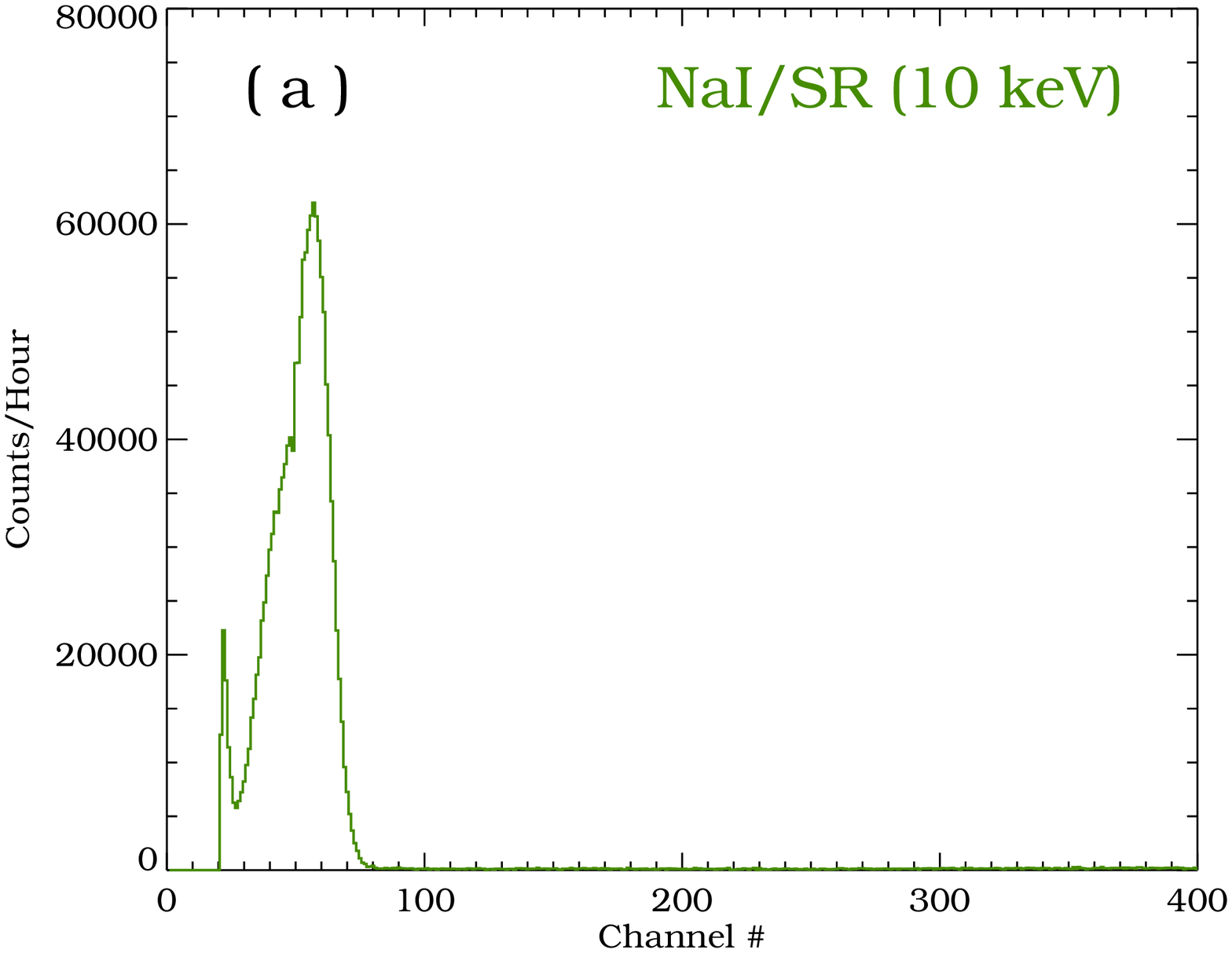} &  					  
\includegraphics[height=42mm,bb=0 10 590 460,clip]{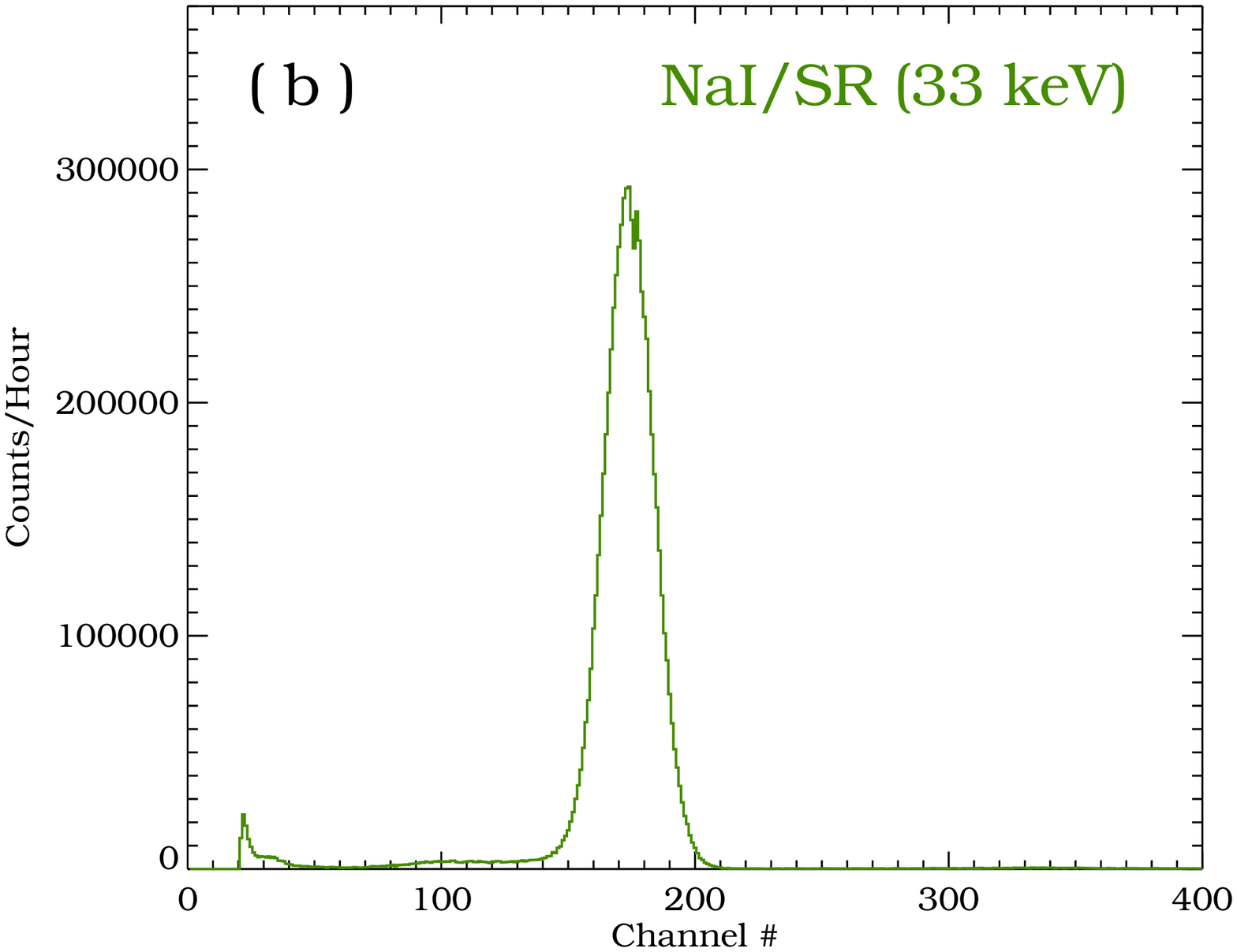} \\
\includegraphics[height=42mm,bb=0 10 590 460,clip]{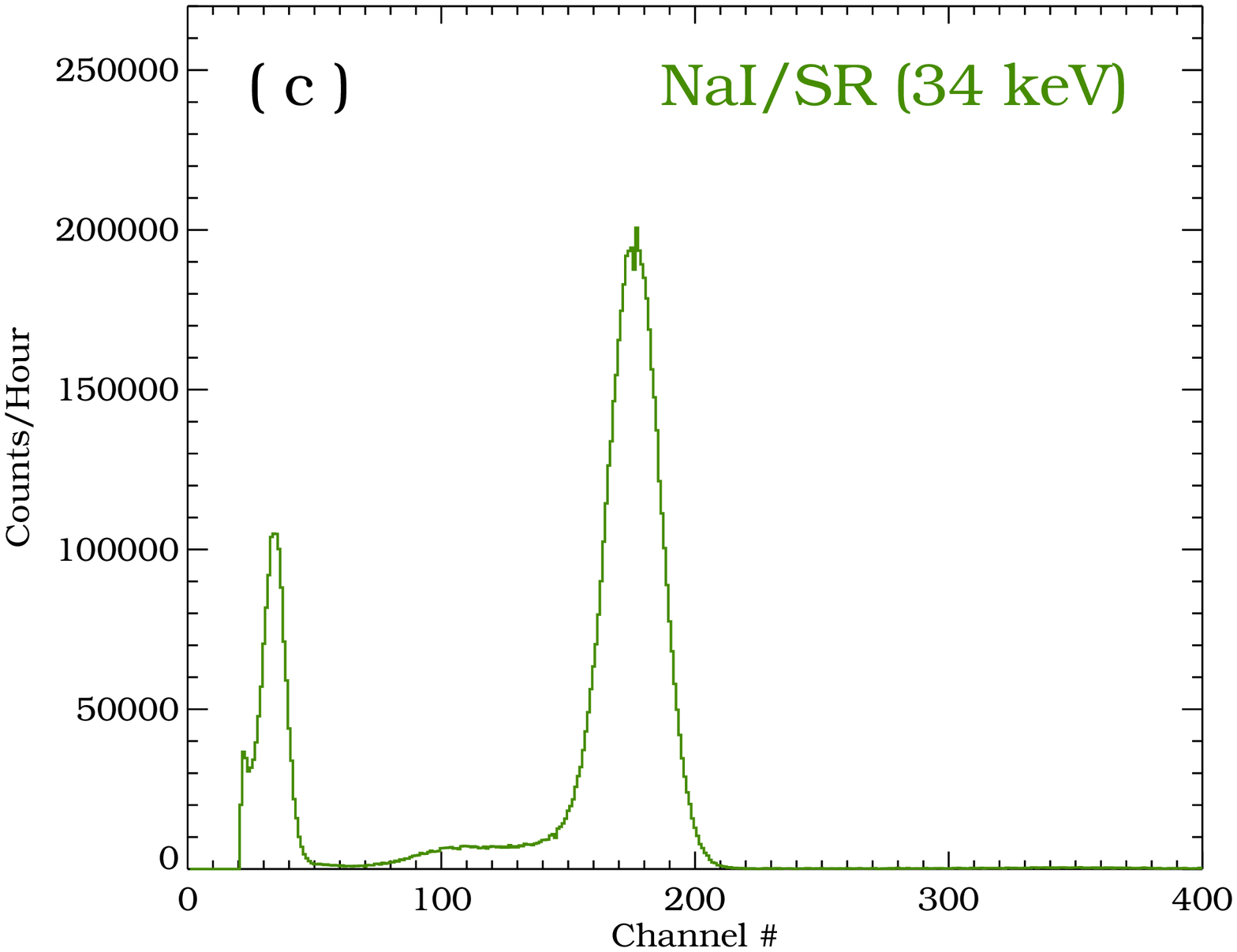} &    				
\includegraphics[height=42mm,bb=0 10 590 460,clip]{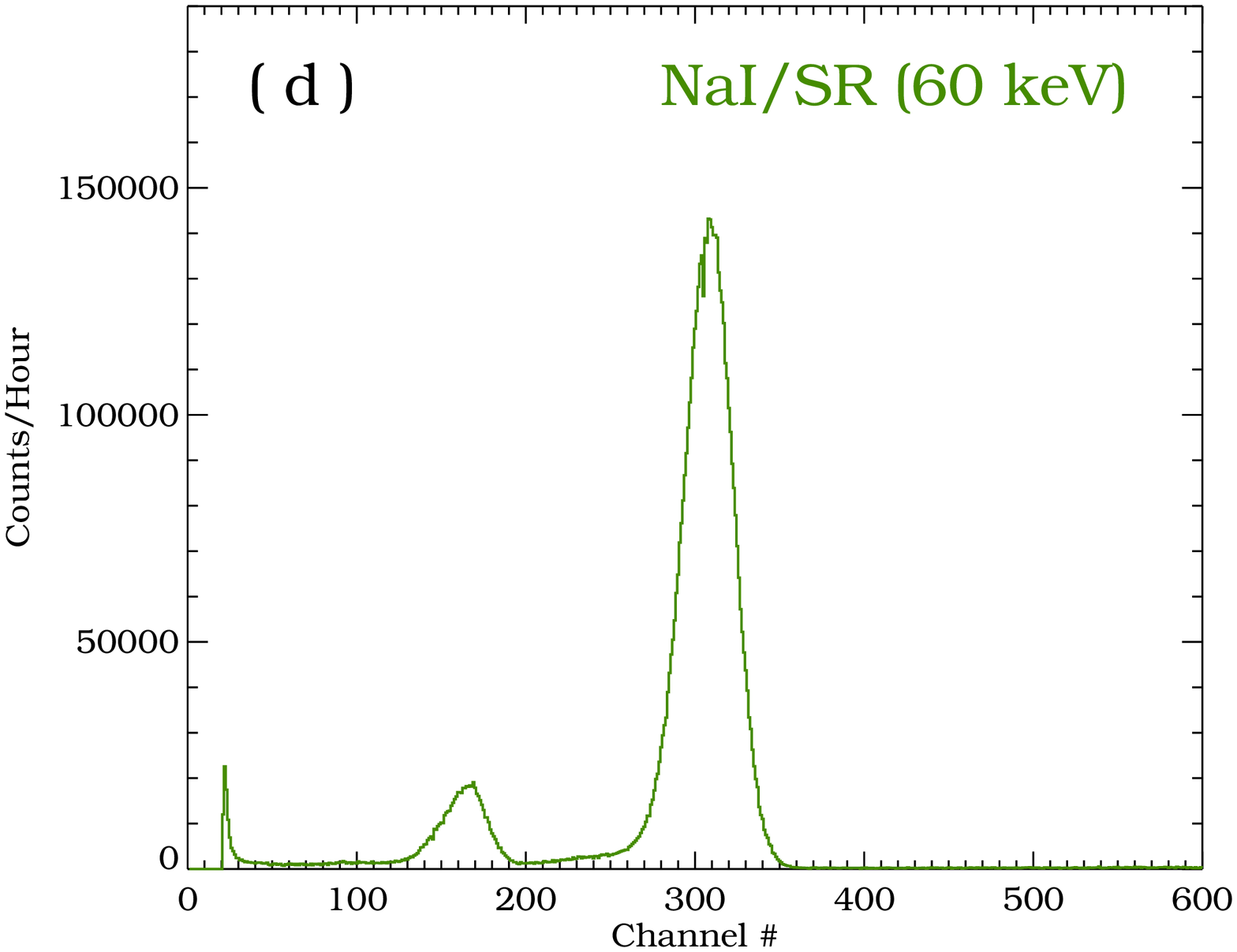}
\end{tabular}
\caption{Spectra measured with monochromatic synchrotron radiation (SR) at PTB/BESSY 
with detector NaI FM 04.
Results for four different photon energies are shown:
(a) 10 keV, 
(b) 33 keV, 
(c) 34 keV, and 
(d) 60 keV. 
The two top panels ({\it a} and {\it b}) display spectra collected
below the Iodine K-edge energy (i.e. < 33.17 keV). Above this energy ({\it panels c} and {\it d}),
the characteristic Iodine escape peak is clearly visible to the left of the full-energy peak
}
\label{figNaIspectra_BESSY}
\end{figure}
%
%
\begin{figure}[p!]
\centering
\begin{tabular}{cc}
\includegraphics[height=42mm,bb=0 10 590 460,clip]{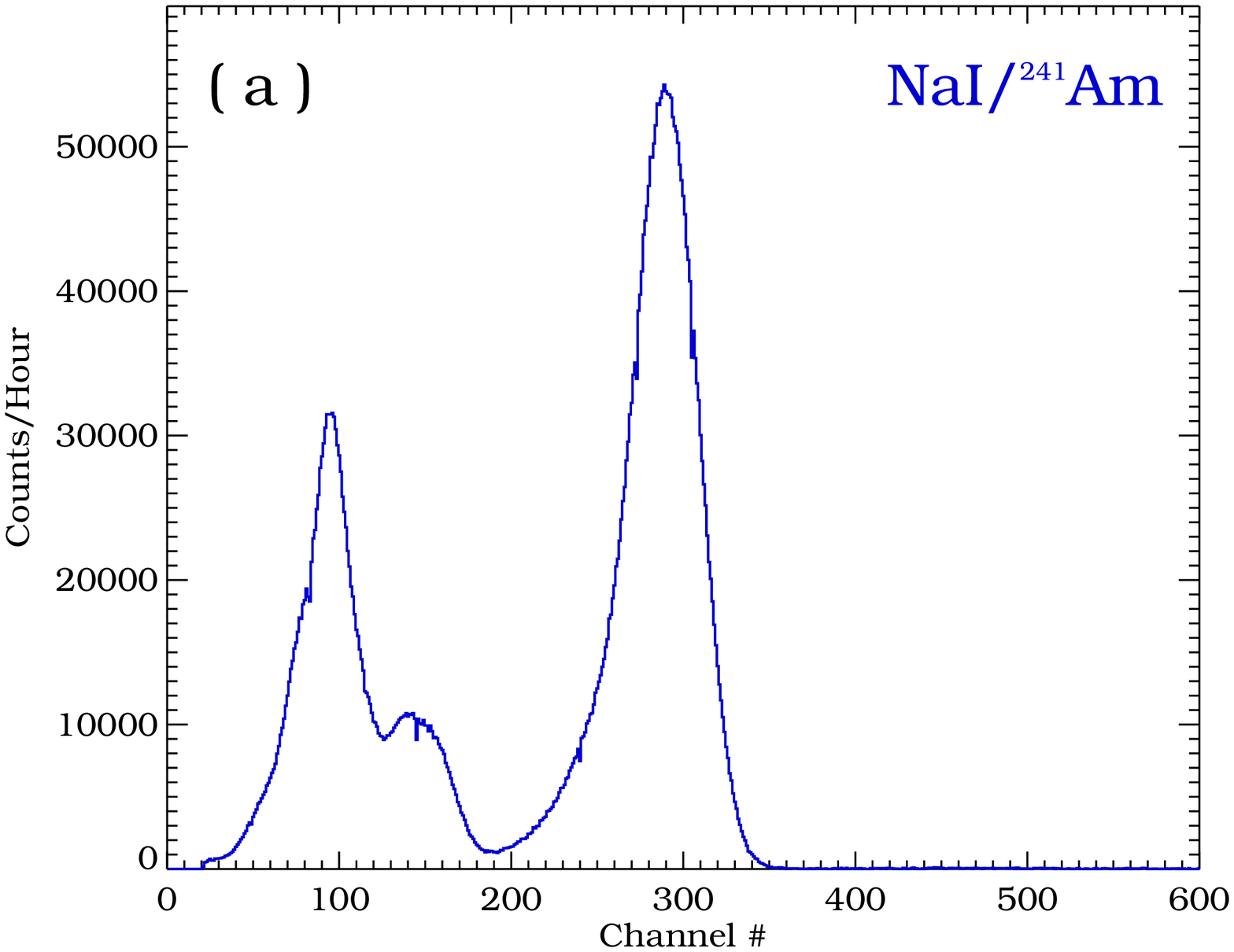}  &    
\includegraphics[height=42mm,bb=0 10 590 460,clip]{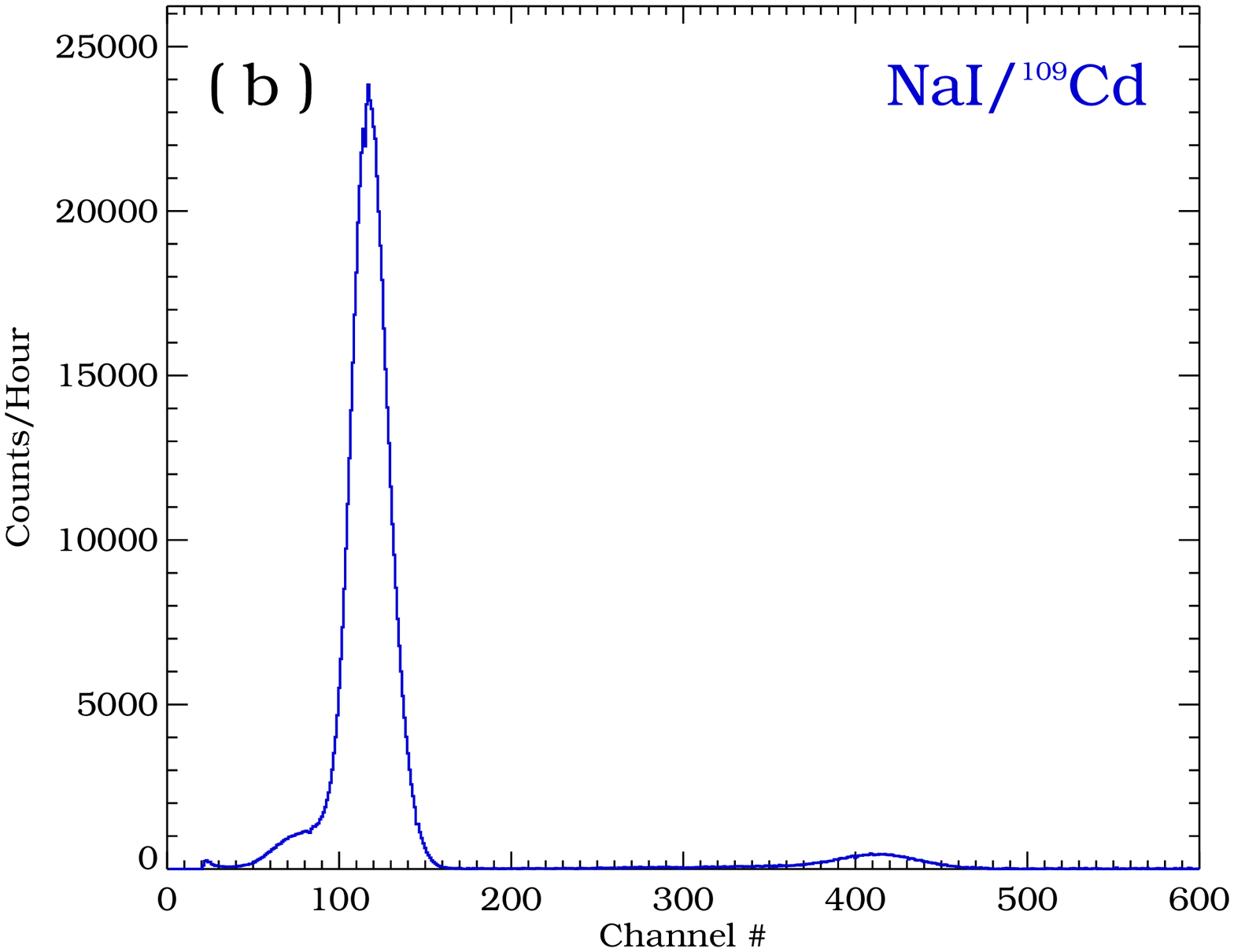}  \\
\includegraphics[height=42mm,bb=0 10 590 460,clip]{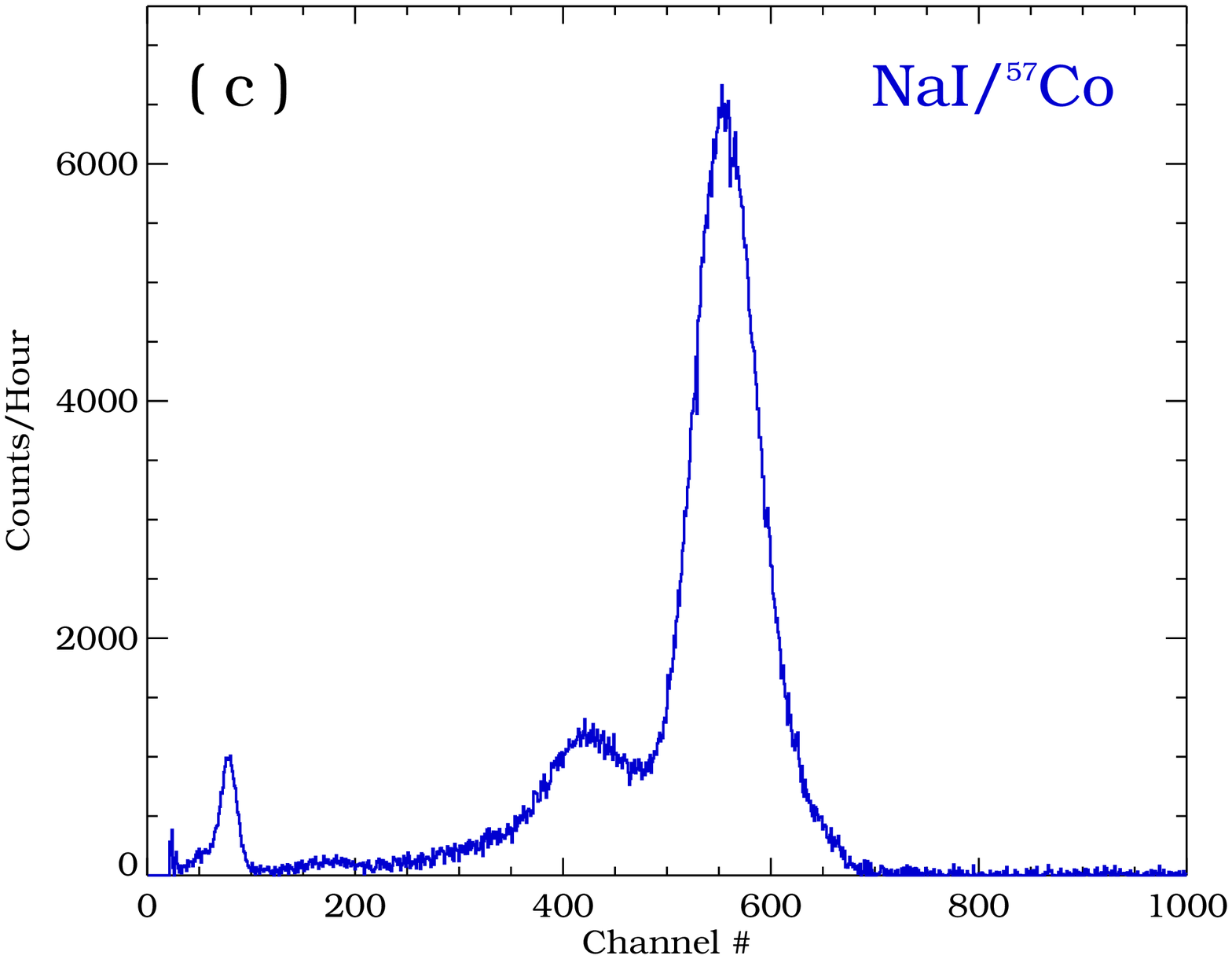}   &    
\includegraphics[height=42mm,bb=0 10 590 460,clip]{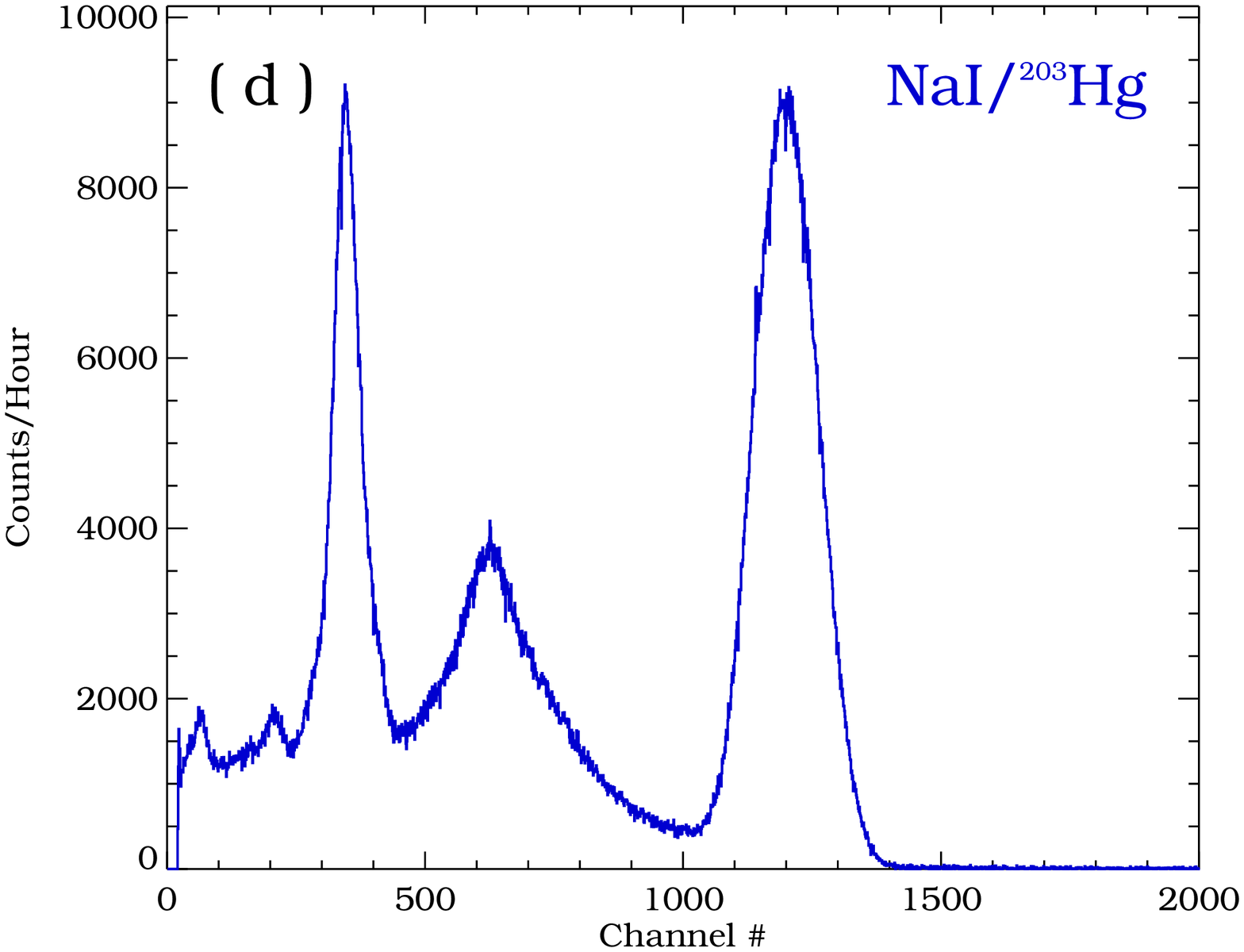}  \\
\includegraphics[height=42mm,bb=0 10 590 460,clip]{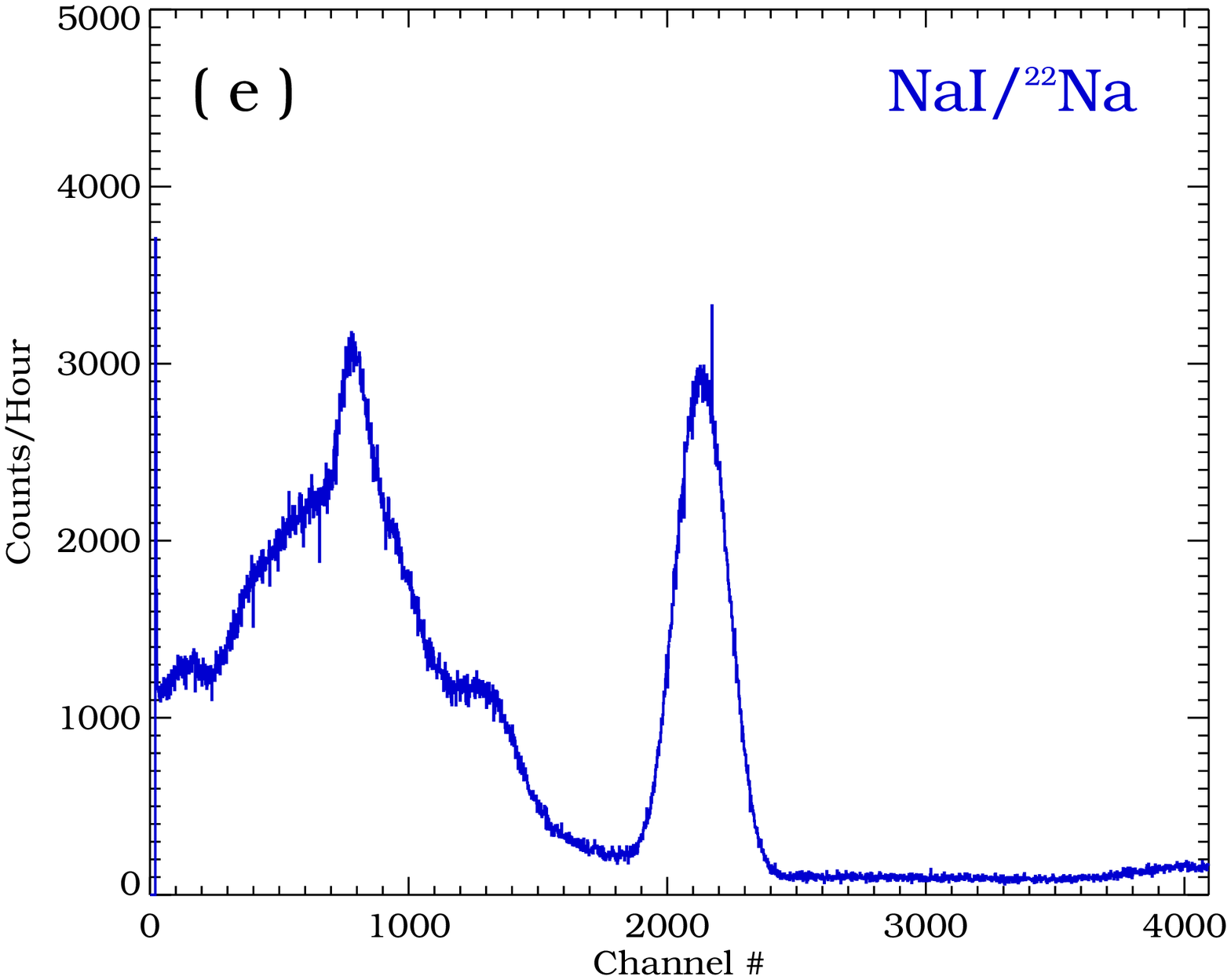}   &  
\includegraphics[height=42mm,bb=0 10 590 460,clip]{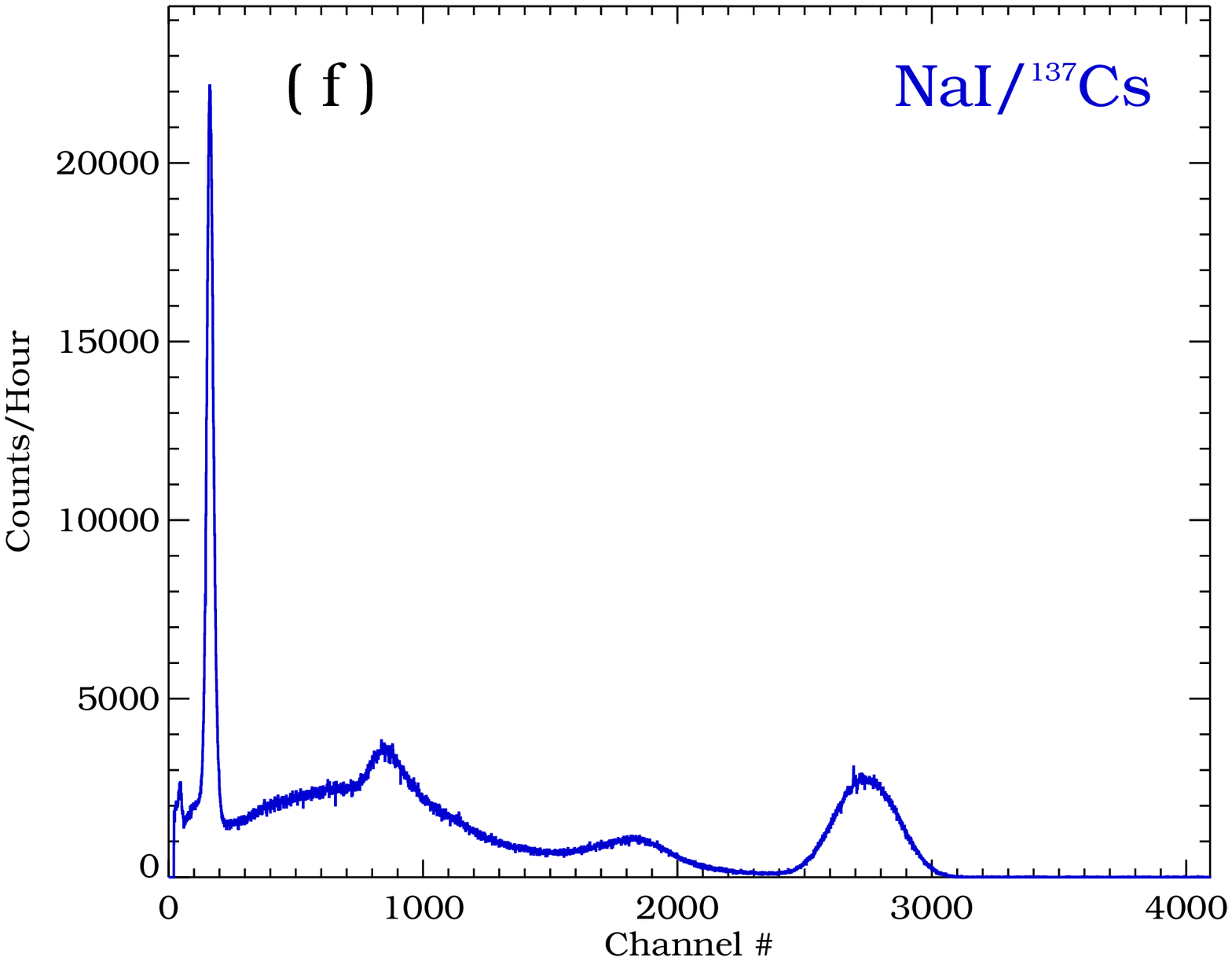}  \\
\includegraphics[height=42mm,bb=0 10 590 460,clip]{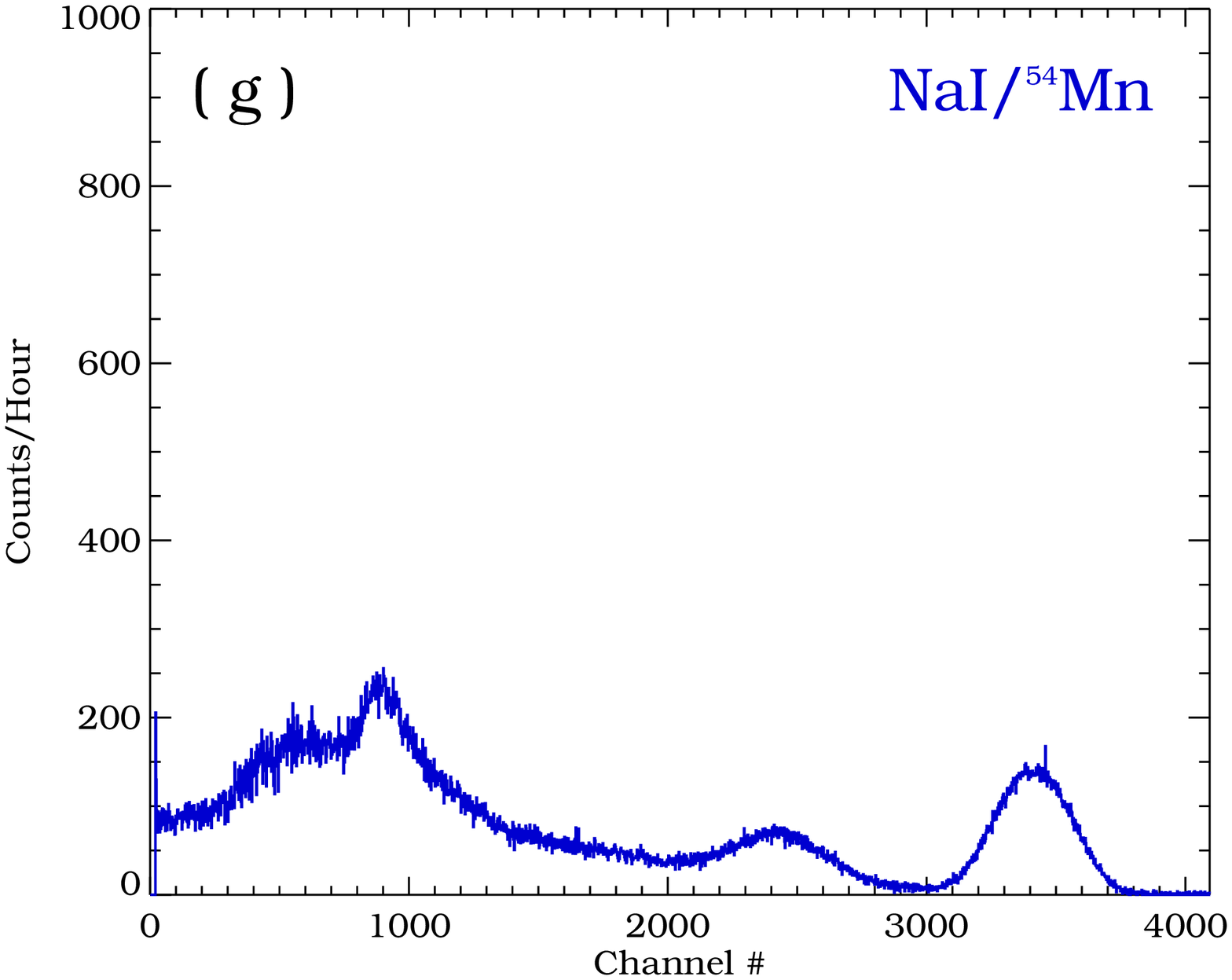}   &   
\includegraphics[height=42mm,bb=0 10 590 460,clip]{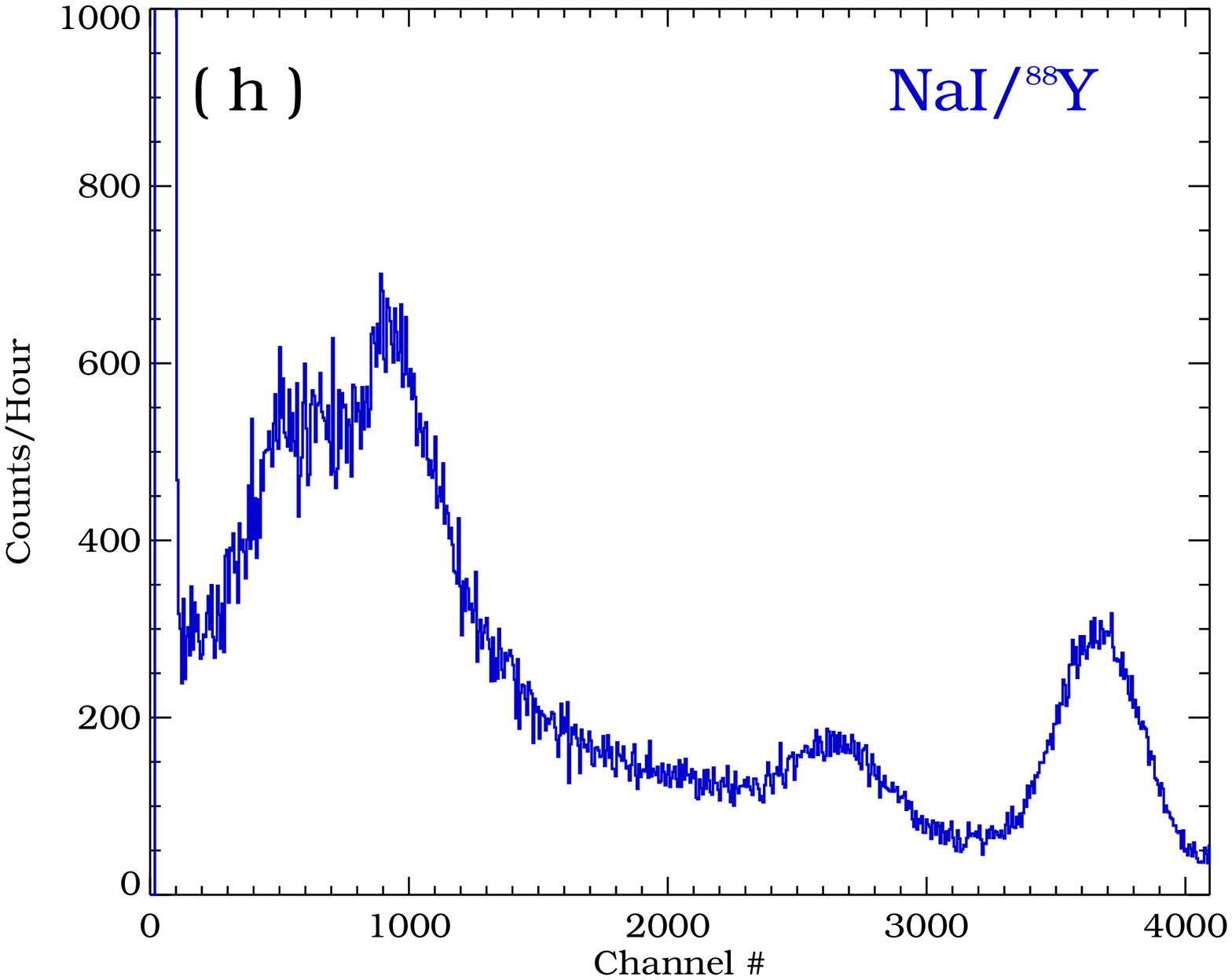}
\end{tabular}
\caption{Normalized background-subtracted spectra measured
at MPE with detector NaI FM 04 from the following radioactive sources:
(a) $^{241}$Am,
(b) $^{109}$Cd,
(c) $^{57}$Co,
(d) $^{203}$Hg,
(e) $^{22}$Na, 
(f) $^{137}$Cs,
(g) $^{54}$Mn, and
(h) $^{88}$Y
}
\label{figNaIspectra}
\end{figure}
%
%
\begin{figure}[p!]
\centering
\begin{tabular}{cc}
\includegraphics[height=42mm,bb=0 10 590 460,clip]{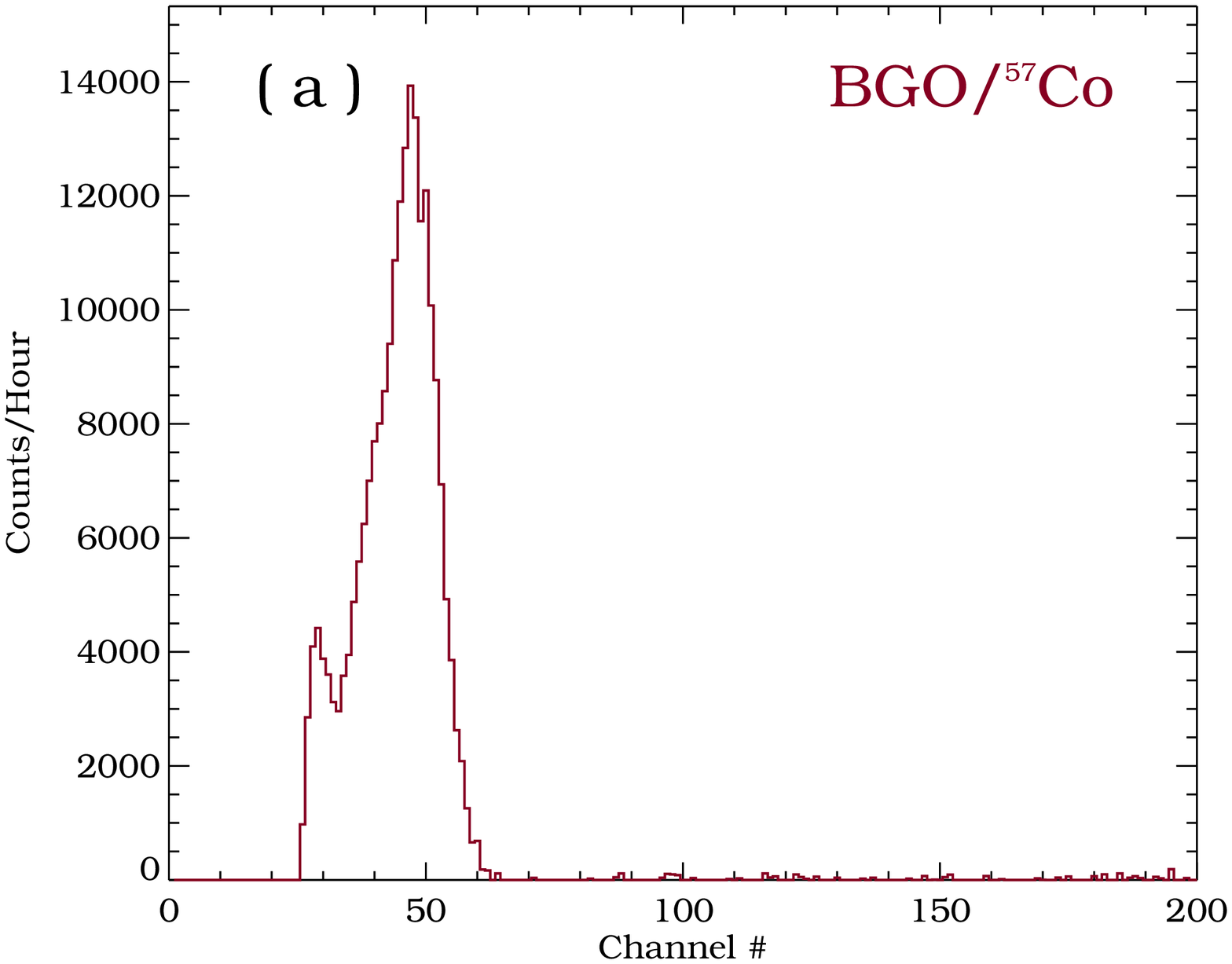}  &            
\includegraphics[height=42mm,bb=0 10 590 460,clip]{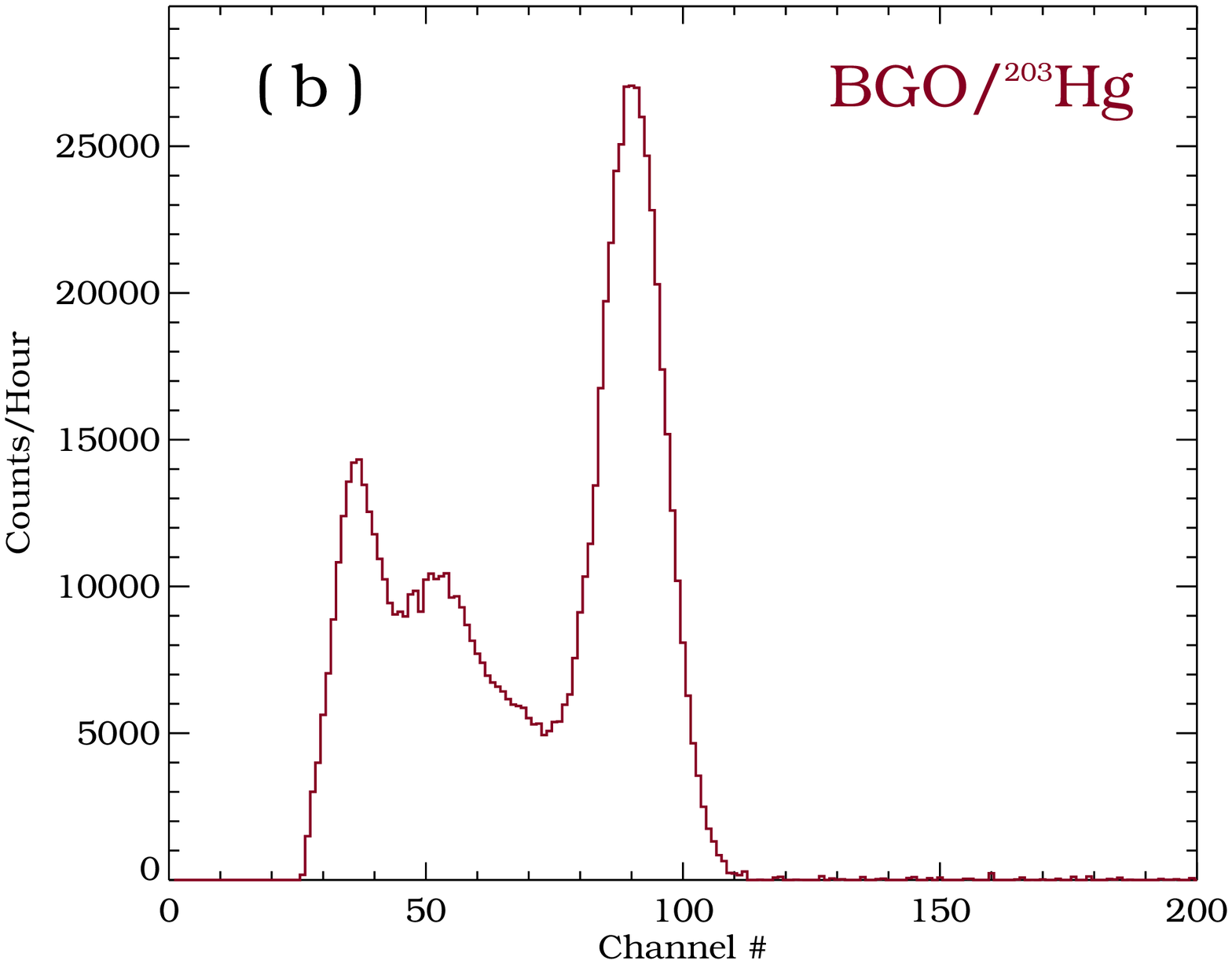}  \\
\includegraphics[height=42mm,bb=0 10 590 460,clip]{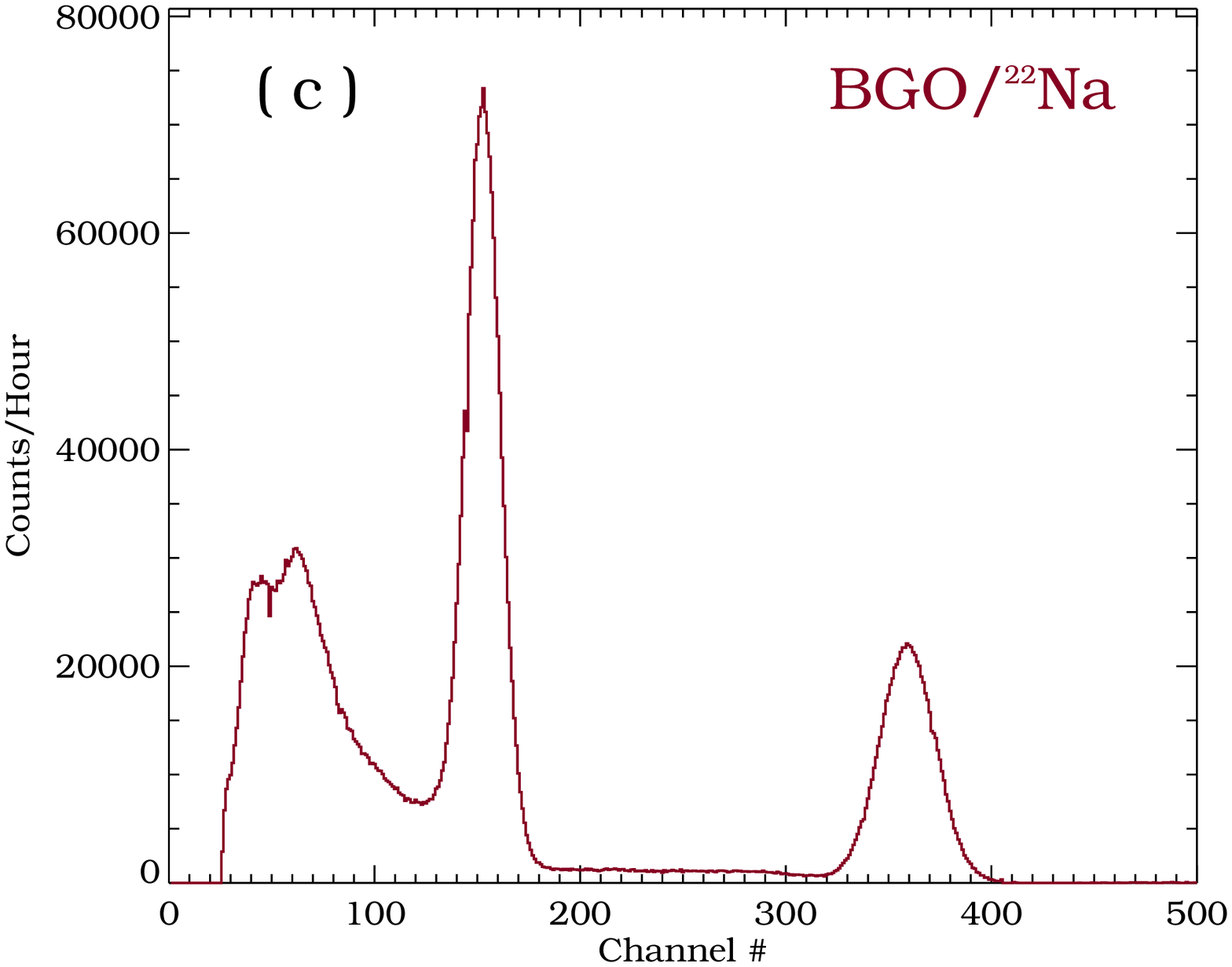}  &    
\includegraphics[height=42mm,bb=0 10 590 460,clip]{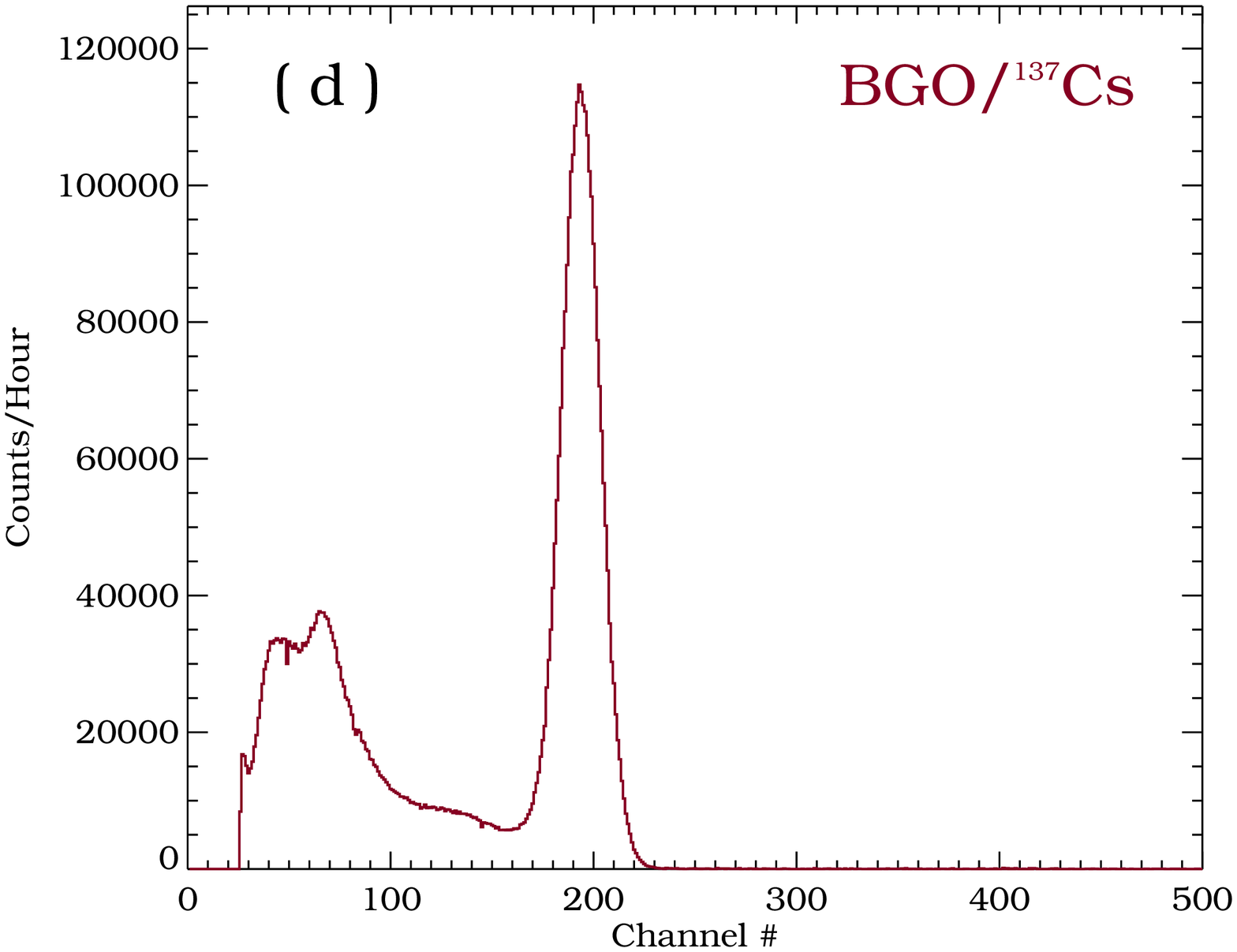}  \\
\includegraphics[height=42mm,bb=0 10 590 460,clip]{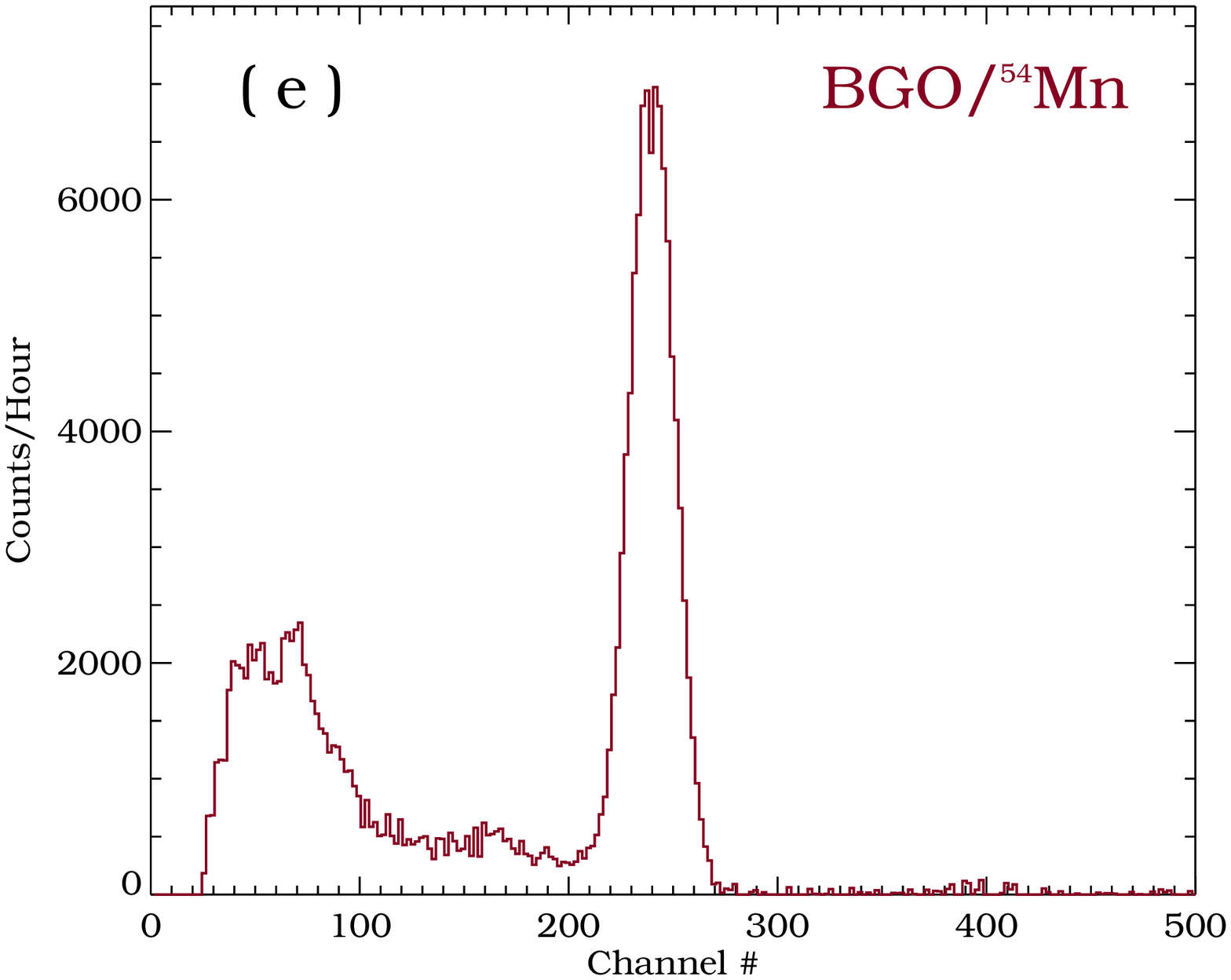}  &      
\includegraphics[height=42mm,bb=0 10 590 460,clip]{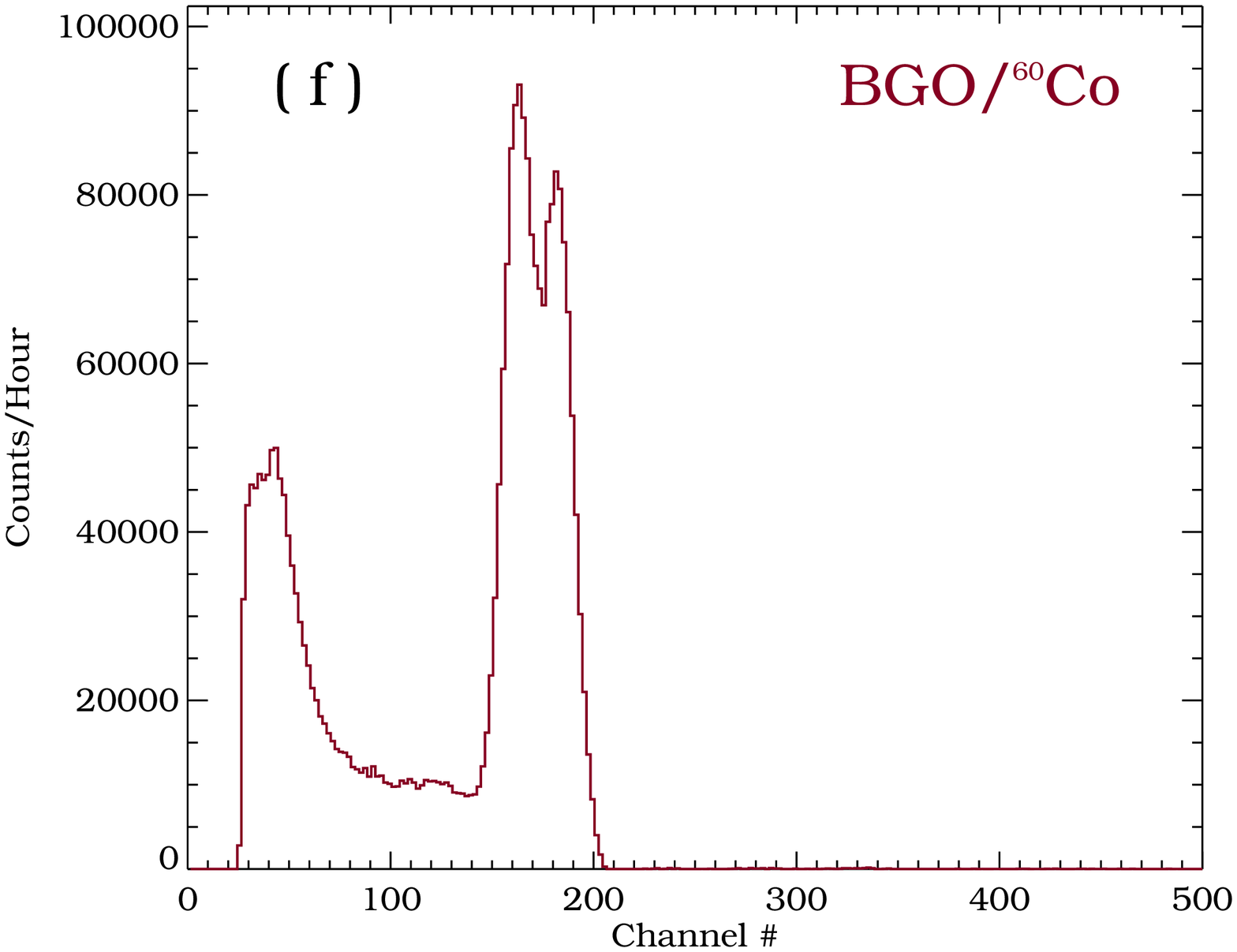}  \\
\includegraphics[height=42mm,bb=0 10 590 460,clip]{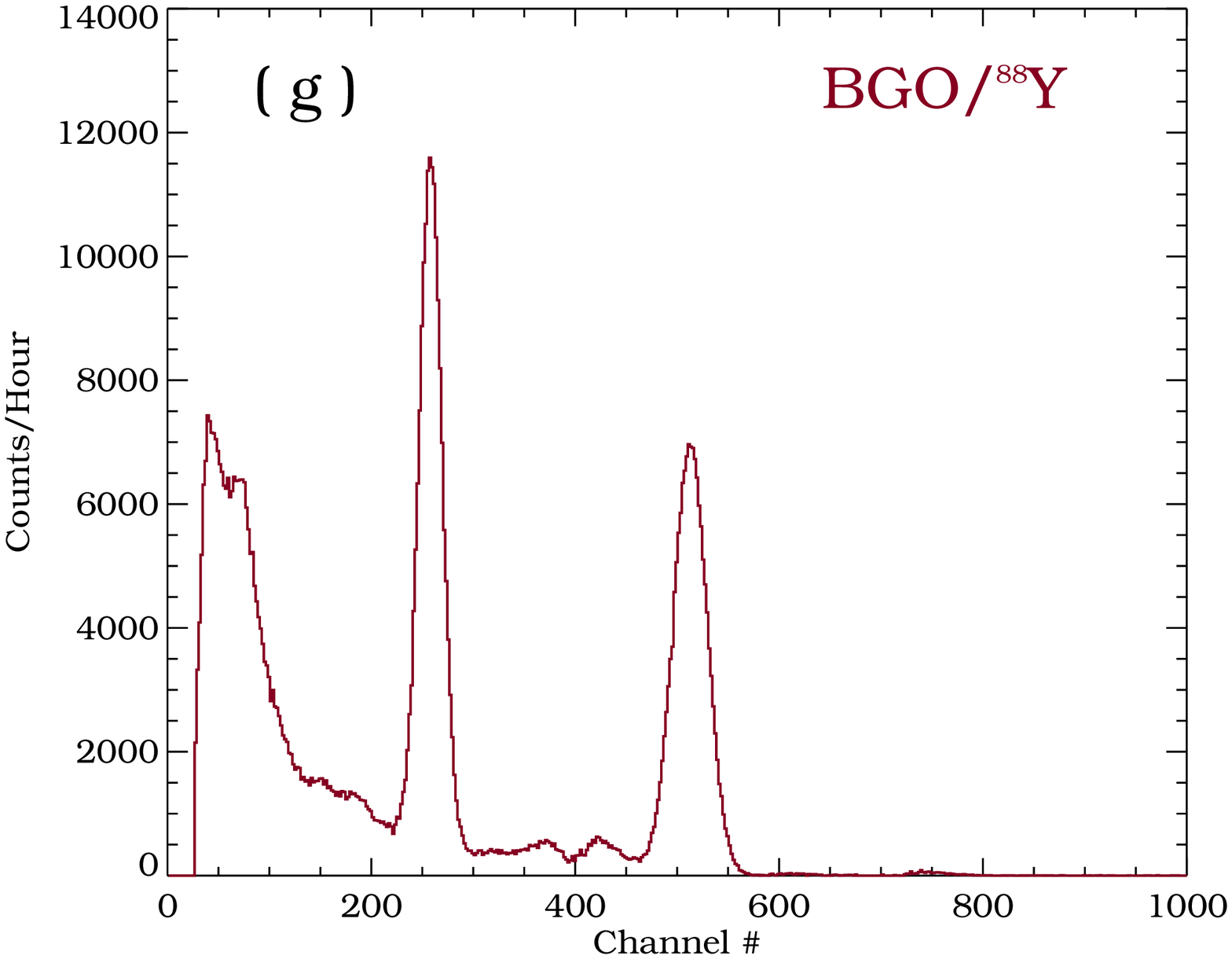}  &     
\includegraphics[height=42mm,bb=0 10 590 460,clip]{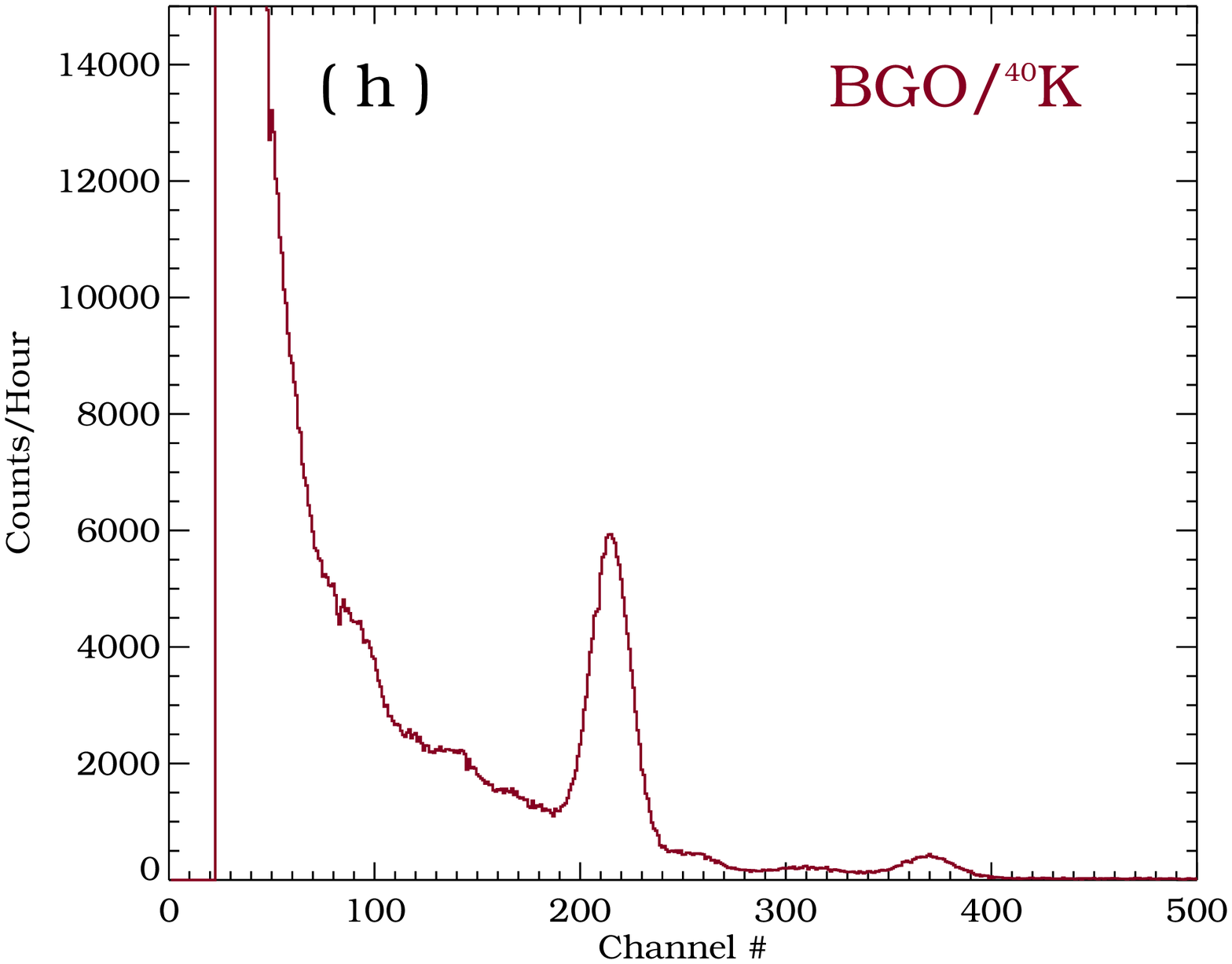}
\end{tabular}
\caption{Normalized background-subtracted spectra measured
at MPE with BGO FM 02 ({\it panels a}$-${\it g}) and EQM ({\it panel h}) from the following radioactive sources:
(a) $^{57}$Co,
(b) $^{203}$Hg,
(c) $^{22}$Na,
(d) $^{137}$Cs,
(e) $^{54}$Mn,
(f) $^{60}$Co,
(g) $^{88}$Y, and
(h) $^{40}$K
}
\label{figBGOspectra1}
\end{figure}
%
%
\begin{figure}[t!]
\centering
\begin{tabular}{cc}
\includegraphics[height=42mm,bb=0 10 590 460,clip]{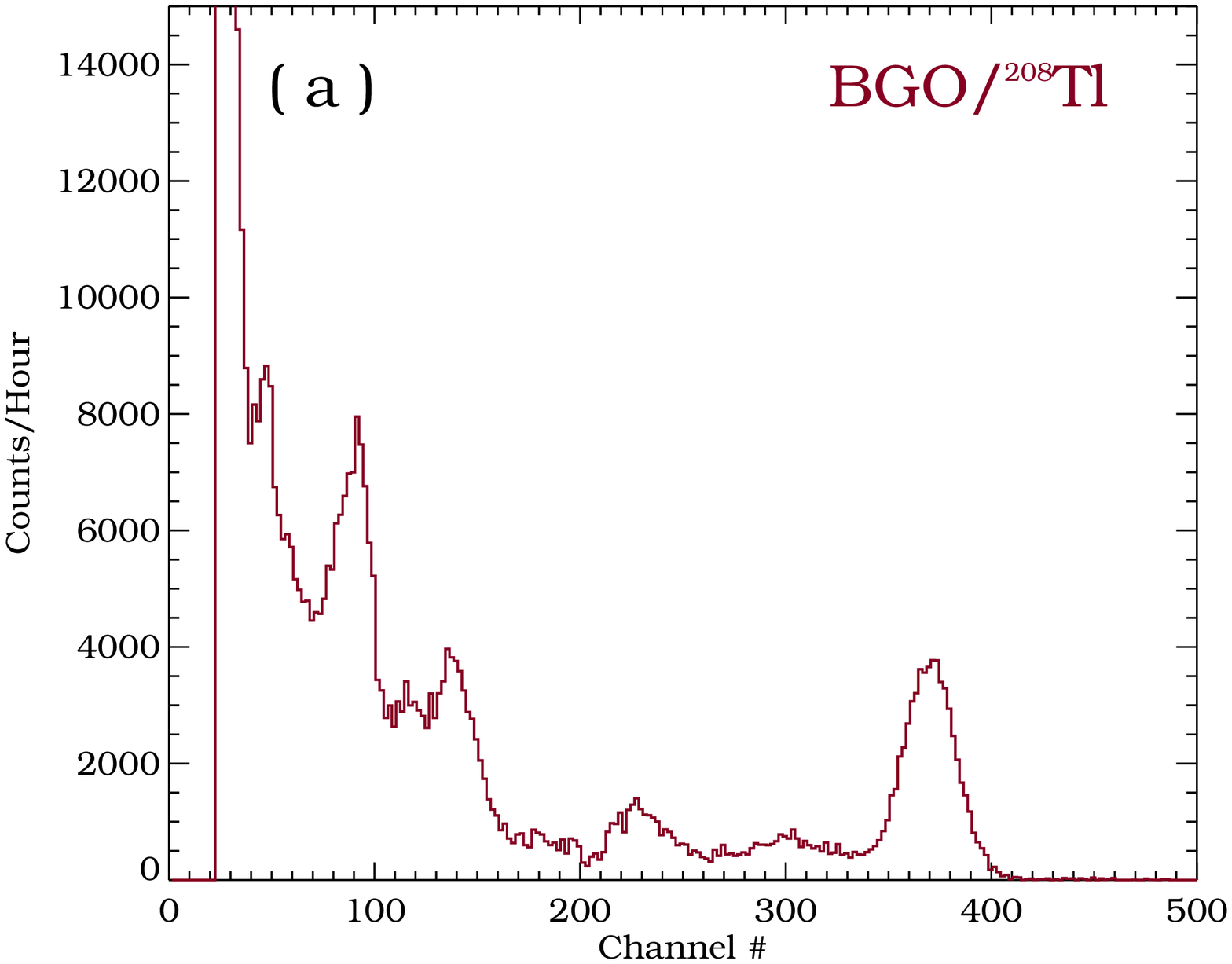}  &   
\includegraphics[height=42mm,bb=0 10 590 460,clip]{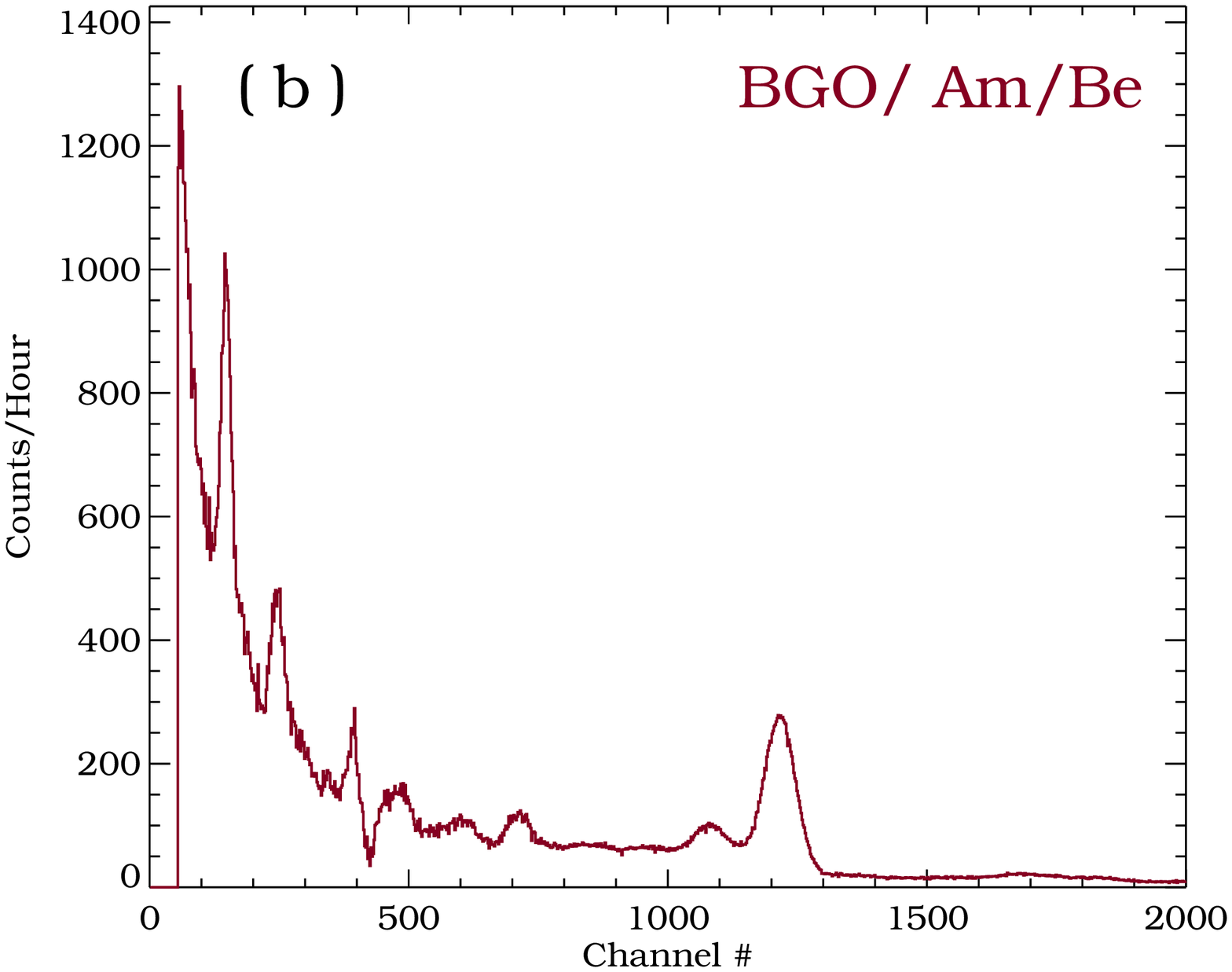}  \\
\multicolumn{2}{c}{\includegraphics[height=42mm,bb=0 10 590 460,clip]{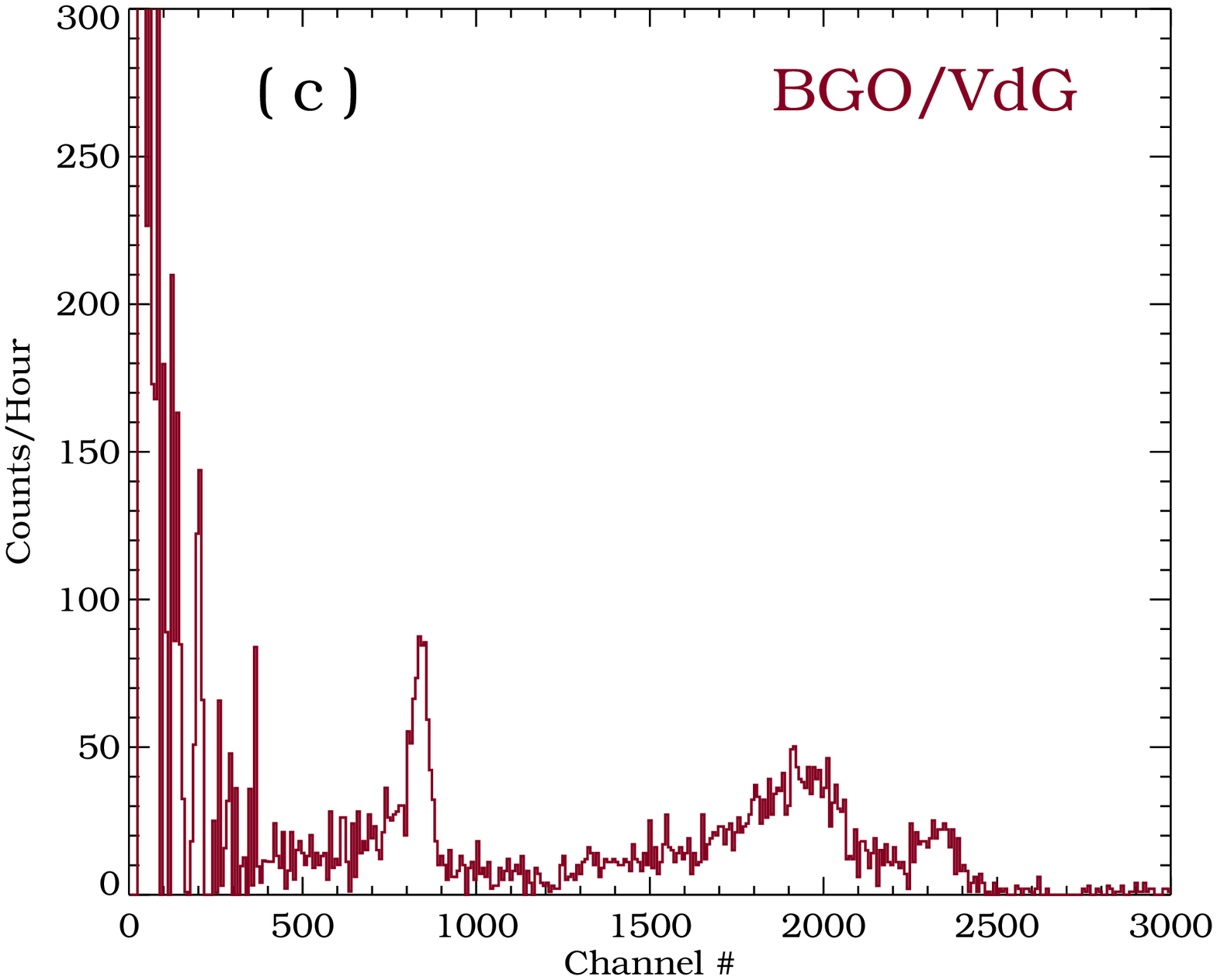}}
\end{tabular}
\caption{Normalized background-subtracted spectra measured at SLAC with BGO EQM
from two radioactive sources:
(a) $^{208}$Tl and
(b) Am/Be.
The spectrum in {\it panel c} was recorded during one of the the Van-de-Graaff (VdG) runs
}
\label{figBGOspectra2}
\end{figure}
%
%
%

Figures from \ref{figNaIspectra_BESSY} to \ref{figBGOspectra2} show a series of 
sample spectra collected with detectors NaI FM 04, BGO FM02 and BGO EQM. 
These particular detectors were chosen arbitrarily to present the whole analysis, 
since it was checked that all other detectors behave in an identical way 
(see Section \ref{Energy Resolution} and Section \ref{Effective_Area}).
In Fig. \ref{figNaIspectra_BESSY}, four calibration runs recorded at PTB/BESSY 
at different photon energies (10, 33, 34 and 60 keV) highlight the appearance of 
an important feature of the NaI spectra. Below the characteristic Iodine (I) K-shell 
binding energy of 33.17 keV (or ``K-edge'' energy), spectra display only the 
full-energy peak, which moves toward higher channel numbers with increasing photon energy
(see {\it panels a} and {\it b}). 
For energies higher than the K-edge energy, a second peak appears to
the left of the full-energy peak (see {\it panels c} and {\it d}), 
which is caused by the escape of characteristic 
X-rays resulting from K-shell transitions ({\it fluorescence} of Iodine). The energy of this 
fluorescence escape peak equals the one of the full-energy peak minus the X-ray 
line energy \cite{BOOK01}. The contributions of the different Iodine K$\alpha$ 
and K$\beta$ fluorescence lines can not be resolved by the detector. 

NaI spectra from radioactive sources recorded at MPE are shown in Fig. \ref{figNaIspectra}. 
A detailed description of the full-energy peaks characterizing every source is given 
in Section \ref{Ana_Photopeak}. Beside full-energy and Iodine escape peaks, spectra from 
high-energy radioactive lines show more features (i.e. see the $^{137}$Cs spectrum in {\it panel f}), 
such as the low-energy X radiation (due to internal scattering of gamma-rays very close to
the radioactive material) at the very left 
of the spectrum, the Compton distribution, which is a continuous distribution
due to primary gamma-rays undergoing Compton scattering within the crystal,
and a backscatter peak at the low-energy end of the Compton distribution.

Similarly, BGO spectra from radioactive sources collected at MPE and SLAC with 
detector FM 02 and BGO EQM are presented in Figures \ref{figBGOspectra1} and \ref{figBGOspectra2}.
The spectrum produced by the Van-de-Graaff proton beam at SLAC, which was measured 
by the spare detector BGO EQM, is shown in {\it panel c} of Fig. \ref{figBGOspectra2}.
\subsubsection{Analysis of the full-energy peak}\label{Ana_Photopeak}
Radioactive lines emerge from the measured spectra as peaks of various shapes 
and with multiple underlying contributions.
Depending on the specific spectrum, one or more Gaussians in the form
\begin{equation}\label{eqGauss}
G(x) = \frac{A}{w}\,\cdot\,\sqrt{\frac{\,4\,\ln 2}{\pi}}\,e^{\,-\,4\,\ln 2}\frac{(x-x_c)^{\;2}}{w^{\;2}}
\end{equation}
were added in order to fit the data. The three free parameters are 
(1) the peak area $\mathbf{A}$,
(2) the peak center $\mathbf{x_c}$, and 
(3) the full width at half maximum $\mathbf{w}$, which is related to the 
standard deviation ($\sigma$) of the distribution through the relation 
$w$~=~2~$\sqrt{\,2 \ln{2}}$~$\cdot$~$\sigma$~$\approx$~2.35~$\cdot$~$\sigma$.
For the analysis of the measured full-energy peaks, further background components 
(linear, quadratic or exponential) and Gaussian components had to be modeled 
in addition to the main Gaussian(s), 
in order to account for non-photo-peak contributions, such as the overlapping 
Compton distributions, the backscattered radiation caused by the presence 
of the uncollimated radioactive source in the laboratory or other
unknown background features.
In the case of PTB/BESSY spectra, asymmetries appearing at the low energy tail of 
the full-energy peak were neglected by choosing a smaller region of interest
and fitting only the right side of the peak.
As already mentioned, no background was modeled under these spectra.

Moreover, since both NaI and BGO detectors are not always able to fully separate two lines 
lying close to each other and thus resulting in a single broadened peak, particular 
constraints between line parameters of single peak components were fixed before 
running the fitting routines. The relation between two line areas (`$A_1$' and `$A_2$') arises 
from the transition probability $P$ of the single line energies (see Table \ref{Tab_Nuclides}, column 4). 
A ratio between areas, K$_{\,area}$ = $P_1/P_2$, was obtained by considering those
probabilities together with the transmission probability for the detector entrance window
and the relative transmission of the photons between source and detector.
Finally, the ratios were cross-checked and determined through
detailed simulations performed for seven double lines measured with NaI and BGO detectors,  
which are listed in Table \ref{TabKarea}. 
\begin{table}[t!]
\centering
\caption{Fit constraints adopted for the analysis of some double peaks
for NaI and BGO detectors. K$_{\,area}$ represents a calculated ratio between
the peak areas of the given lines}
\begin{tabular}{cccc}
\hline\noalign{\smallskip}
Nuclide      &     Detector    &   Double Line Energy (keV)     &   K$_{\,area}$   \\
\noalign{\smallskip}\hline\noalign{\smallskip}
$^{109}$Cd   &      NaI        &       22.1    $-$     25  			&  4.88    \\
$^{137}$Cs   &      NaI        &       32.06   $-$     36.6   	&  5.33    \\
$^{57}$Co    &      NaI        &      122.06   $-$    136.47   	&  8.14    \\
$^{60}$Co    &      BGO        &     1173.23   $-$    1332.49   &  0.99    \\
$^{8}$Be     &      BGO        &     5619      $-$    6130      &  0.20    \\
$^{8}$Be     &      BGO        &    14075      $-$   14586    	&  0.29    \\
$^{8}$Be     &      BGO        &    17108      $-$   17619    	&  0.27    \\
\noalign{\smallskip}\hline
\end{tabular}
\label{TabKarea}
\end{table}
%

\begin{figure}[t!]
\centering
\begin{tabular}{cc}
\includegraphics[height=42mm,bb=0 13 590 460,clip]{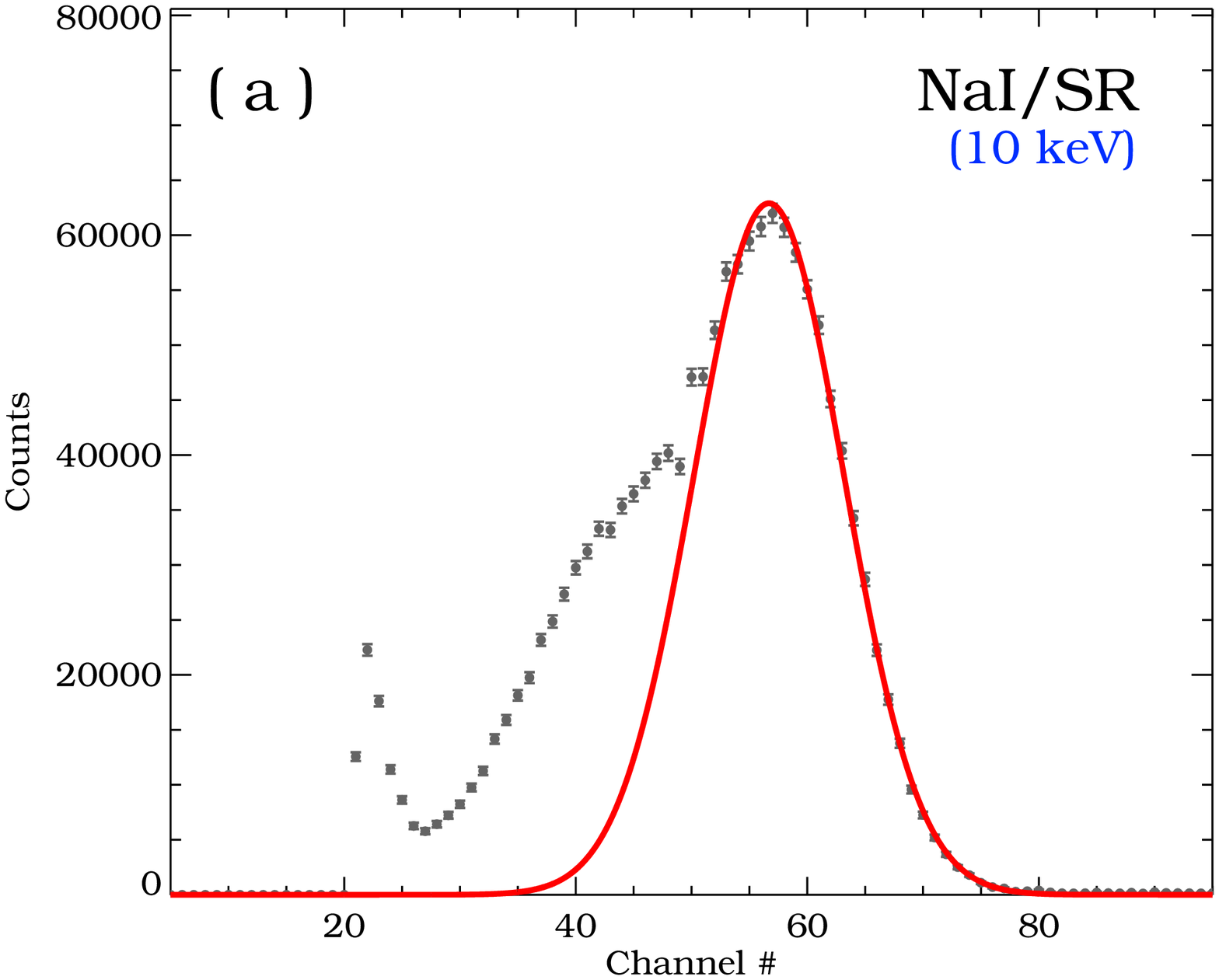}  &				
\includegraphics[height=42mm,bb=0 13 590 460,clip]{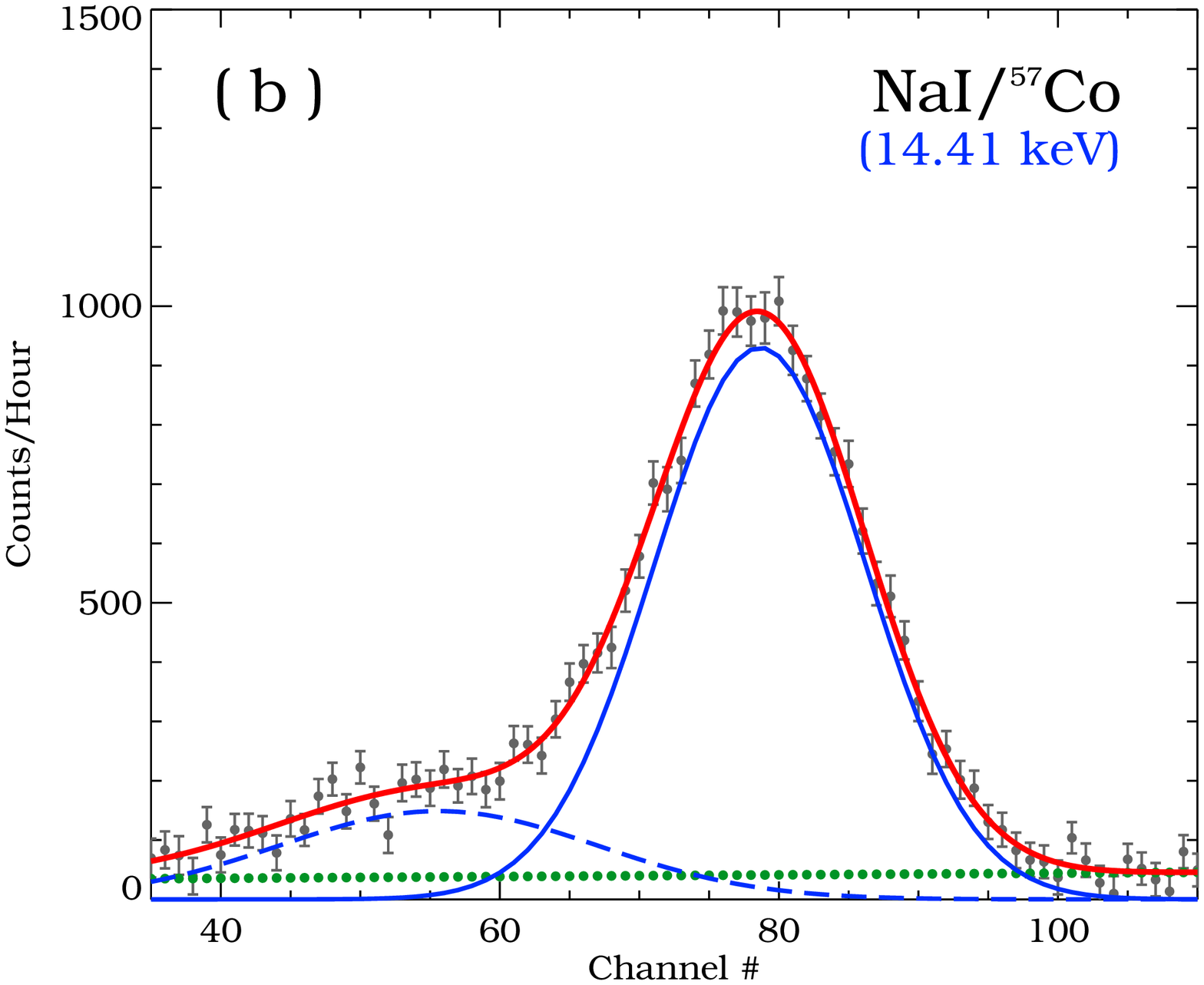}  \\
\includegraphics[height=42mm,bb=0 13 590 460,clip]{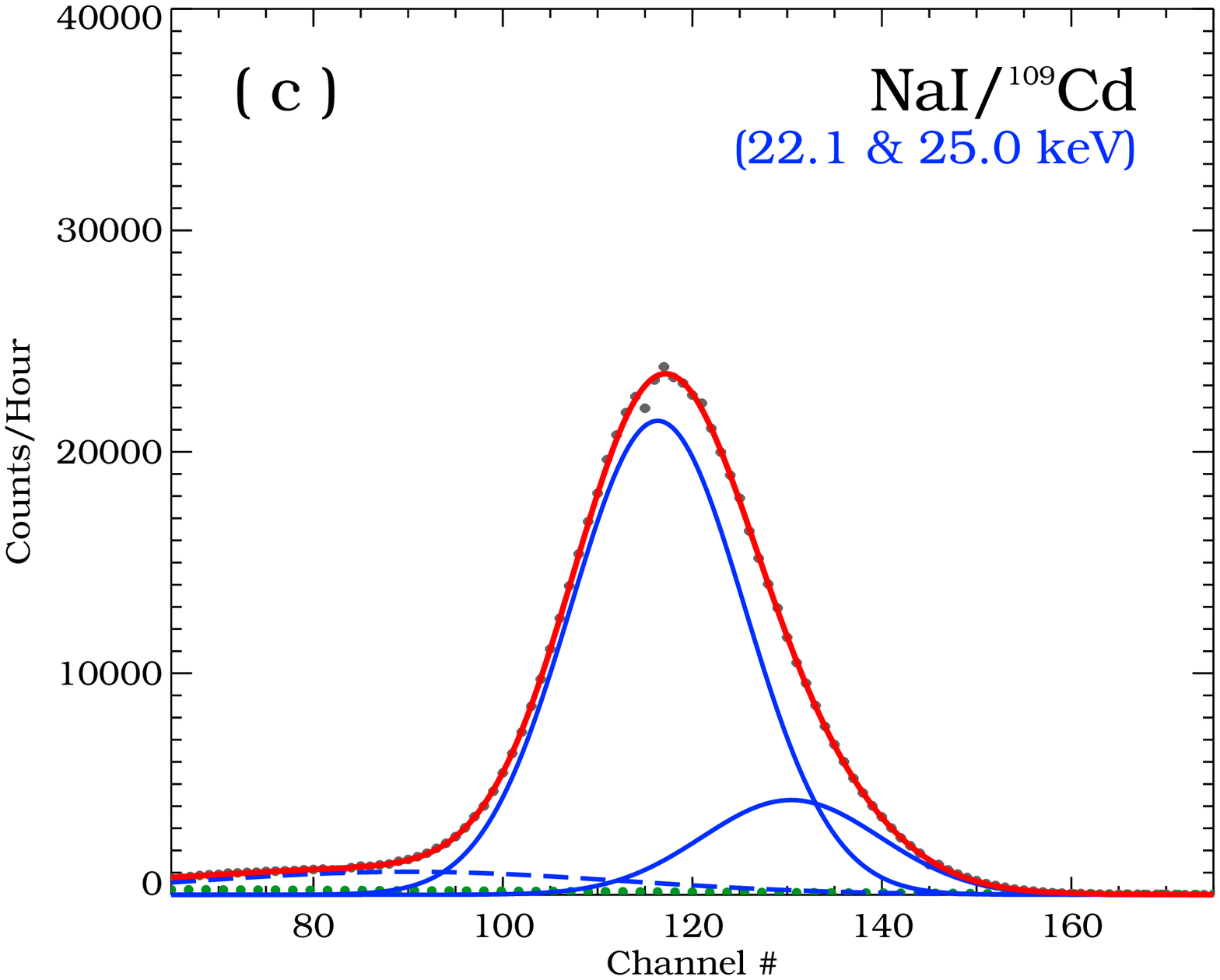}  &								
\includegraphics[height=42mm,bb=0 13 590 460,clip]{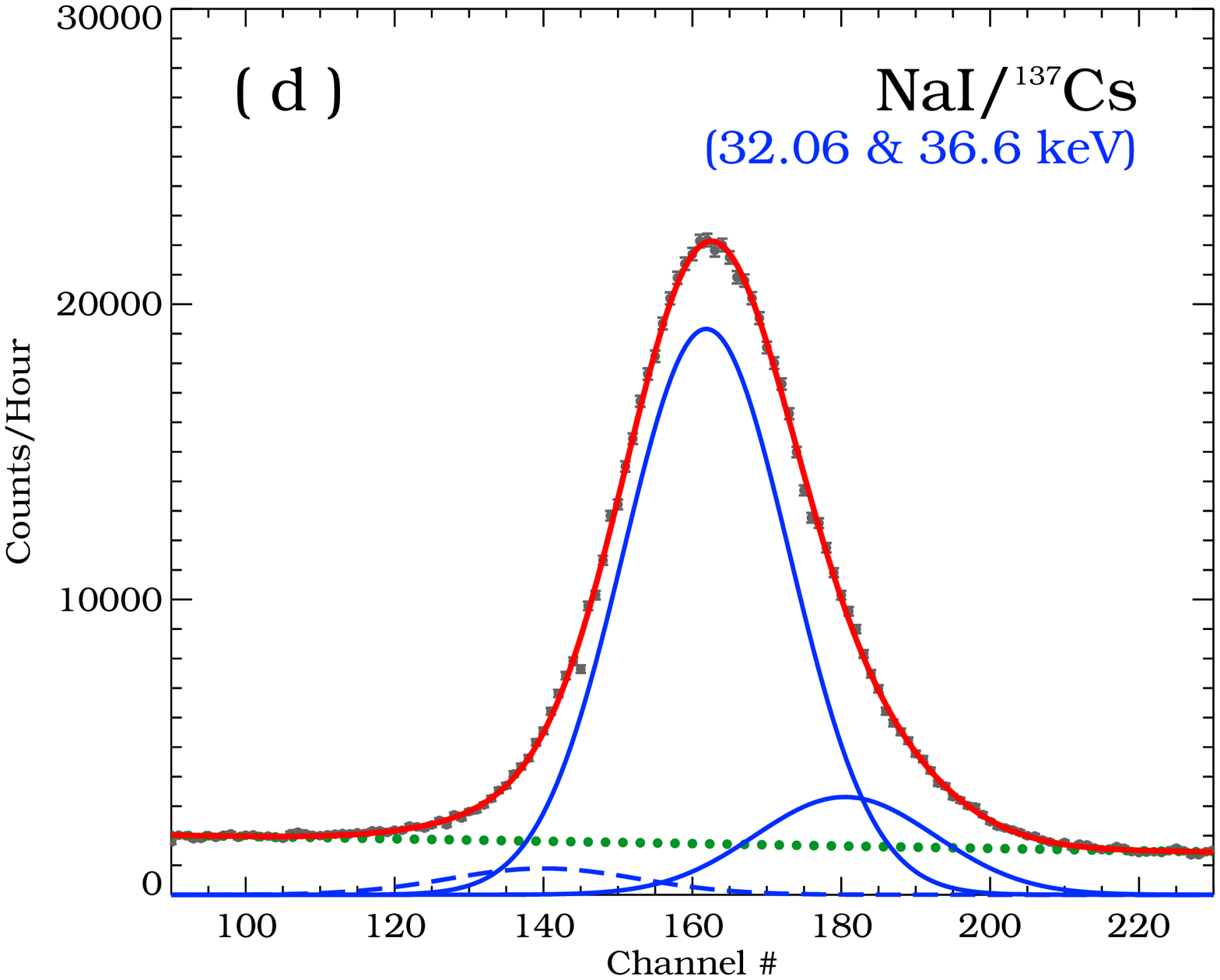}  \\
\includegraphics[height=42mm,bb=0 13 590 460,clip]{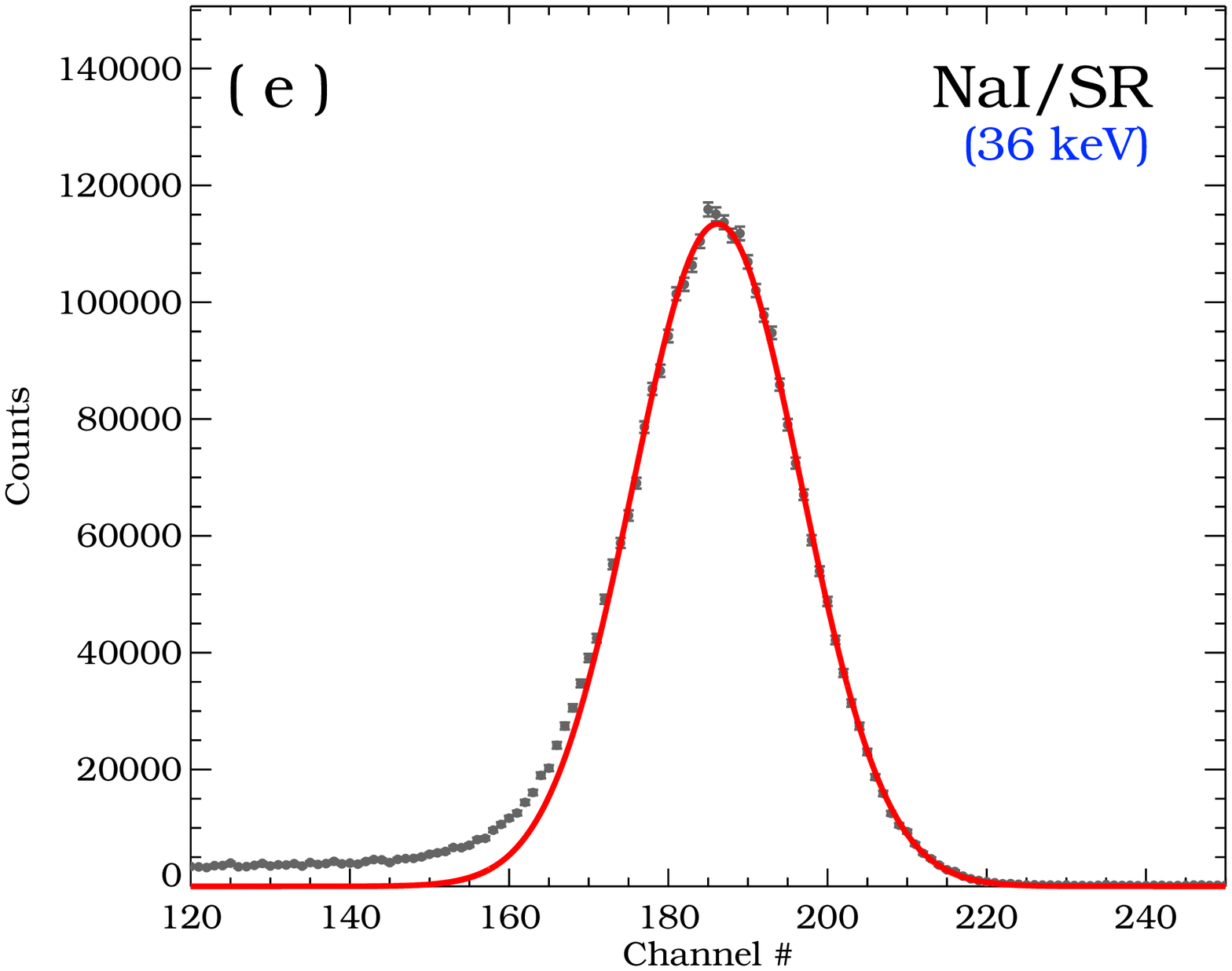}  &						
\includegraphics[height=42mm,bb=0 13 590 460,clip]{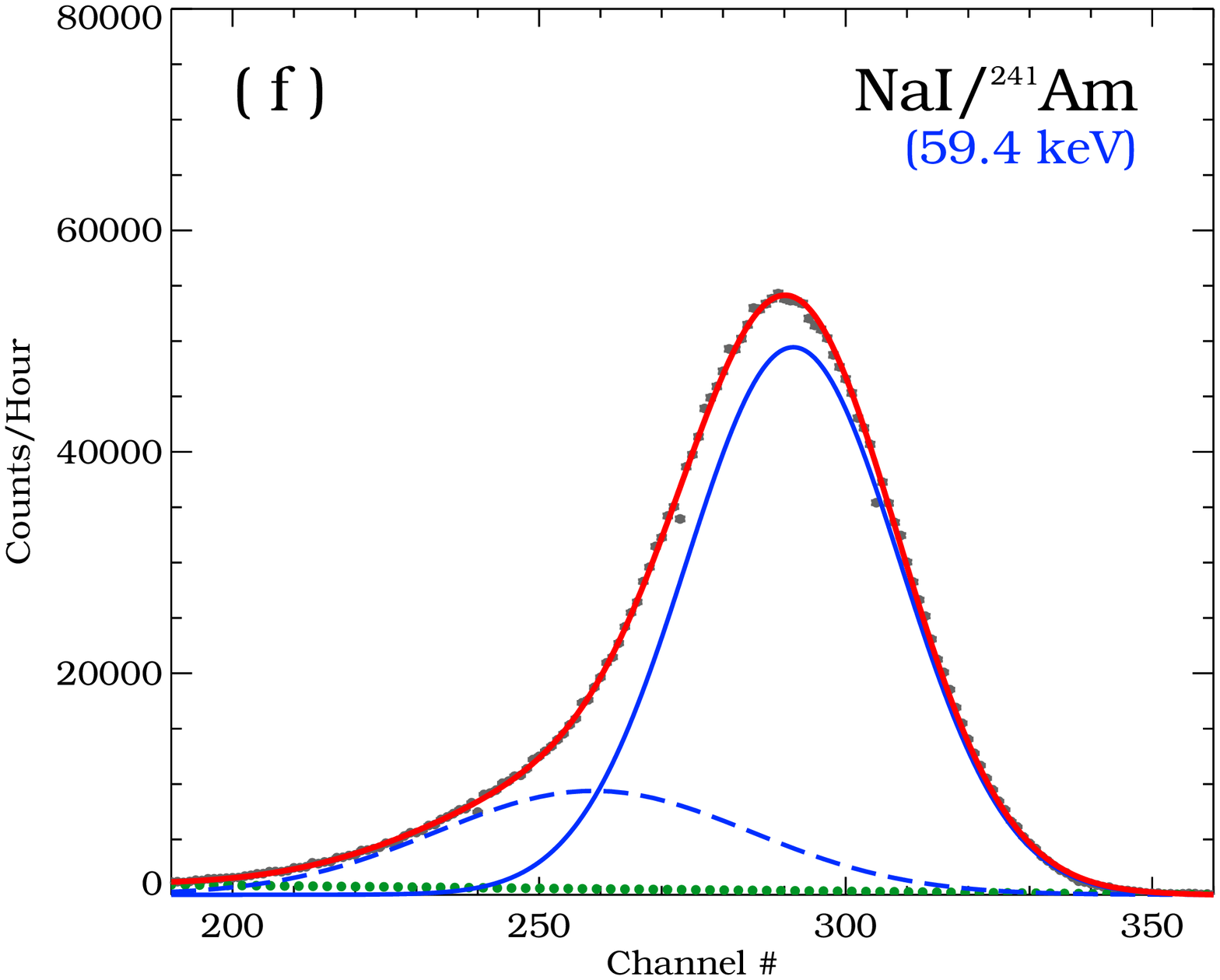}  \\
\end{tabular}	
\caption{Full-energy peak analysis of NaI lines.
Data points (in {\it black}) are plotted with statistical errors.
Line fits ({\it solid red curves}) arise from the superposition of different components:
(i) one (or more) Gaussian functions describing the full-energy peak(s) ({\it solid blue curves});
(ii) secondary Gaussian functions modeling the Iodine escape peaks or other
unknown background features ({\it dashed blue curves});
(iii) a constant, linear, quadratic or exponential function accounting
for background contributions ({\it dotted green curves}).
For PTB/BESSY line analysis ({\it panels a} and {\it e}),
the background contributions could be neglected and only the fit to the
full-energy peak was performed starting from 4 to 10 channels before
the maximum
}
\label{NaI_all_1}
\end{figure}

An important consideration when fitting mathematical functions to these data
is that the calculated statistical errors of the fit parameters are always within 0.1~\%\ in the case
of line areas and FWHM, or even 0.01~\%\ in the case of line-centers. 
Such extreme precisions cause very high chi-square values in subsequent analysis,
as in the determination of the channel-energy relation, which extends over an
entire energy decade in the case of NaI detectors. Moreover, it was noticed that
by slightly changing the initial fitting conditions, such as the region of interest around
the peak or the type of background, parameter values suffered from substantial
changes with respect to a precedent analysis. This effect is particularly strong
in the analysis of multiple peaks, were more Gaussians and background 
functions are added and the number of free parameters increases. In order to
account for this effects and to get a more realistic evaluation of 
the fit parameter errors, we decided to analyse several times one spectrum per source 
(measured at normal incidence by detectors NaI FM 04 and BGO FM 02), each time
putting different initial fitting conditions. This procedure was repeated several
times (usually $\sim$10-20 times, i.e.
until the systematic contribution was not further increasing and
a good chi-square value of the individual fit was produced), thus obtaining
a dataset of fit parameters and respective errors.
For each error dataset, standard deviations ($\sigma$) were calculated, resulting in
values of the order of 1~\%\ for line areas and FWHMs and of 0.1~\%\ for line-centers,
and were finally added to the fit error, thus obtaining realistic errors.

\begin{figure}[p!]
\centering
\begin{tabular}{cc}
\includegraphics[height=42mm,bb=0 13 590 460,clip]{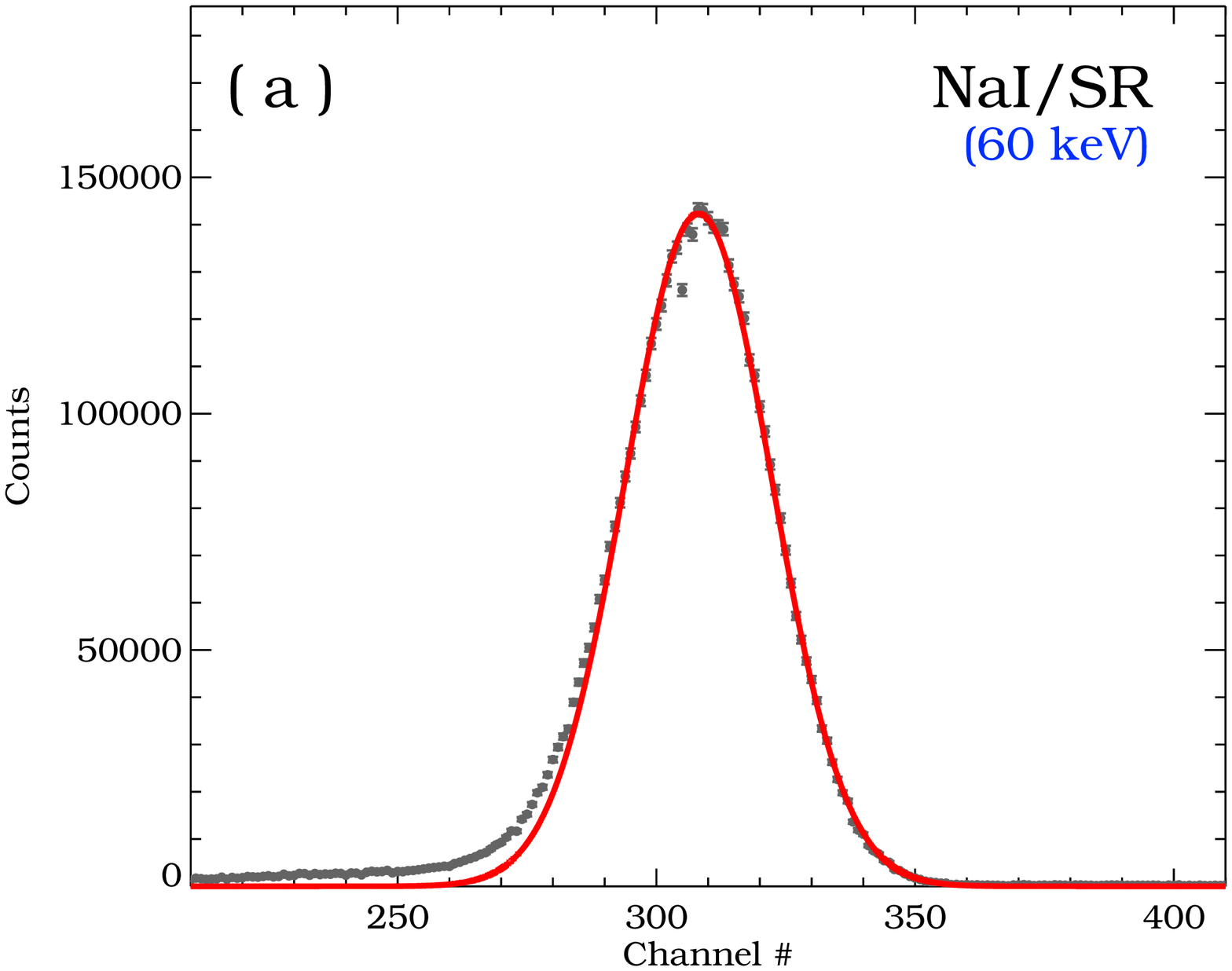}  &					
\includegraphics[height=42mm,bb=0 13 590 460,clip]{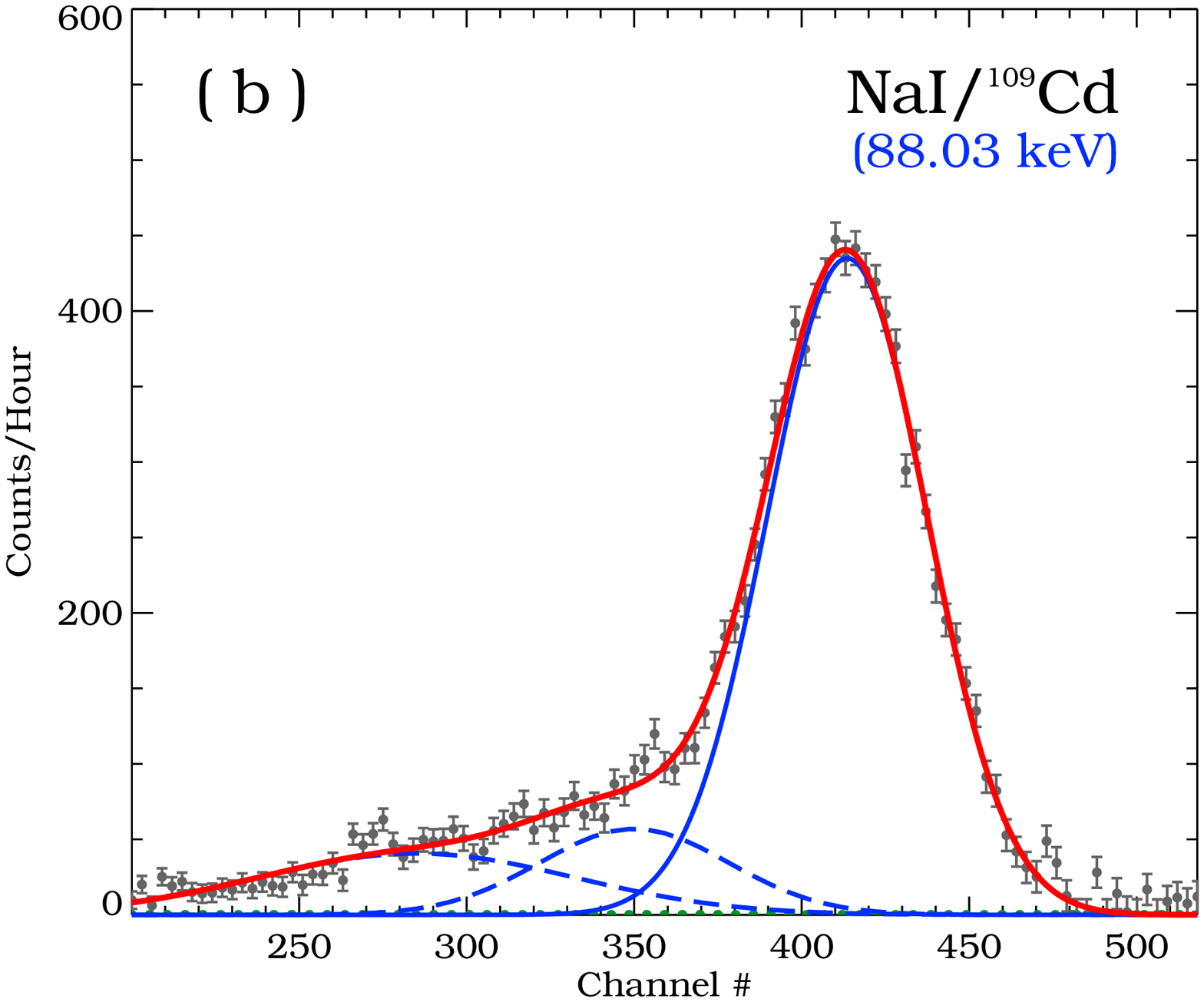}  \\
\includegraphics[height=42mm,bb=0 13 590 460,clip]{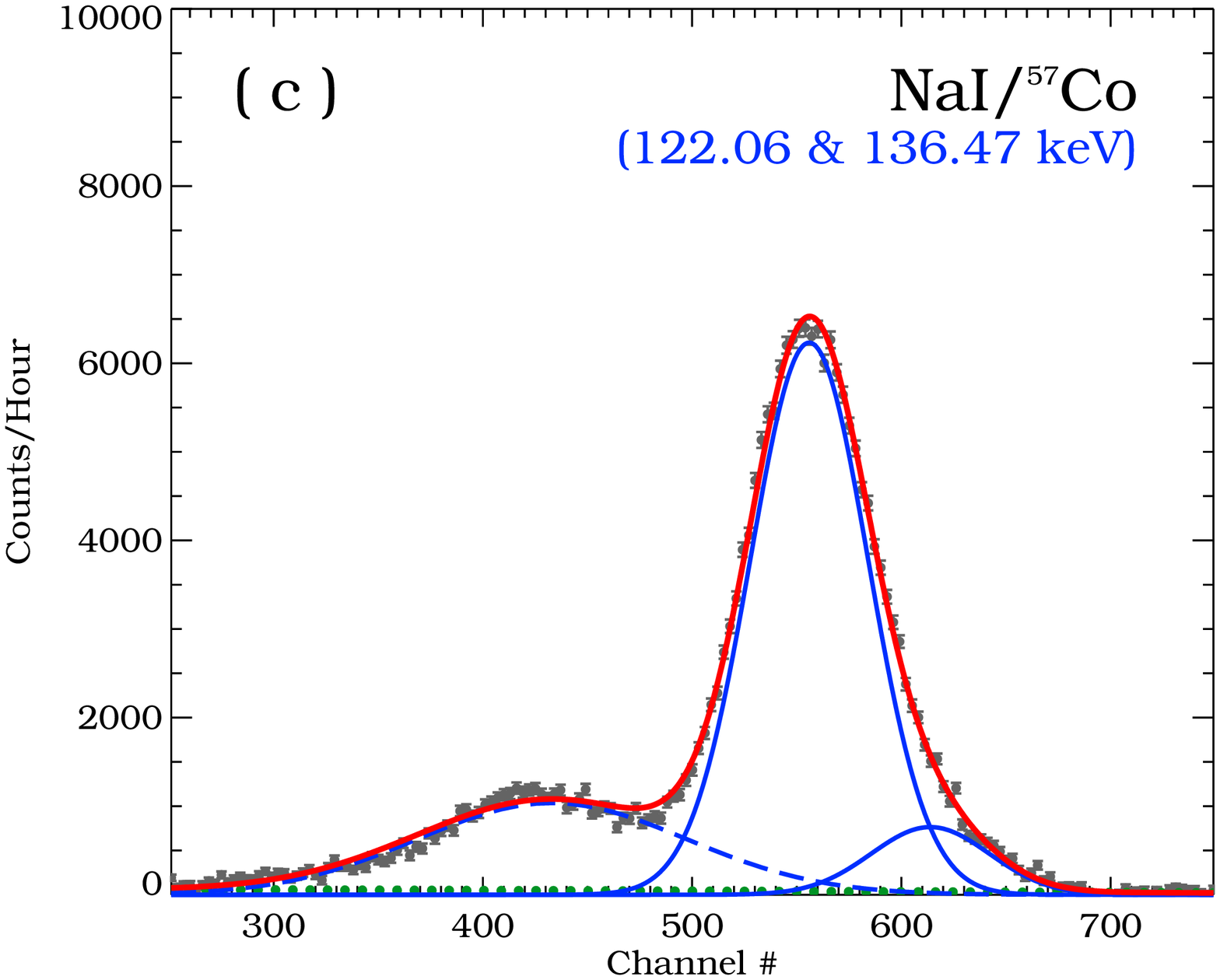}   &								
\includegraphics[height=42mm,bb=0 13 590 460,clip]{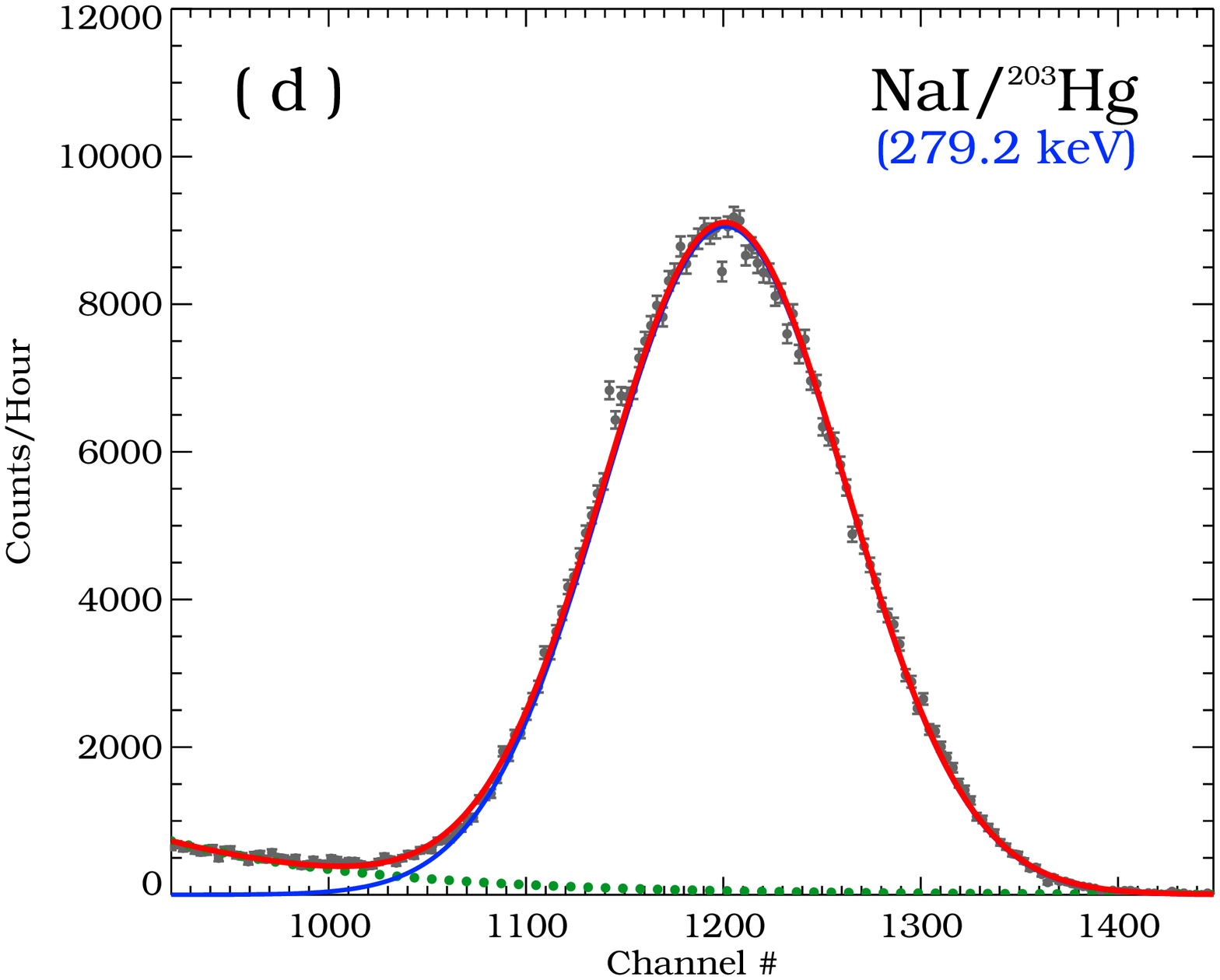}   \\
\includegraphics[height=42mm,bb=0 13 590 460,clip]{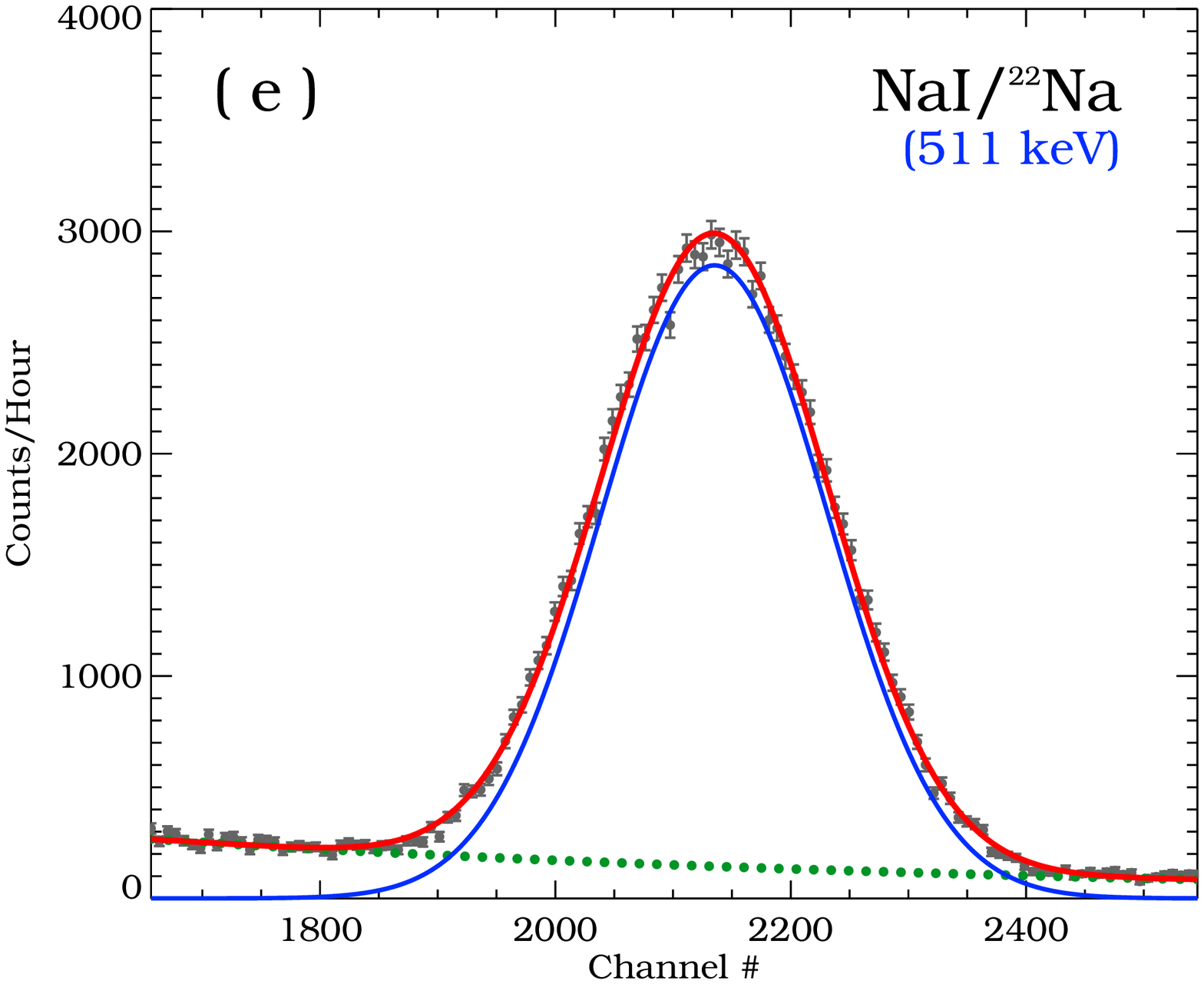}   & 									
\includegraphics[height=42mm,bb=0 13 590 460,clip]{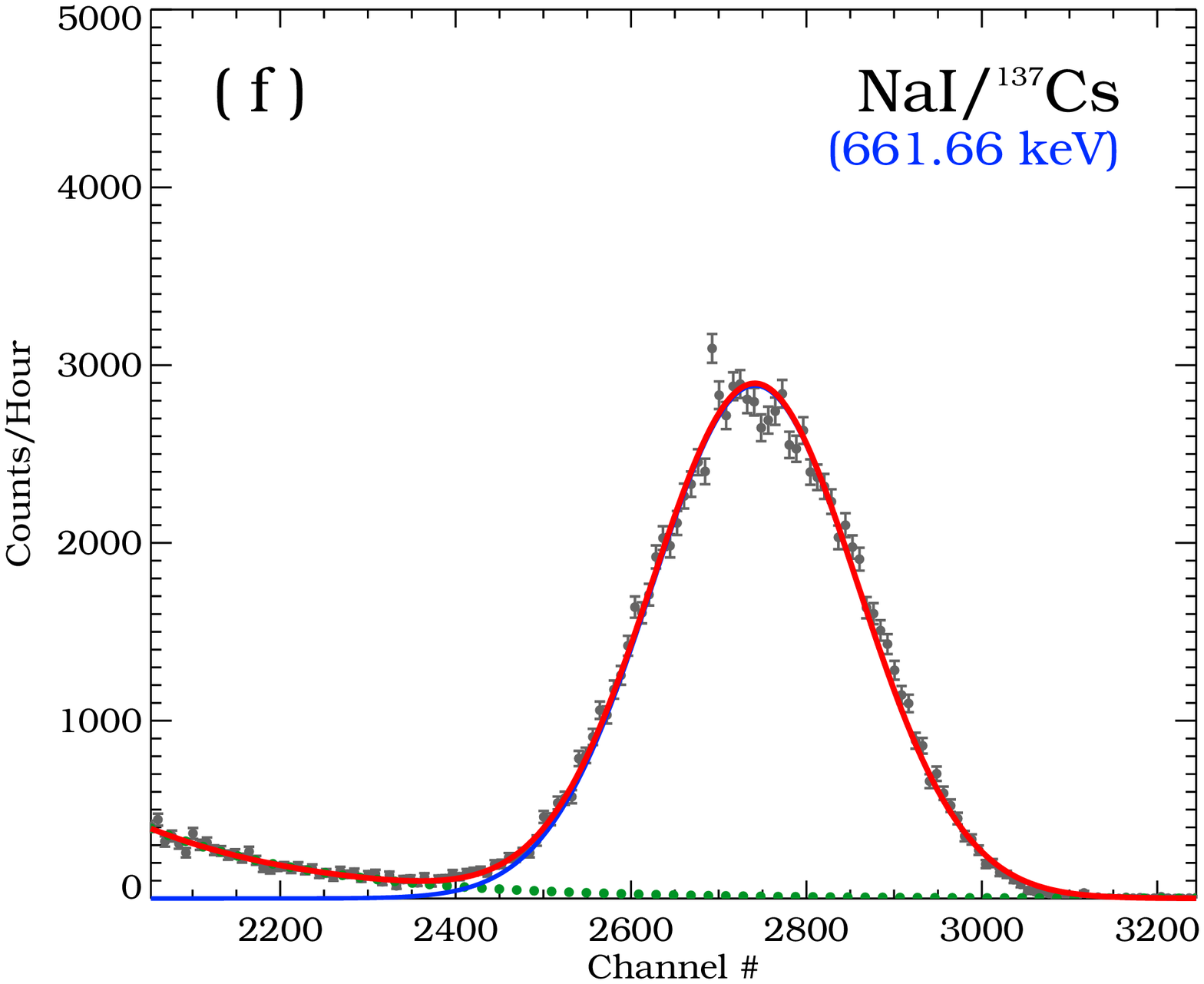}   \\ 
\includegraphics[height=42mm,bb=0 13 590 460,clip]{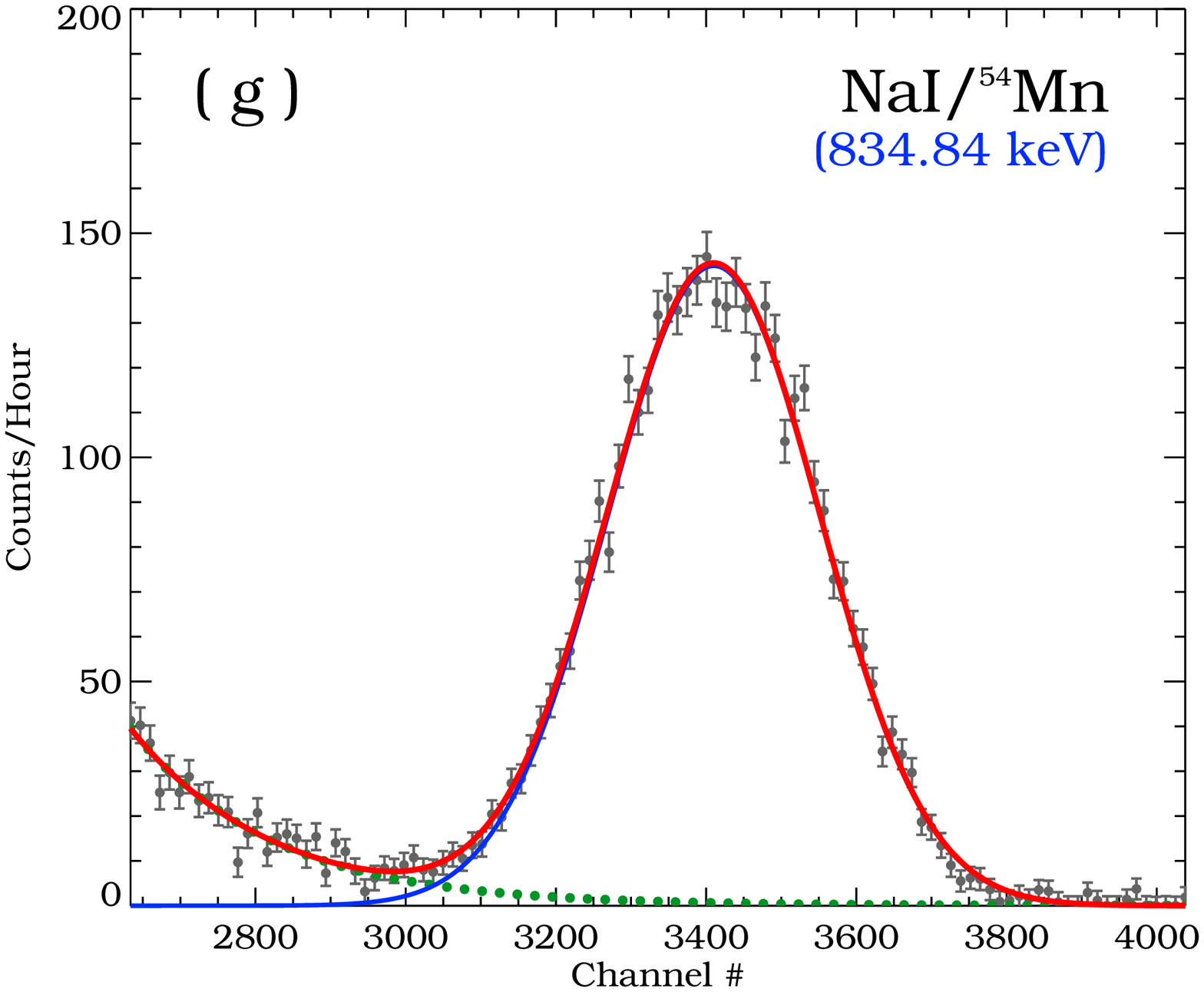}   &									
\includegraphics[height=42mm,bb=0 13 590 460,clip]{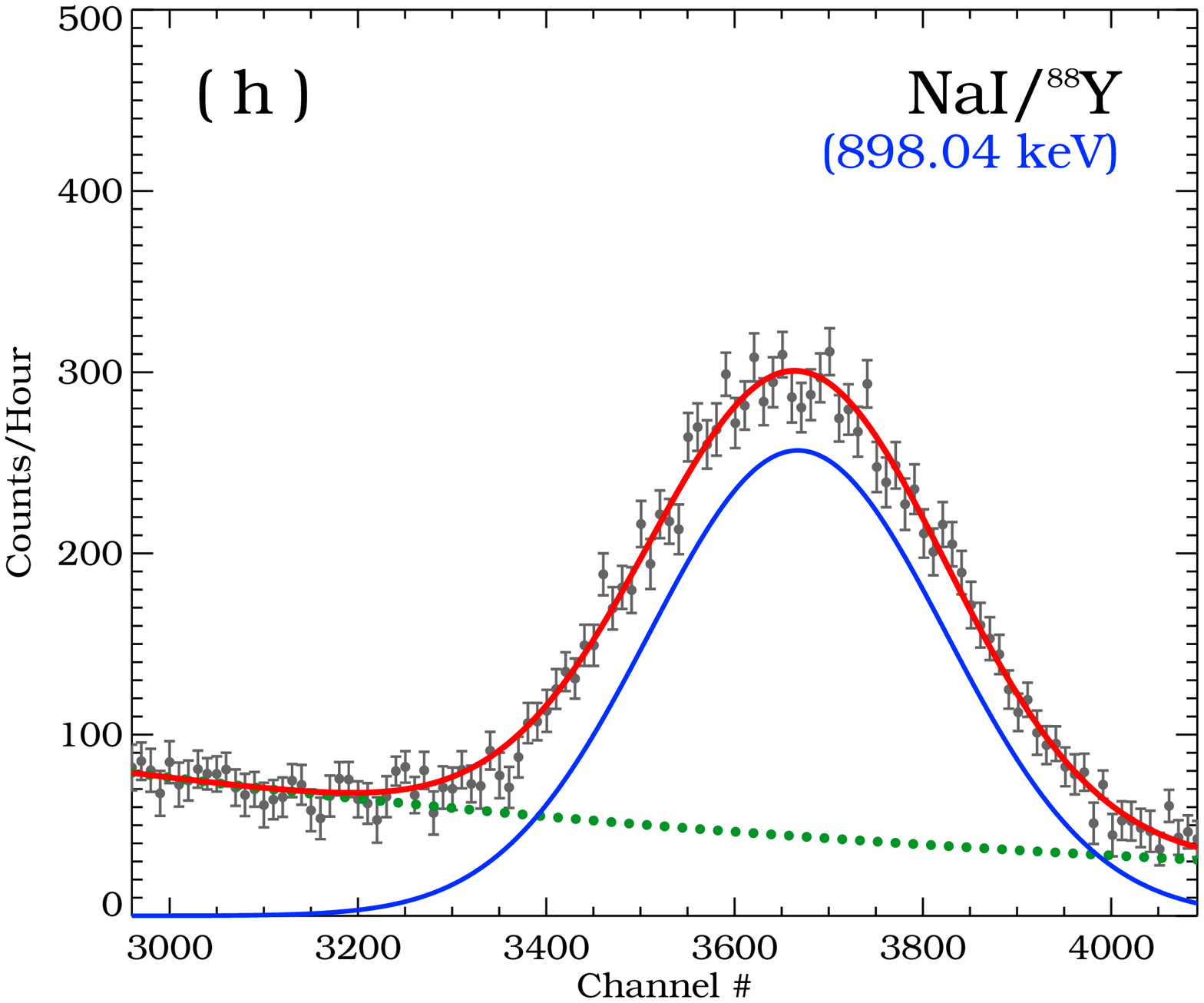}
\end{tabular}
\caption{
Full-energy peak analysis of NaI lines.
Data points (in {\it black}) are plotted with statistical errors.
Line fits ({\it solid red curves}) arise from the superposition of different components:
(i) one (or more) Gaussian functions describing the full-energy peak(s) ({\it solid blue curves});
(ii) secondary Gaussian functions modeling the Iodine escape peaks or other
unknown background features ({\it dashed blue curves});
(iii) a constant, linear, quadratic or exponential function accounting
for background contributions ({\it dotted green curves}).
For PTB/BESSY line analysis ({\it panels a} and {\it e}),
the background contributions could be neglected and only the fit to the
full-energy peak was performed starting from 4 to 10 channels before
the maximum
}
\label{NaI_all_2}
\end{figure}

%
%
\begin{figure}[p!]
\centering
\begin{tabular}{cc}
\includegraphics[height=42mm,bb=0 13 590 460,clip]{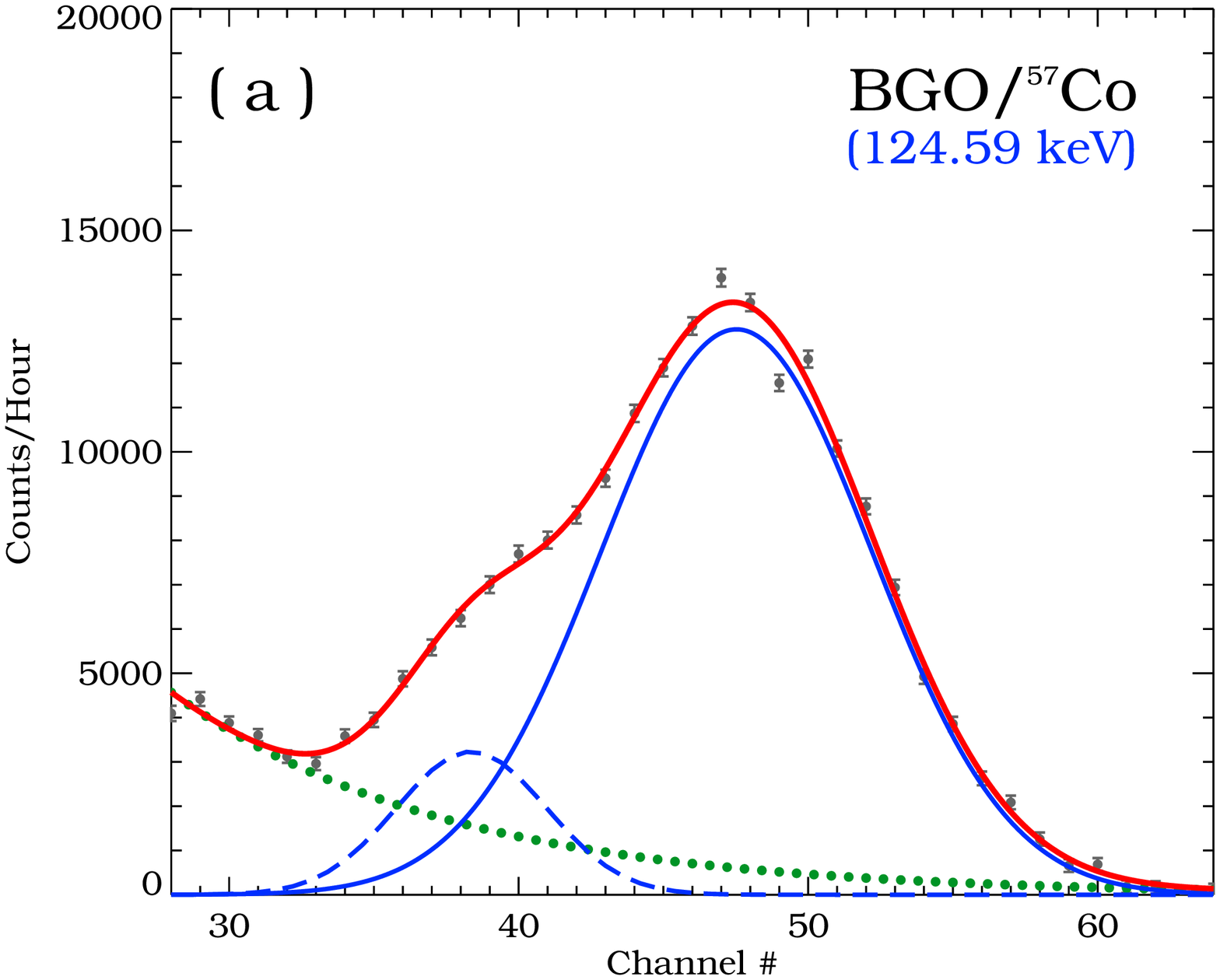}  & 
\includegraphics[height=42mm,bb=0 13 590 460,clip]{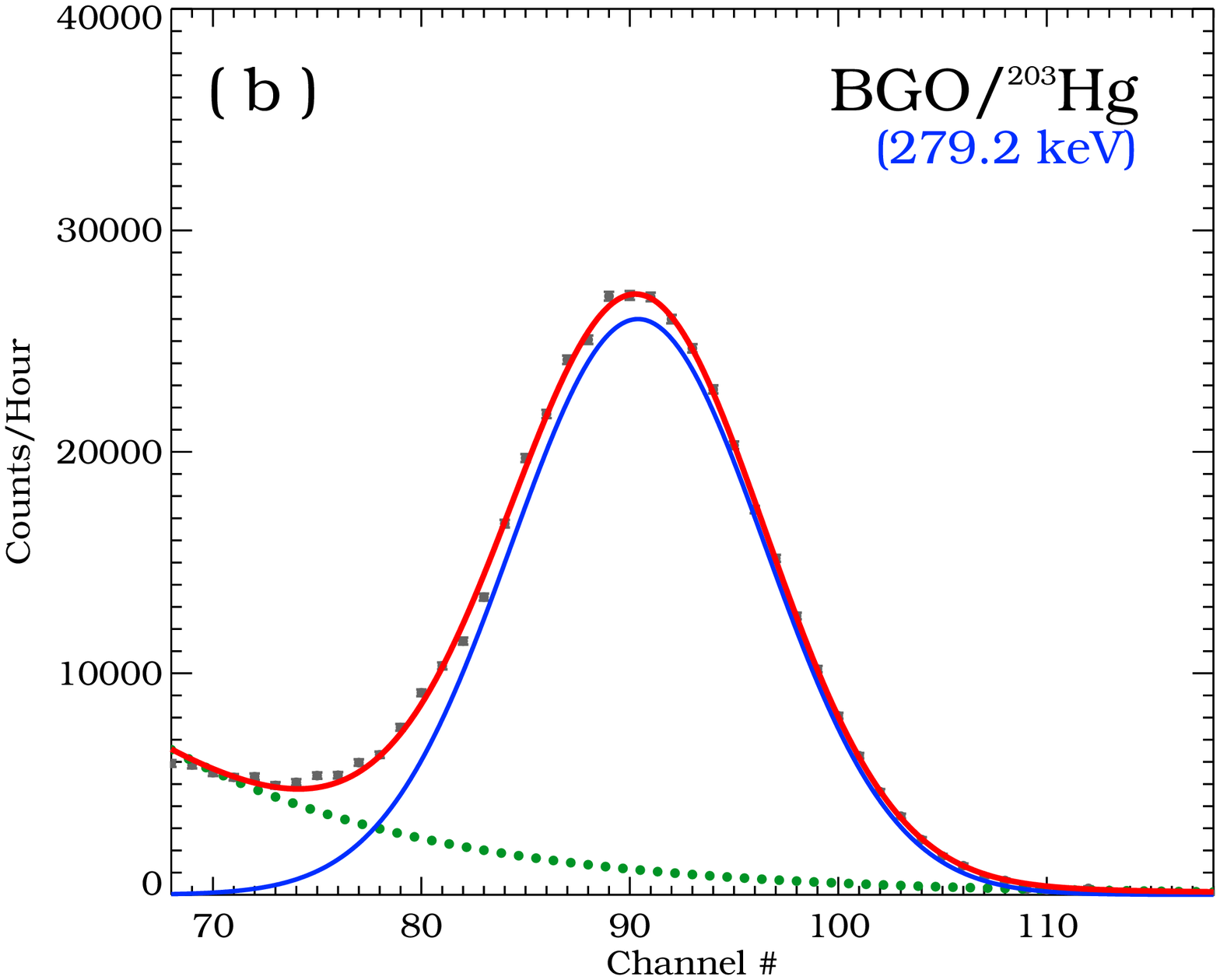}  \\
\includegraphics[height=42mm,bb=0 13 590 460,clip]{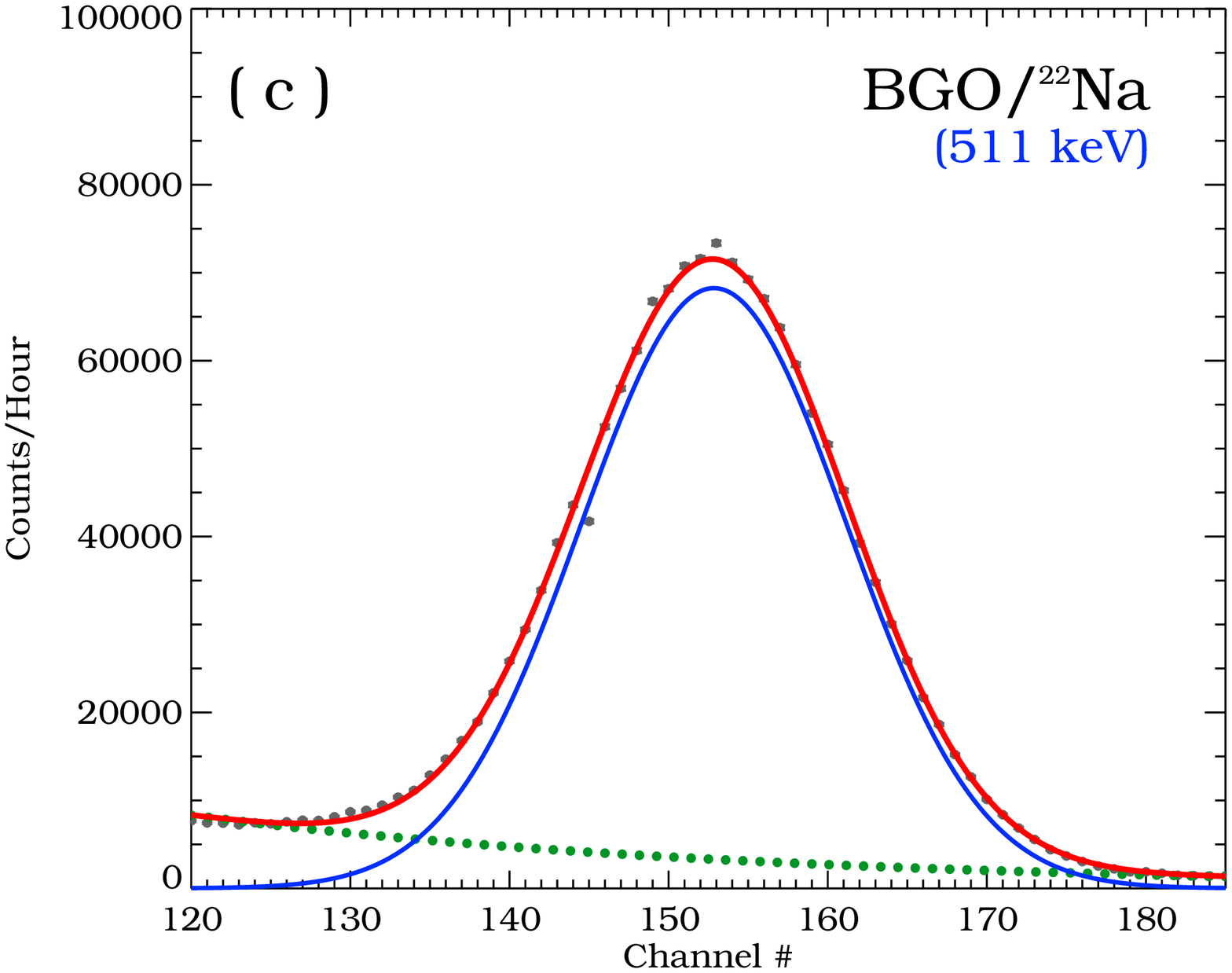}  &                   
\includegraphics[height=42mm,bb=0 13 590 460,clip]{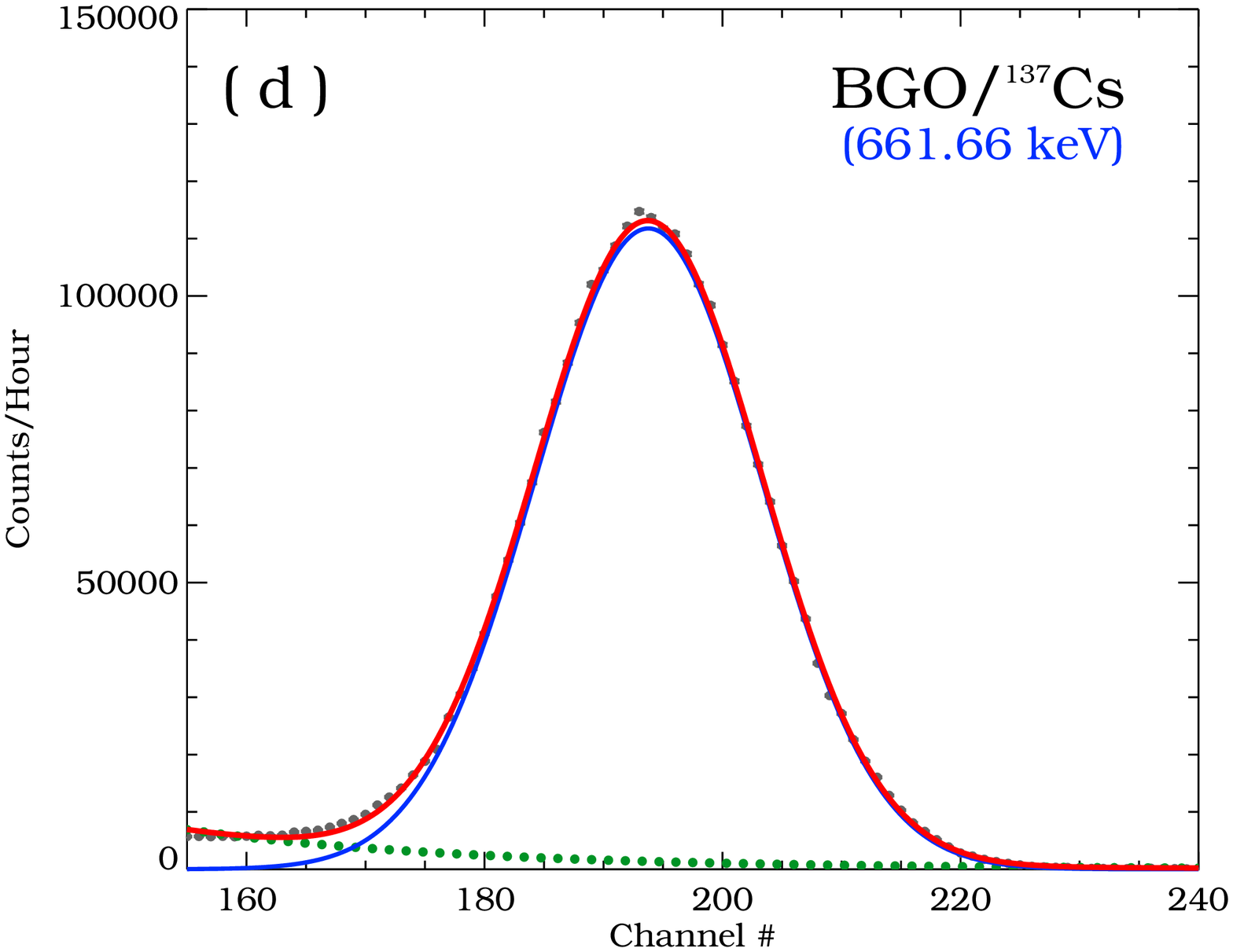}  \\
\includegraphics[height=42mm,bb=0 13 590 460,clip]{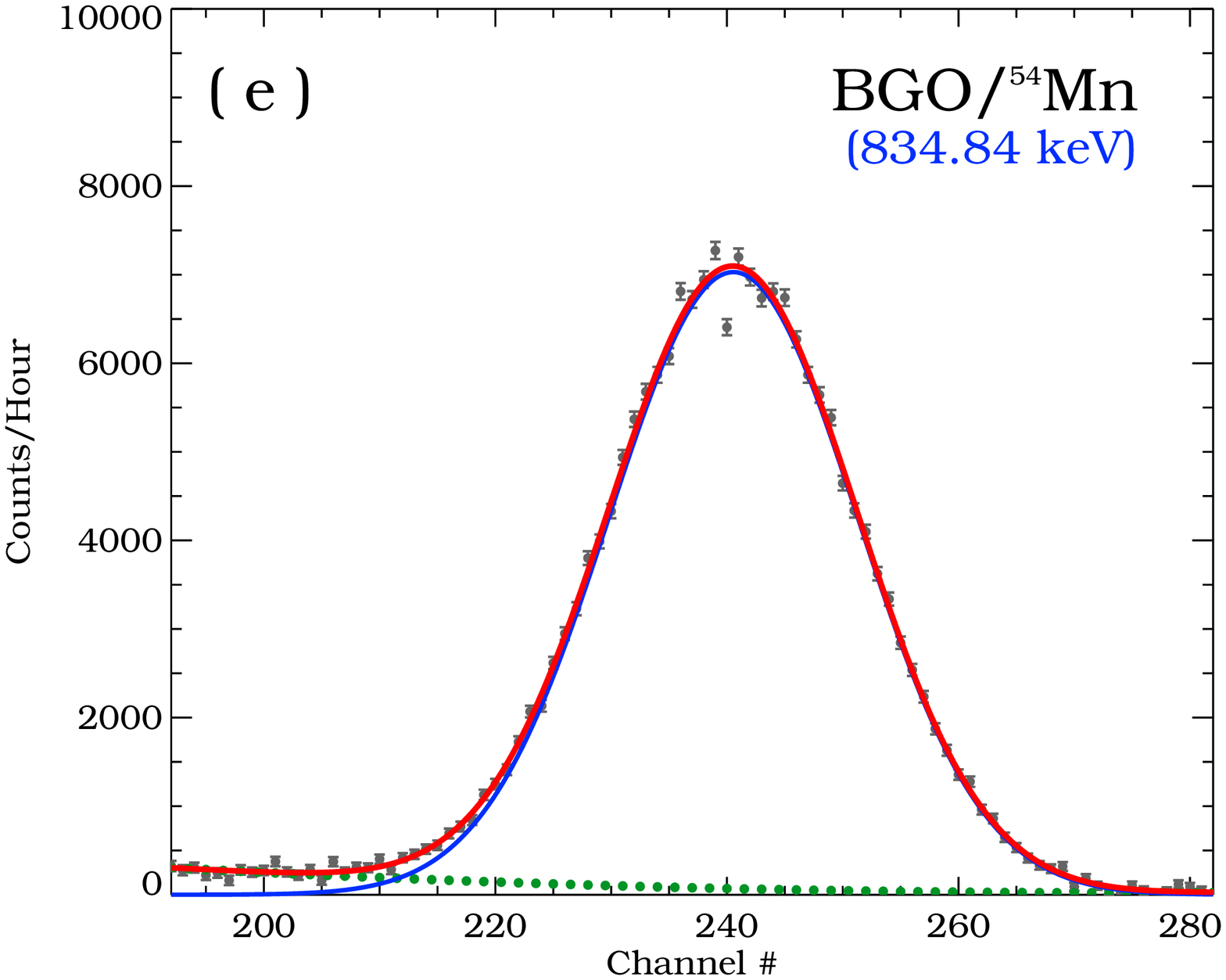}  &                  
\includegraphics[height=42mm,bb=0 13 590 460,clip]{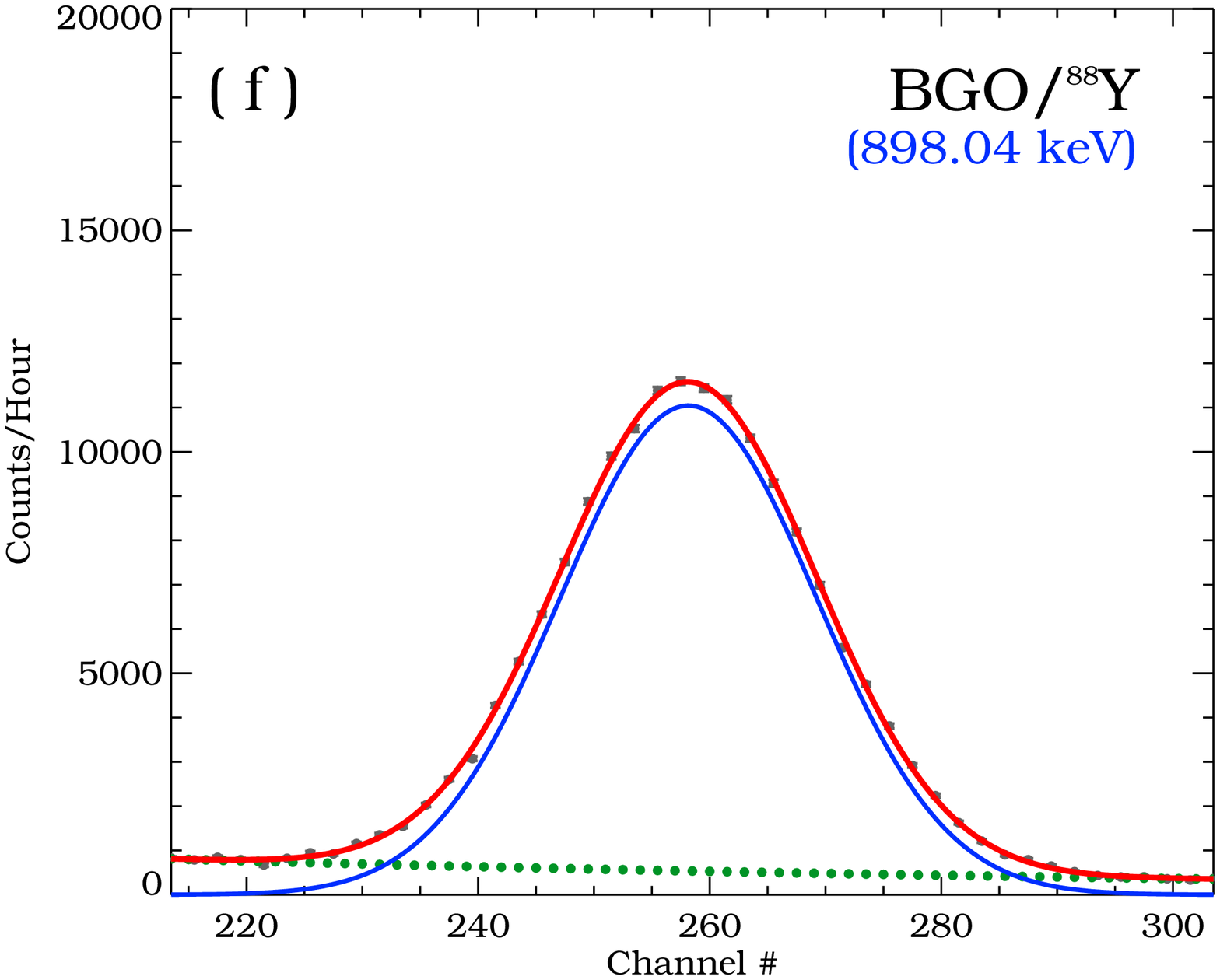}  \\
\includegraphics[height=42mm,bb=0 13 590 460,clip]{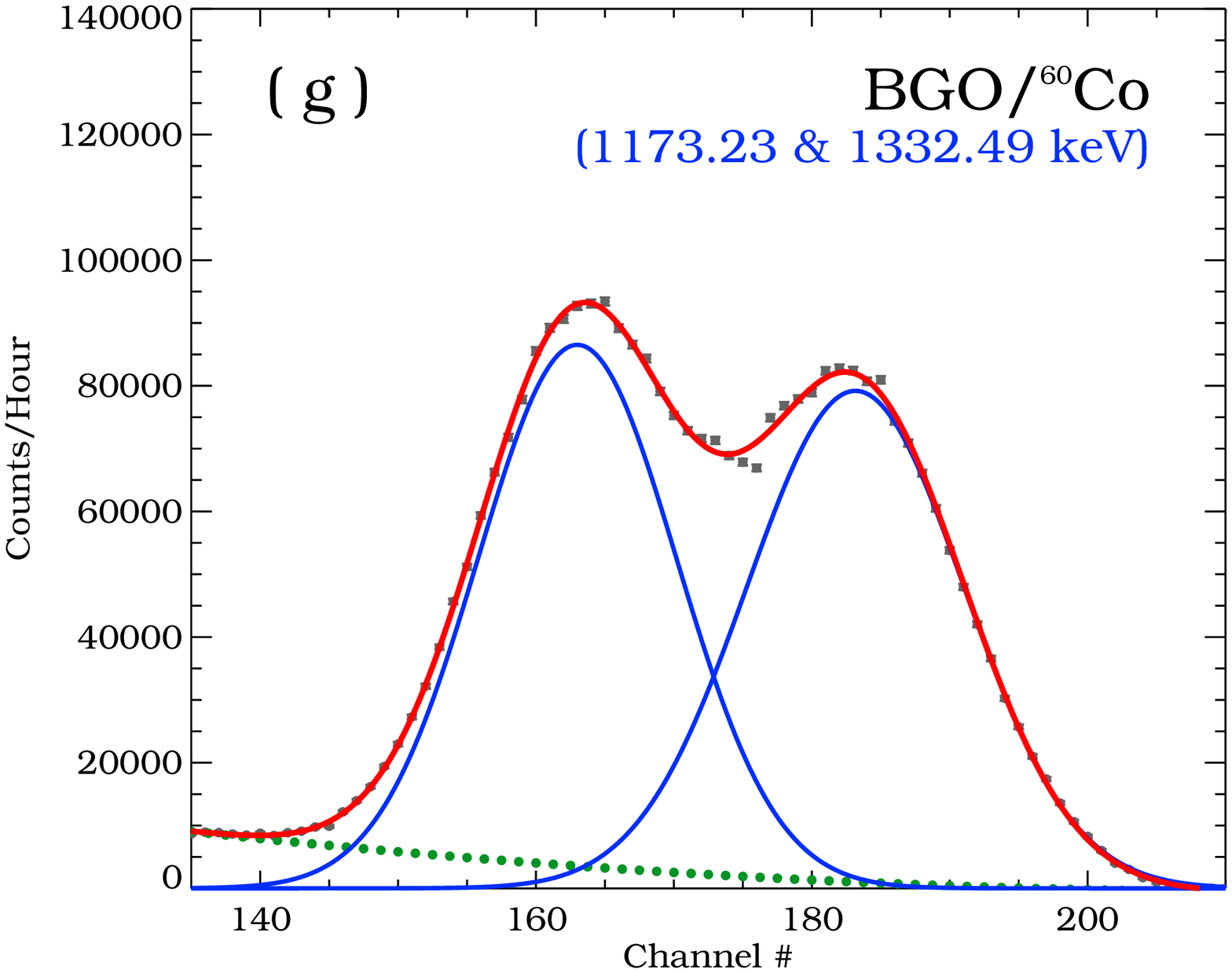}  &                        
\includegraphics[height=42mm,bb=0 13 590 460,clip]{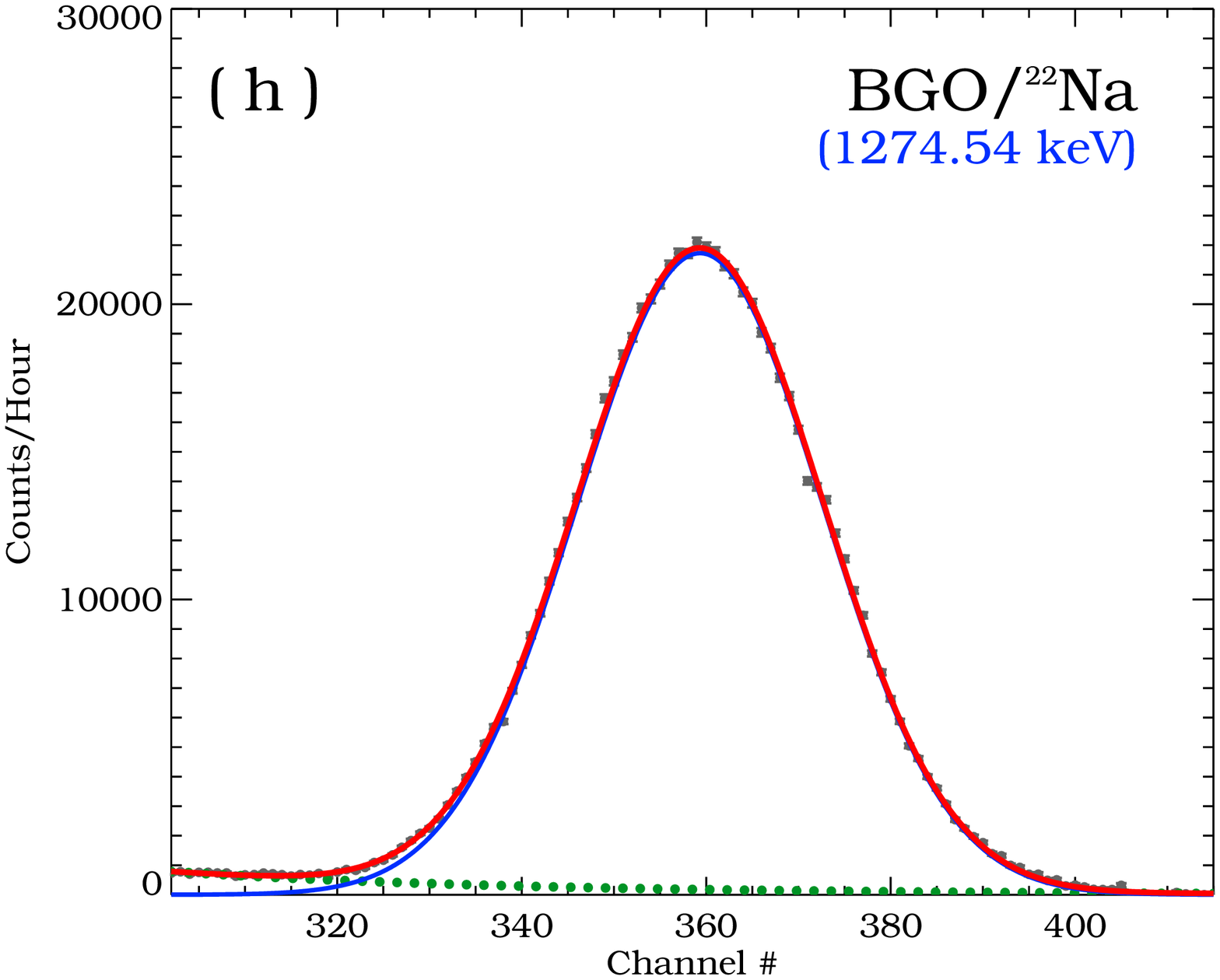}
\end{tabular}
\caption{
Full-energy peak analysis of BGO lines.
Data points (in {\it black}) are plotted with statistical errors.
Line fits ({\it solid red curves}) arise from the superposition of different components:
(i) one (or more) Gaussian functions describing the full-energy peak(s)
and the pair production escape peaks ({\it solid blue curves});
(ii) a constant, linear, quadratic or exponential function accounting
for background contributions ({\it dotted green curves})
}
\label{BGO_all_1}
\end{figure}
%
%
\begin{figure}[p!]
\centering
\begin{tabular}{cc}
\includegraphics[height=42mm,bb=0 13 590 460,clip]{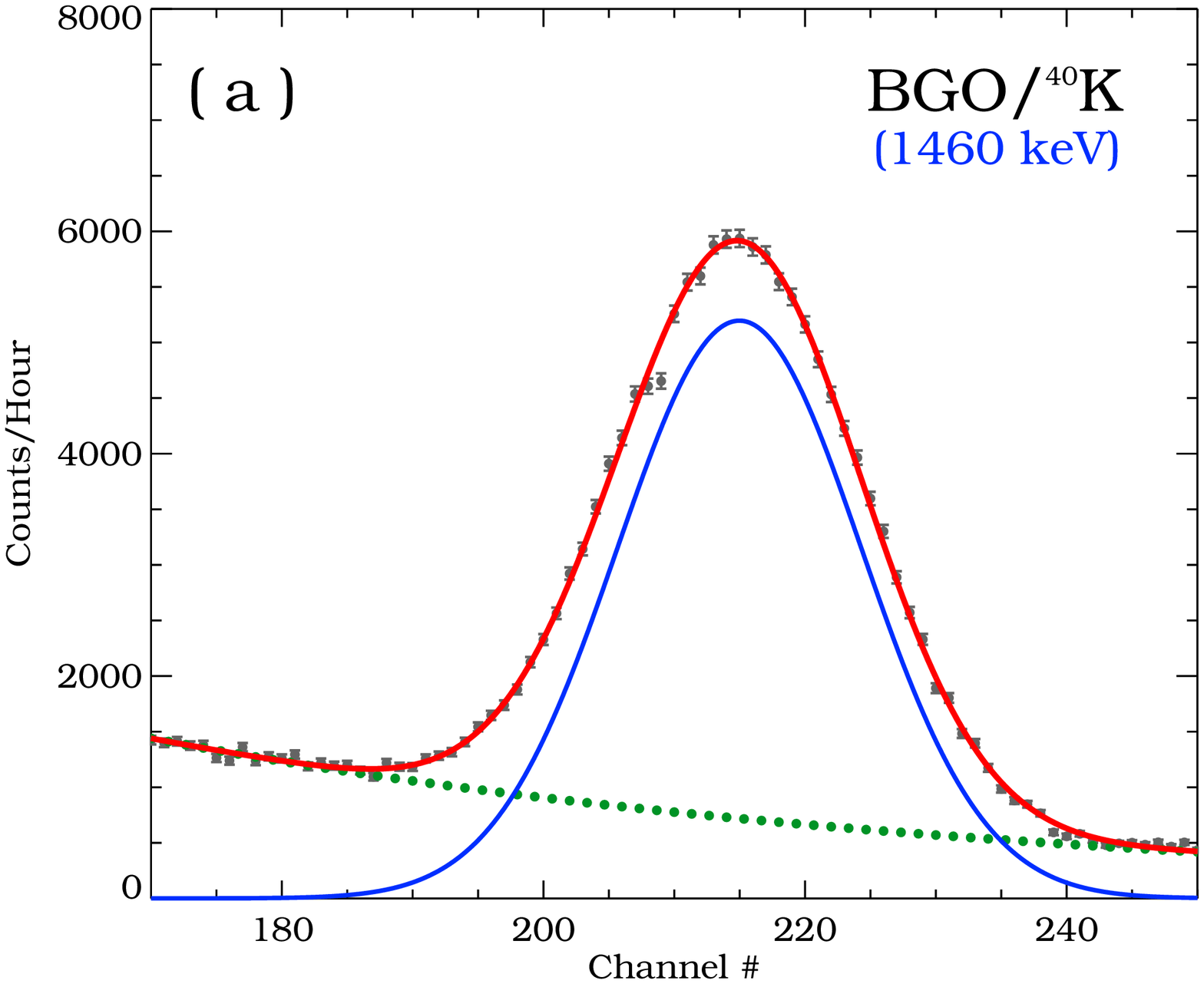}   &                
\includegraphics[height=42mm,bb=0 13 590 460,clip]{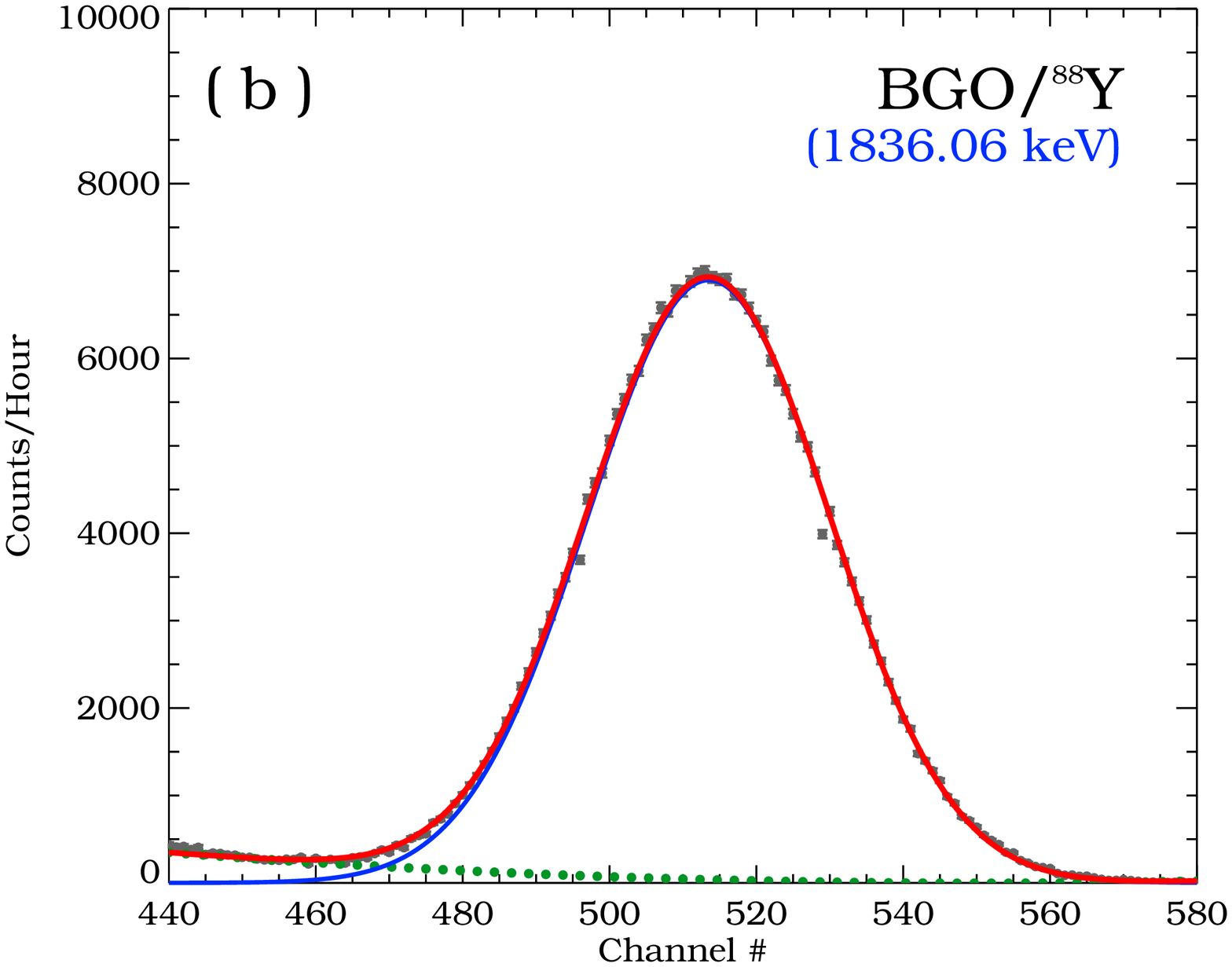}   \\
\includegraphics[height=42mm,bb=0 13 590 460,clip]{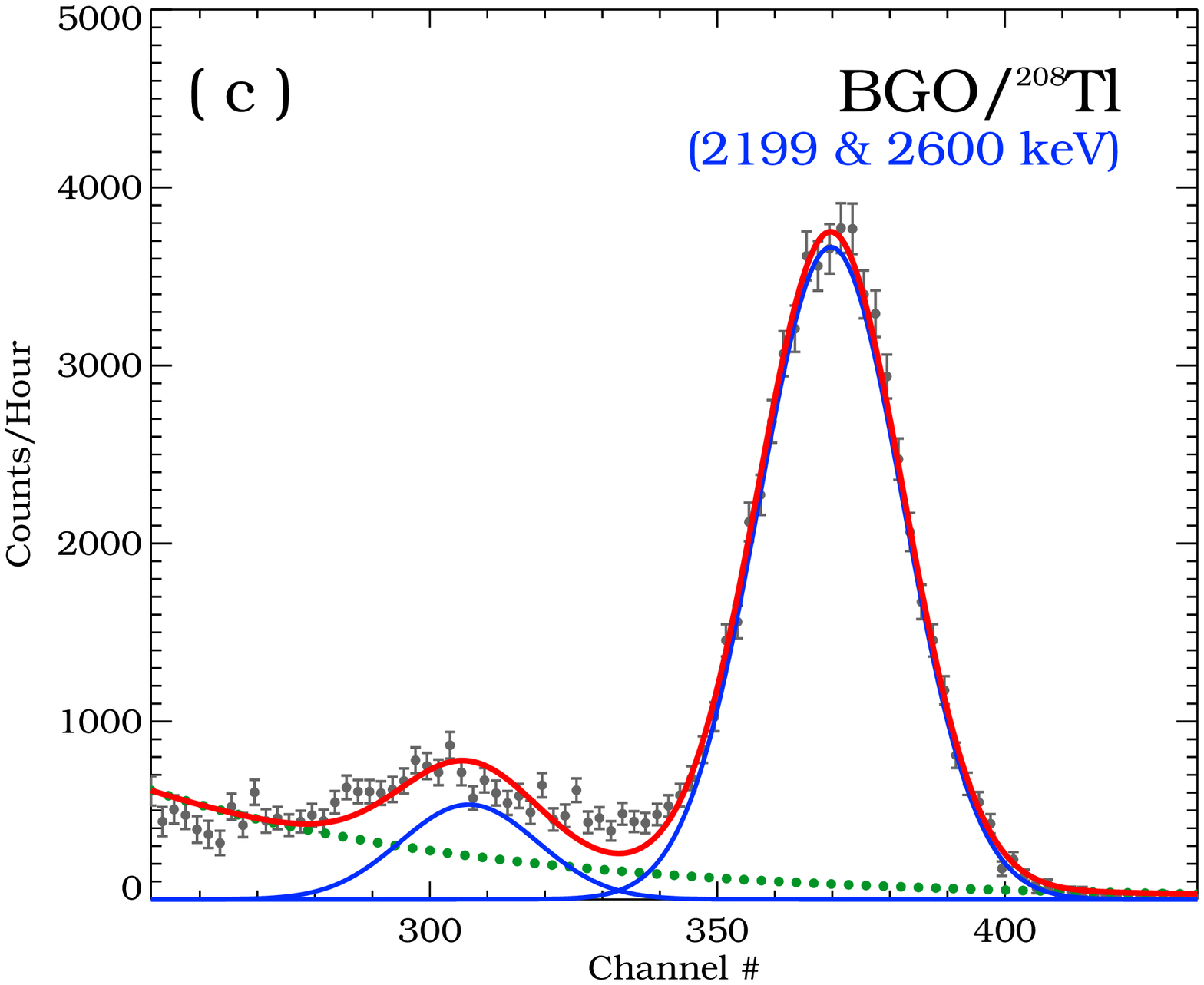}   &            
\includegraphics[height=42mm,bb=0 13 590 460,clip]{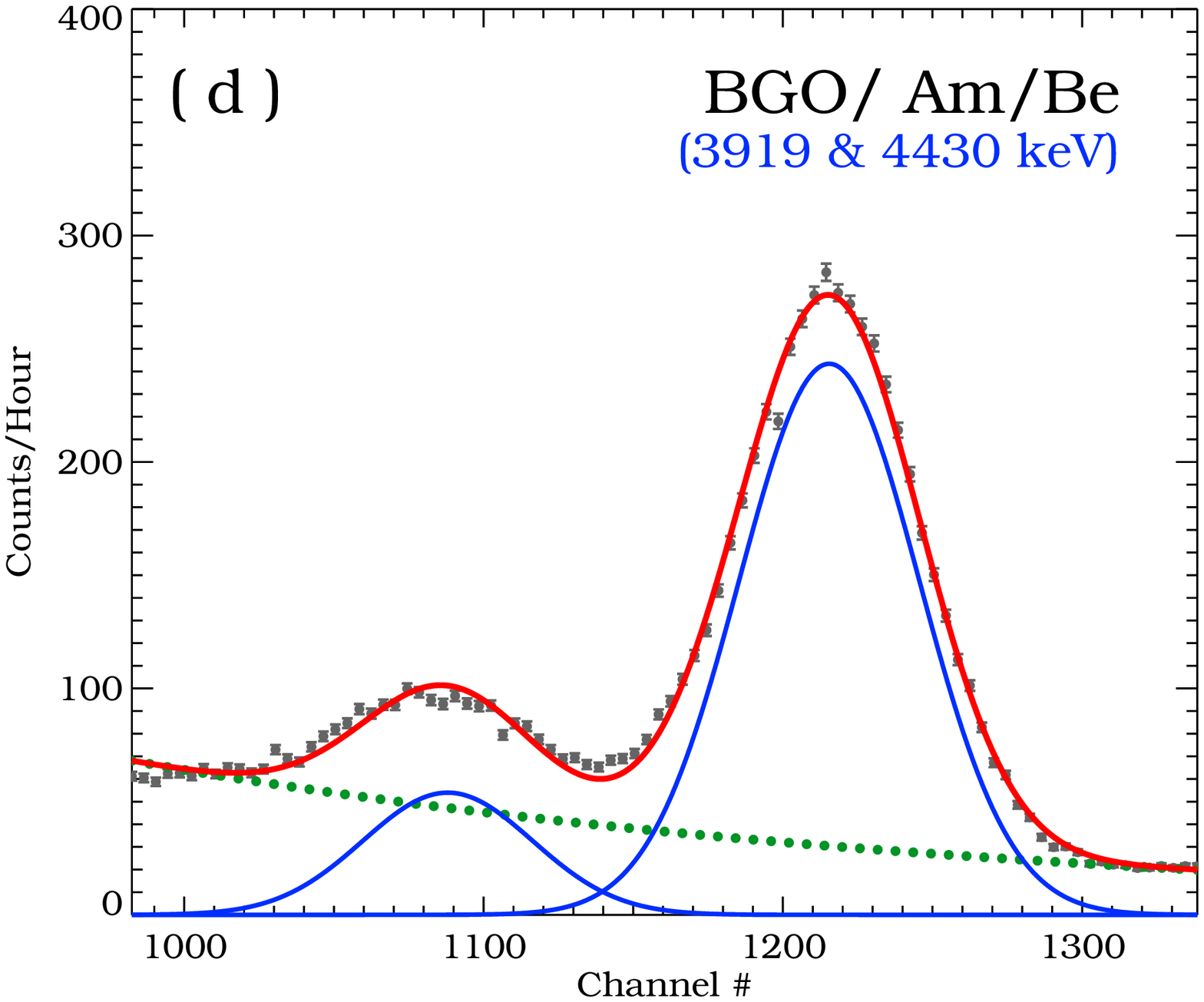}   \\
\includegraphics[height=42mm,bb=0 13 590 460,clip]{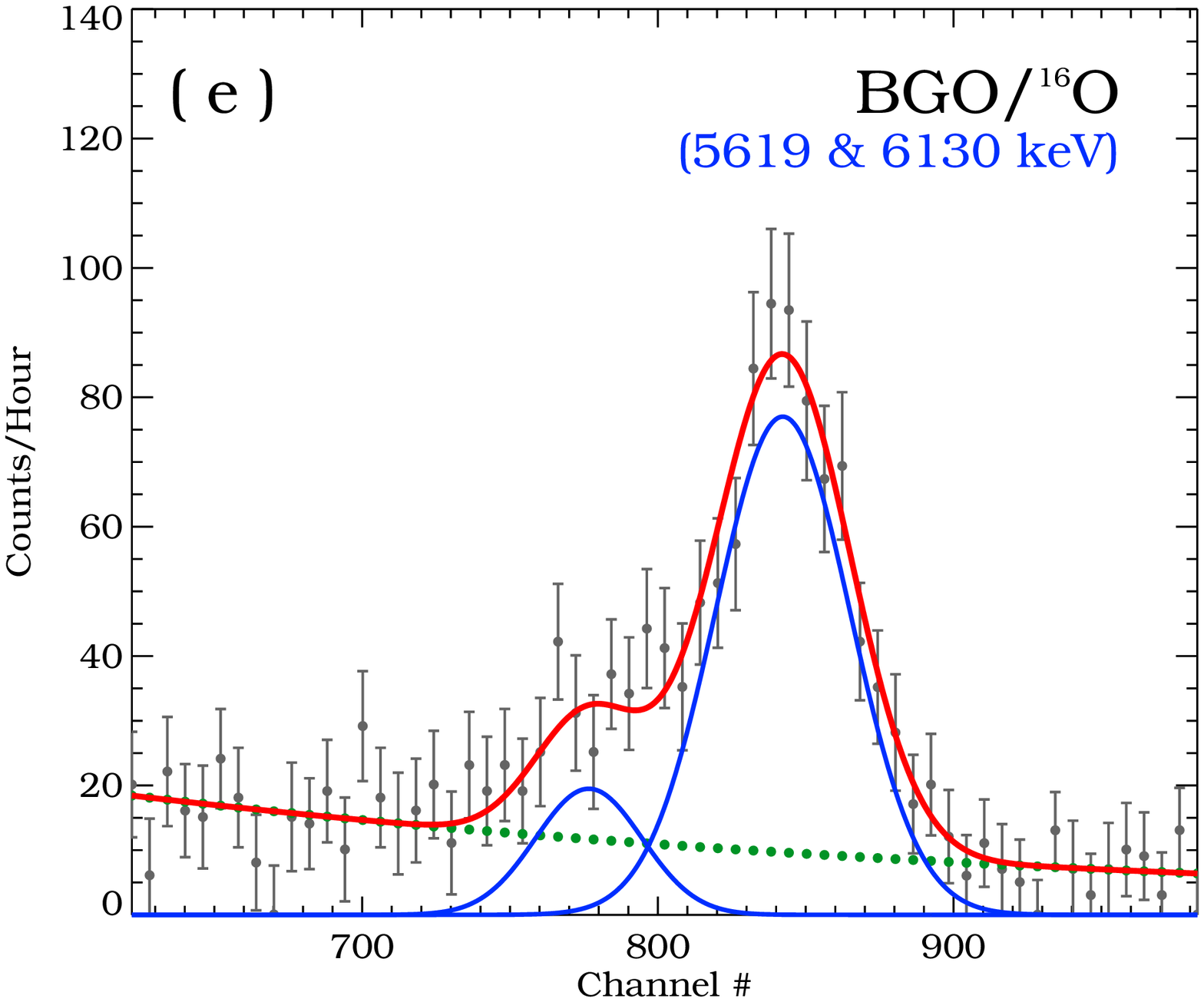}   &          
\includegraphics[height=42mm,bb=0 13 590 460,clip]{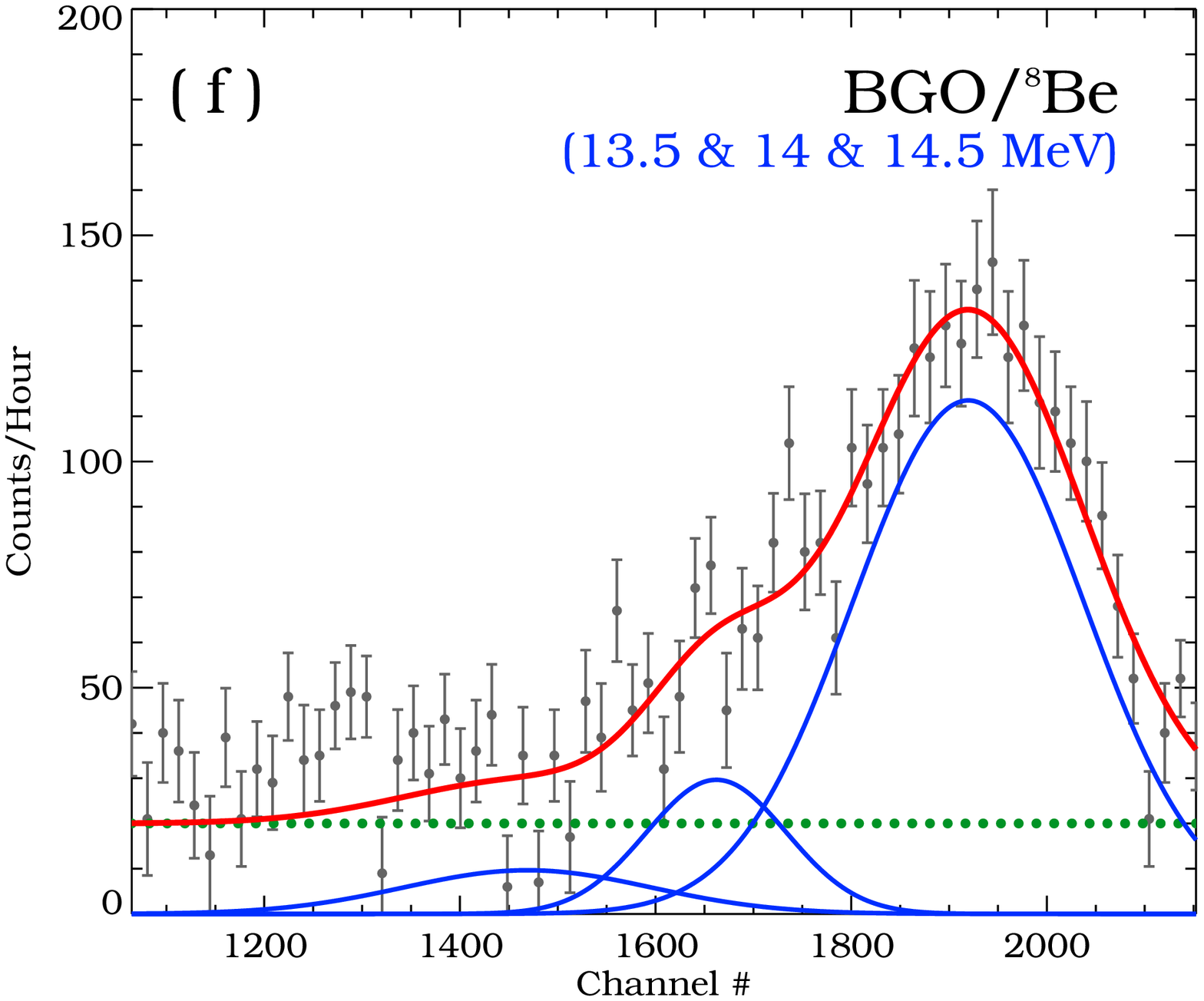}   \\
\multicolumn{2}{c}{\includegraphics[height=0.22\textheight,bb=0 13 590 460,clip]{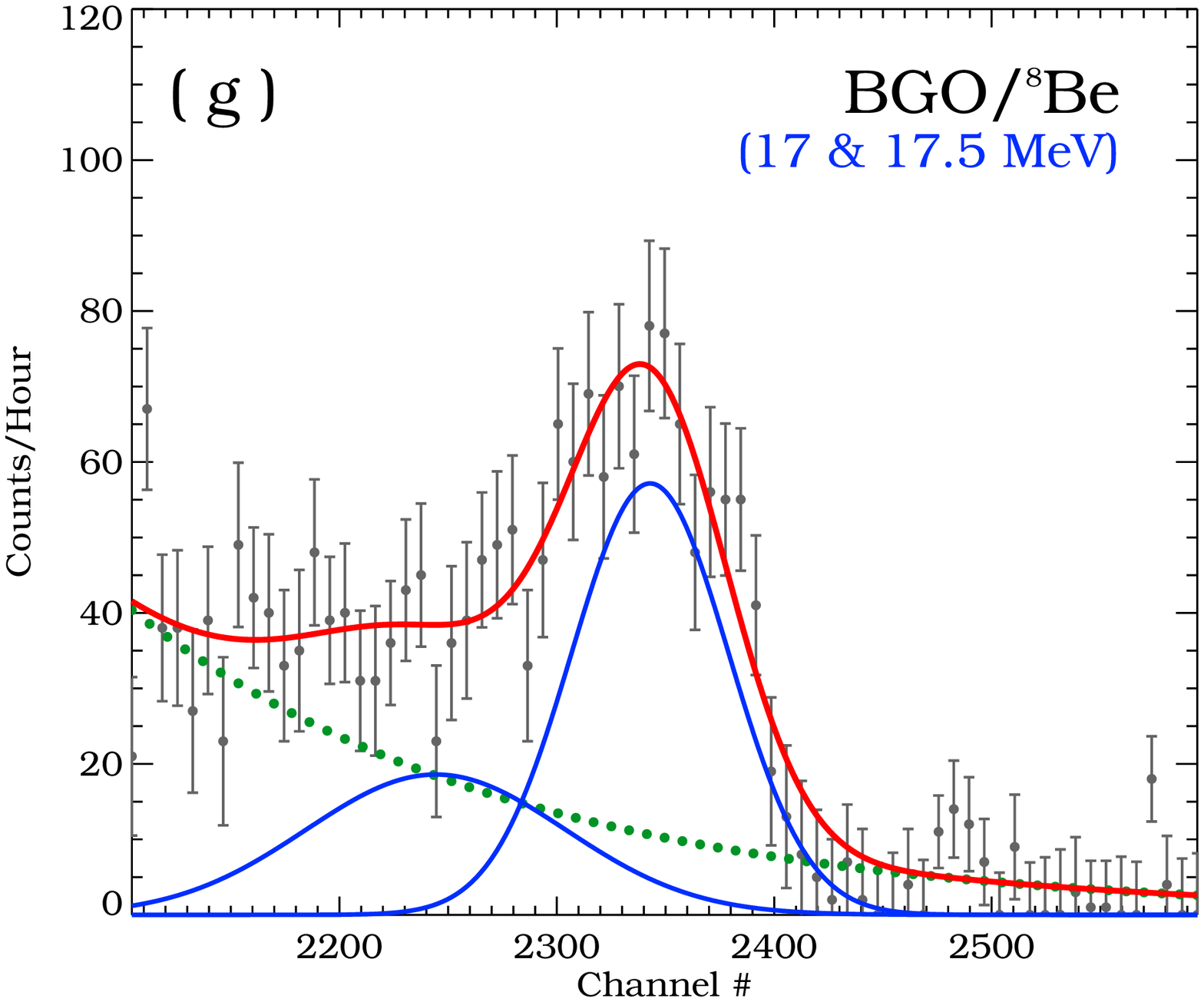}}    
\end{tabular}
\caption{
Full-energy peak analysis of BGO lines.
Data points (in {\it black}) are plotted with statistical errors.
Line fits ({\it solid red curves}) arise from the superposition of different components:
(i) one (or more) Gaussian functions describing the full-energy peak(s)
and the pair production escape peaks ({\it solid blue curves});
(ii) a constant, linear, quadratic or exponential function accounting
for background contributions ({\it dotted green curves})
}
\label{BGO_all_2}
\end{figure}
%
%

The fitting results for 17 lines measured by NaI FM 04 are presented 
graphically in Figures \ref{NaI_all_1} and \ref{NaI_all_2}. In each panel the 
fitted line energies are given in the top right corner. 
Fits to the data are shown in {\it red}. Gaussian 
components, describing the full-energy peaks, and
background components are shown as {\it solid blue} and 
{\it dotted green curves}, respectively. 
{\it Dashed blue curves} represent either background 
(Fig. \ref{NaI_all_1}, {\it panels b, c, d}, and {\it f}) or
Iodine escape peaks (Fig. \ref{NaI_all_2}, {\it panels b} and {\it c})
For energies above 279.2 keV (Fig. \ref{NaI_all_2}, {\it panels d} to {\it h}), 
the Iodine escape peak is no longer fitted as an extra
component but is absorbed by the full-energy peak.
The tails observed at energies lower than 20 keV
(Fig. \ref{NaI_all_1}, {\it panels a} and {\it b})
are supposed to be due to scattering from the entrance
window materials or to some L-shell escape X-rays.

Results for 16 fitted BGO lines are shown in Figures \ref{BGO_all_1} and \ref{BGO_all_2}.
Each full-energy peak was modeled with a single Gaussian ({\it solid blue curves}) over an exponential background
({\it dotted green curves}). The line from $^{57}$Co at 124.59 keV (Fig. \ref{BGO_all_1}, {\it panel a}) lies outside 
the nominal BGO energy range (150 keV $-$ 30 MeV) and shows a strong asymmetric broadening on the left of
the full-energy peak, which can be described by an additional Gaussian component ({\it dashed blue curve}).
Panels {\it a, c, e, f} and {\it g} of Fig. \ref{BGO_all_2} show fitted lines from spectra taken at SLAC with BGO EQM.
For some spectra, energies of the first and second pair production escape peaks, which lie 
$\sim$~511 keV and $\sim$~1 MeV below the full-energy peak, respectively,
are reported in the top right of each plot. Some of these secondary lines were 
included in the determination of the BGO channel-energy relation (see Section \ref{BGO_CE}).
%
%
%
%
\begin{figure}[b!]
\centering
\includegraphics[width=84mm,bb=20 10 575 500,clip]{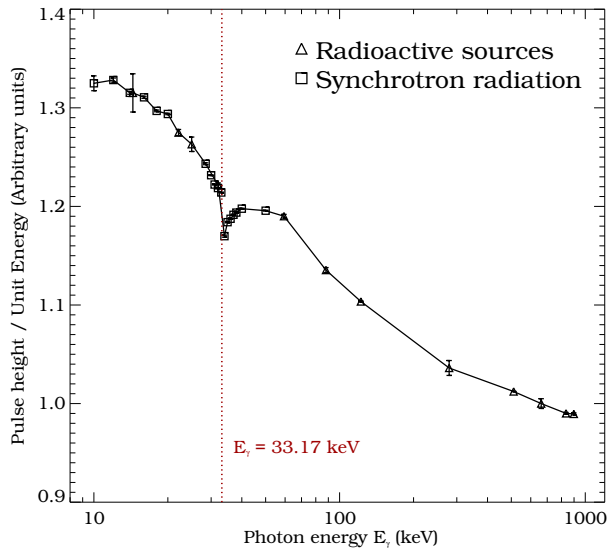}
\caption{The differential linearity measured for NaI FM 04, normalized to unity at 661.66 keV}
\label{figNaI_nonlin}
\end{figure}
%
\subsection{Channel-Energy Relation}\label{CE Relation}
\subsubsection{NaI Nonlinear Response}\label{NaI CE}
Several decades of experimental studies of the response of NaI(Tl) to
gamma rays have indicated that the scintillation efficiency mildly varies with 
the deposited energy \cite{ENG56,IRE61,PRE69,MOS02}. Such nonlinearity must be correctly taken into
account when relating the pulse-height scale (i.e. the channel numbers) to gamma-ray energies.
Fig. \ref{figNaI_nonlin} shows the pulse height per unit energy 
(normalized to a value of unity at 661.66 keV) versus incident 
photon energy E$_{\gamma}$ as measured by detector NaI FM 04. The data points include
radioactive source measurements performed at MPE ({\it triangles})
together with additional low-energy measurements taken at PTB/BESSY between 10 and 60 keV
({\it squares}). In this case, nonlinearity clearly appears as a dip in the plot 
at a characteristic energy corresponding to the K-shell binding energy in Iodine, i. e. 33.17 keV 
(as previously mentioned in Section \ref{Proc_Runs}).  
Photoelectrons ejected by incident gamma-rays just above the K-shell absorption edge
have very little kinetic energy, so that the response drops. Just below this energy, however, 
K-shell ionization is not possible and L-shell ionization takes place. 
Since the binding energy is lower, the ejected photoelectrons 
are more energetic, which causes a rise in the response.
%
%
%
\begin{figure}[t!]
\centering 
\begin{tabular}{cc}
\includegraphics[width=55mm,bb=25 10 556 491,clip]{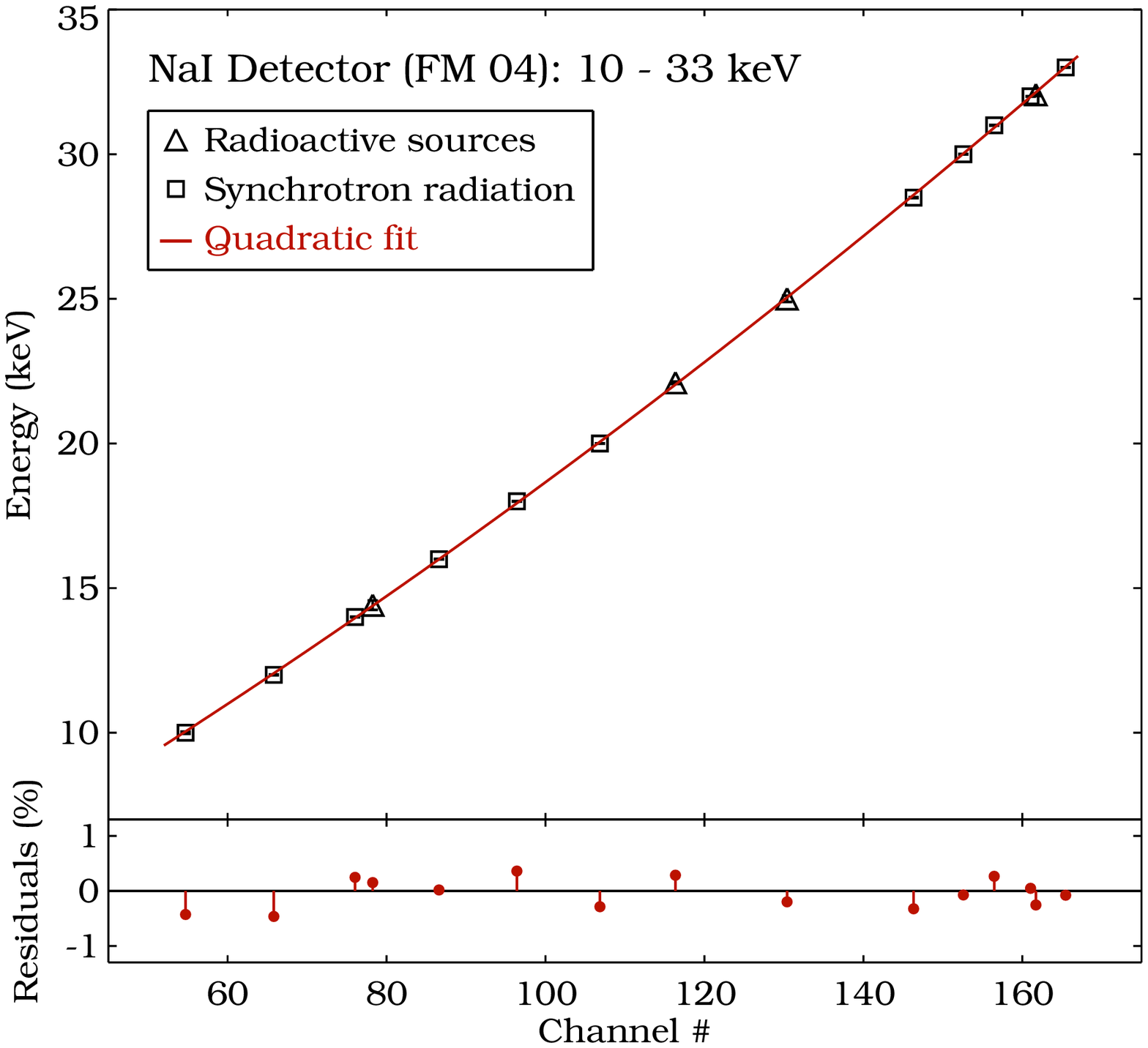}   & 
\includegraphics[width=55mm,bb=15 10 556 491,clip]{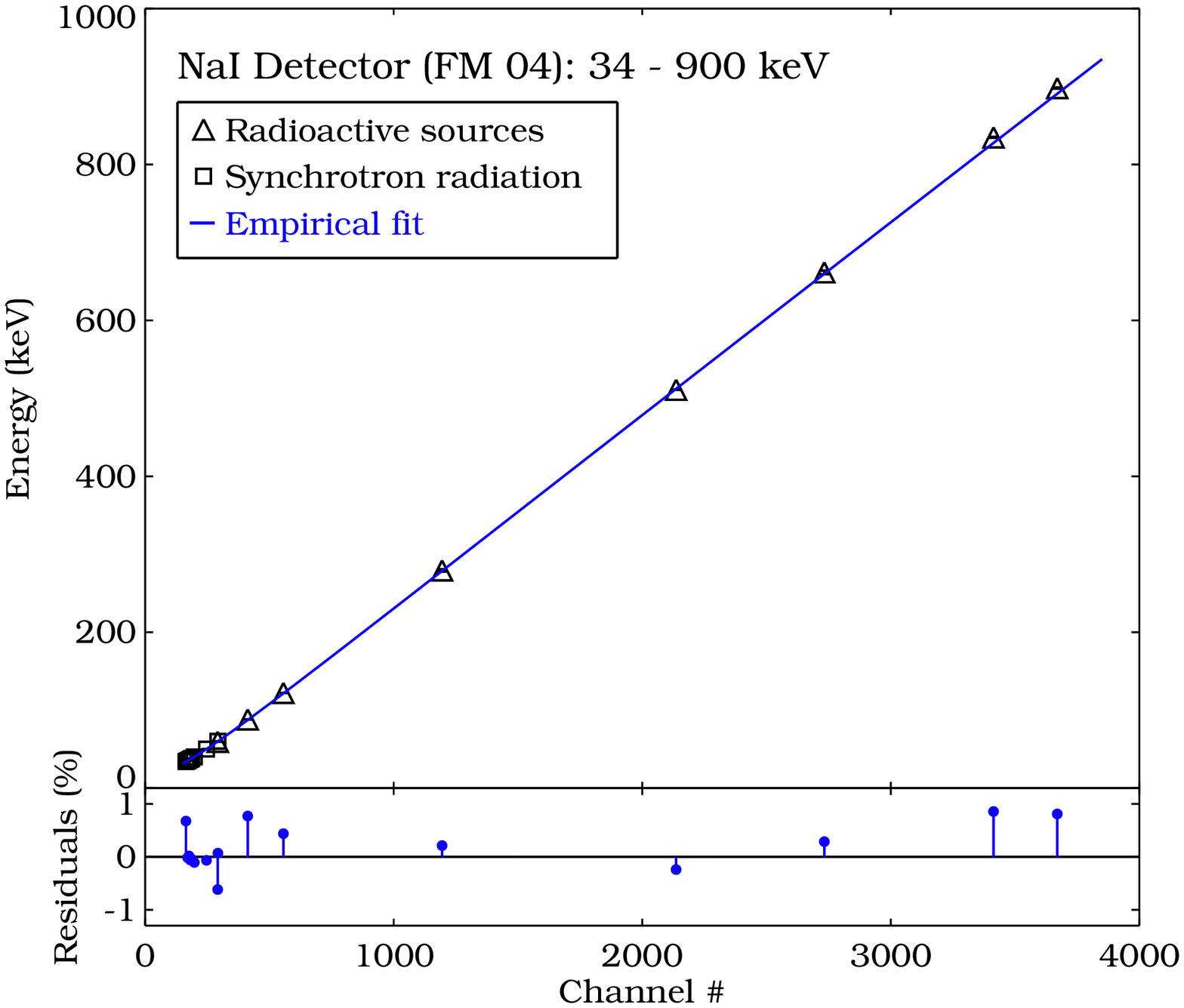}
\end{tabular}
\caption{
Channel-energy relation calculated below ({\it left panel}) and 
above ({\it right panel}) the K-edge energy for detector NaI FM 04. 
MPE data points ({\it triangles}) and PTB/BESSY data points ({\it squares}) 
are fitted together with a second degree polynomial below 33.17 keV 
({\it left panel, red curve}) and with the empirical 
function (see Eq. \ref{fitfunc}) above 33.17 keV ({\it right panel, blue curve}).
Residuals to the fits are given in the panel under the respective plots.
For the quadratic fit below the K-edge, following fit parameters were obtained: 
$a$ = 1.08 $\pm$ 0.05, $b$ = 0.1495 $\pm$ 0.0009, c = (2.64 $\pm$ 0.04) $\cdot$ 10$^{-4}$,
with a reduced $\chi^{\,2}$ of 23. In the case of the empirical fit above the K-edge, the fit parameters are
a = 94.7 $\pm$ 1.4,  b = 2.73 $\pm$ 0.09, c = 0.2306 $\pm$ 0.0008, d = -26.2 $\pm$ 0.4,
with a reduced $\chi^{\,2}$ of 60
}
\label{figNaI_FM04_low-high}
\end{figure}

The addition of measurements taken at PTB/BESSY with four NaI detectors 
(see Section \ref{Sec_BESSY}) for computing the NaI response
is particularly necessary in the region around the K-edge energy, since the 
radioactive sources used at MPE only sample it with four lines, three of which 
(22.1, 25 and 32.06 keV) belong to double peaks and the first line from 
$^{57}$Co at 14.41 keV shows asymmetries and broadening (see Fig. \ref{NaI_all_1}, {\it panel a}). 
From the collected PTB/BESSY data, line fitting results 
were obtained for 19 spectra collected at energies
between 10 and 60 keV. Corrections of gain settings between the detectors 
during the two different calibration campaigns were carefully taken into account.
%
%
\begin{figure}[t!]
\centering
\begin{tabular}{cc}
\includegraphics[height=53mm,bb=30 10 556 491,clip]{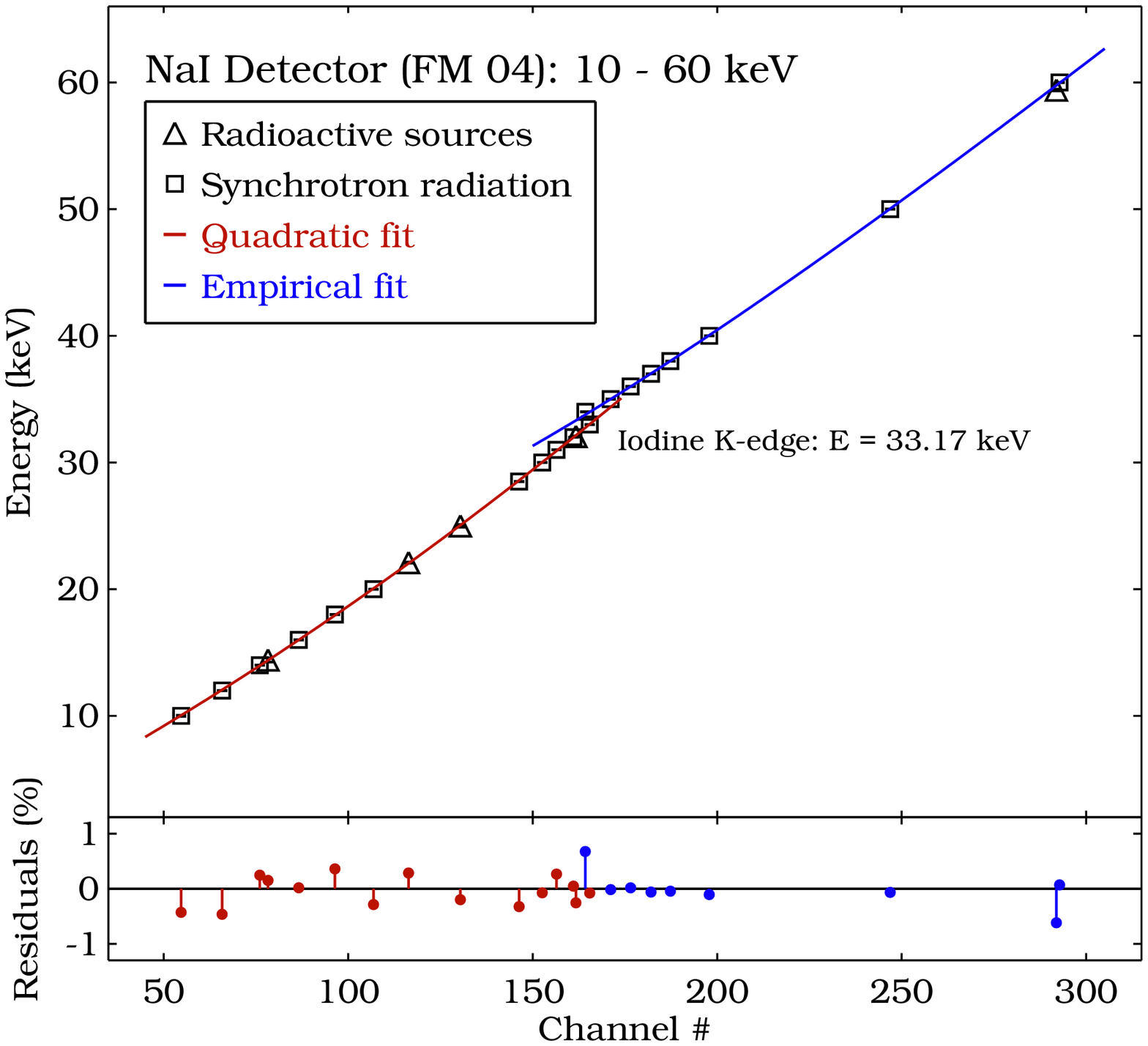}   & 
\includegraphics[height=53mm,bb=117 162 504 549,clip]{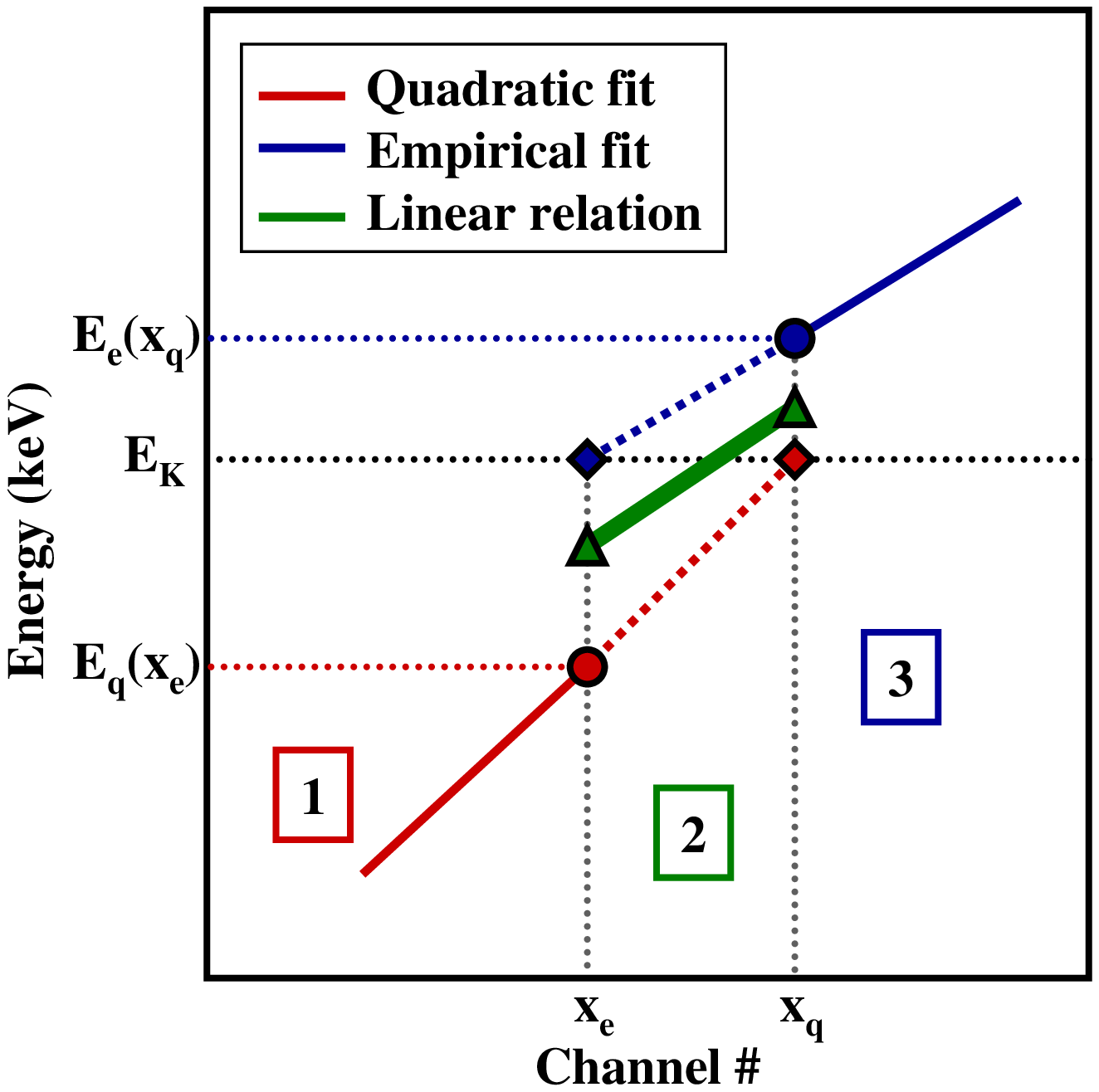}
\end{tabular}
\caption{
{\it On the left}: Channel-energy relation around the Iodine K-edge for detector NaI FM 04.
Data points from radioactive source lines ({\it triangles}) and from synchrotron radiation ({\it squares}) 
are fitted together with a quadratic function below the K-edge energy ({\it red curve})
and with the empirical function above the K-edge energy ({\it blue curve}).
Residuals to the fits are given in the panel under the plot. 
{\it On the right}: 
Schematic representation of the channel-energy relation around the 
Iodine K-edge energy. In region 1, for $x$ < $x_e$, the quadratic relation is adopted
({\it red curve}). In region 3, for $x$ > $x_q$, the 
empirical function is adopted ({\it blue curve}).
The ambiguity arises at channel $x_e$ (region 2), which could in principle be equally 
described by both relations, i.e. we could assign to it either an energy $E_q(x_e)$
({\it red circle}) or the K-edge energy $E_K$ ({\it blue diamond}).
In order to avoid this ambiguity, a linear relation ({\it green curve}) 
has been derived in order to assign an average energy value to each channel
in region 2 ({\it green triangles})
}
\label{Schema}
\end{figure}

In order to compute a valuable NaI channel-energy relation, the energy range was initially
split into two regions, one below and one above the  K-edge energy. 
For E < 33.17 keV, data were fitted with a second degree polynomial (parabola), while
for E > 33.17 keV the following empirical function was adopted:
\begin{equation}\label{fitfunc}
E(x_{\,c}) = a \; + \; b  \cdot \sqrt{x_{\,c}} \; + \; c \cdot x_{\,c} \; + \; d \cdot \ln{x_{\,c}} \; \; ,
\end{equation}
where $E$ is the line energy in keV and $x_{\,c}$ is the line-center position in channels.

Fig. \ref{figNaI_FM04_low-high} shows an example of the channel-energy relation calculated for detector FM 04
in the low-energy ({\it left panel}) and in the high-energy range ({\it right panel}). 
Radioactive sources data ({\it triangles}) and PTB/BESSY data ({\it squares}) are fitted together with a second degree 
polynomial ({\it left panel, red curve}) and with the empirical function of Eq. 
\ref{fitfunc} ({\it right panel, blue curve}). 
Analysis of all detectors shows similar results. In particular, all calculated relations give fit residuals below 1~\%,
as required. The obtained fit parameters are reported in the figure's caption.
\subsubsection{The Iodine K-edge region}\label{Iodine_Reg}
By taking a closer look to the region around the K-edge energy (see Fig. \ref{Schema}, {\it left panel}), 
the discrepancy between low and high-energy fits becomes clearly visible. Thus,
assigning a unique energy to every channel is a more delicate issue. 
A possible way to solve such an ambiguity is to 
divide the channels domain into three parts, as shown in Fig. \ref{Schema}, {\it right panel}.
$x_e$ and $x_q$ represent the channels were the empirical and quadratic fit, respectively, assume the
value of $E_K$ = 33.17 keV ({\it red} and {\it blue diamonds}). For all channels in the interval $x_e$ < $x$ < $x_q$, 
a linear relation was calculated in order to assign an average energy to each channel:
The {\it green triangles} represent the average energies of the two energies calculated with
both relations ({\it diamonds} and {\it circles}).
For the analysis of laboratory calibrations, we calculated a K-edge region (region 2) 
of about five to seven channels for every NaI detector. 
On orbit however, due to a much smaller number of channels
(128), the flight DPU groups these channels into one ``transition'' 
channel (per detector), which is calculated through lookup tables.
%
%
\begin{figure}[t!]
\centering
\begin{tabular}{cc} 
\includegraphics[height=42mm,bb=0 10 590 458,clip]{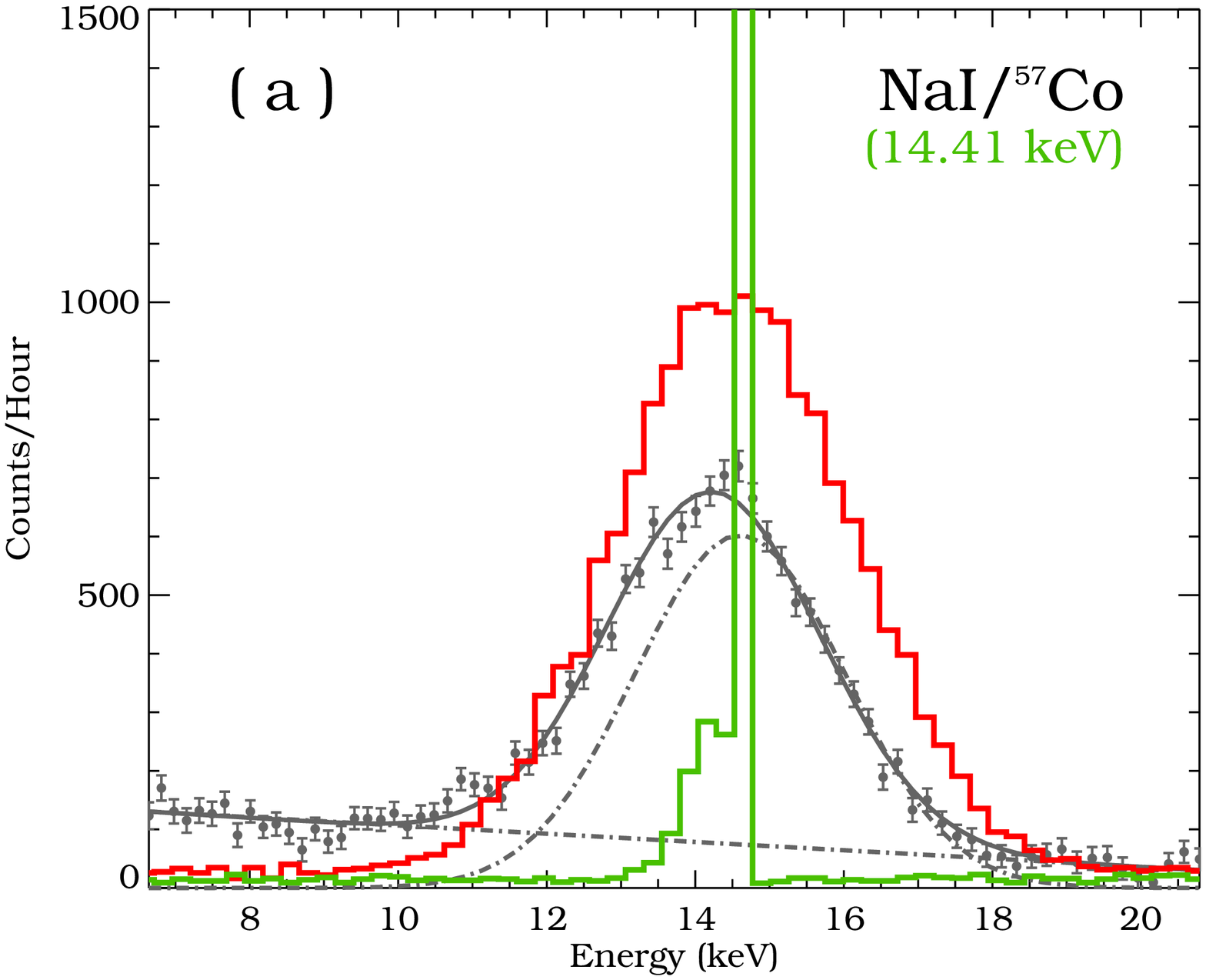}  & 
\includegraphics[height=42mm,bb=0 10 590 458,clip]{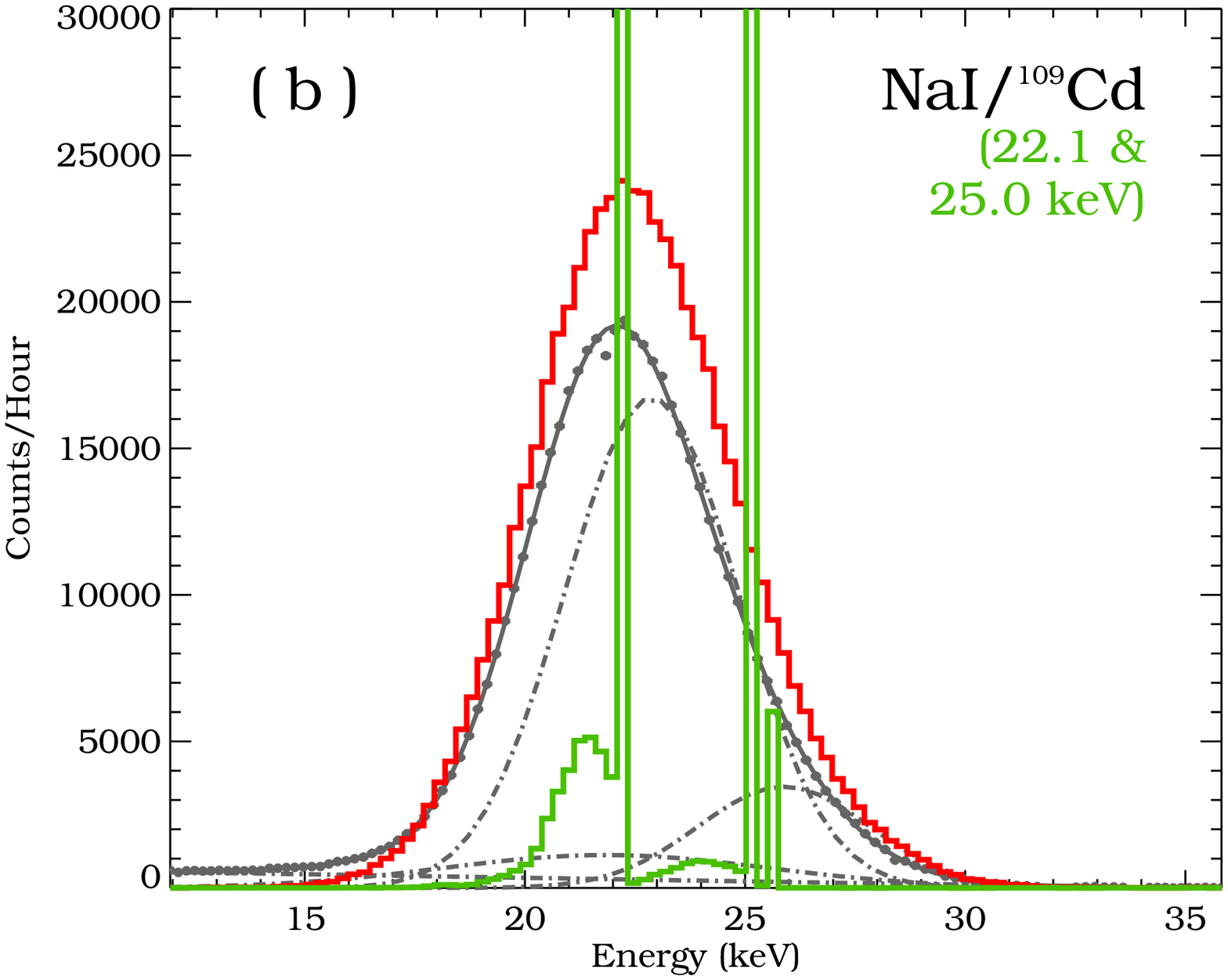}  \\
\includegraphics[height=42mm,bb=0 10 590 458,clip]{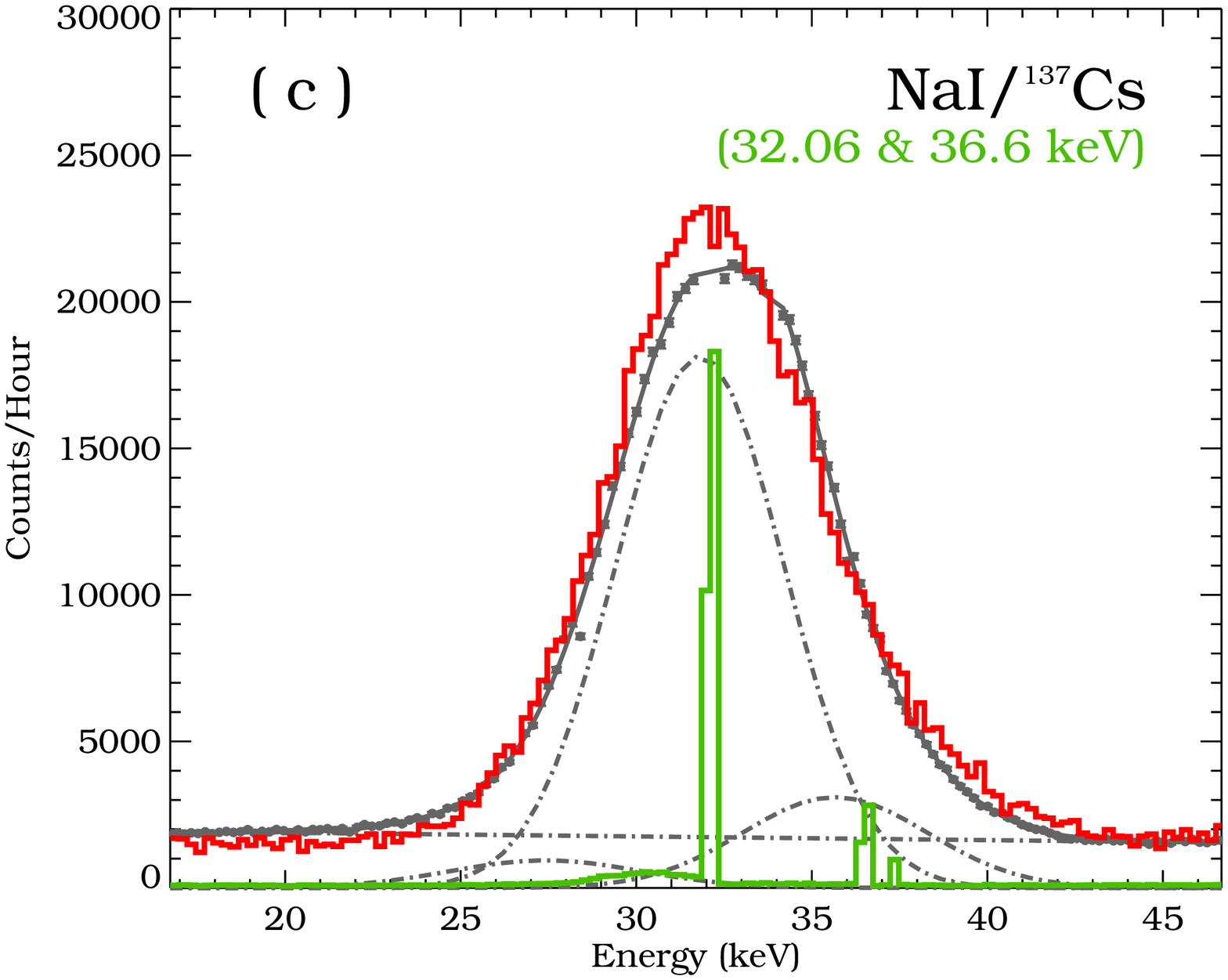}  &            
\includegraphics[height=42mm,bb=0 10 590 458,clip]{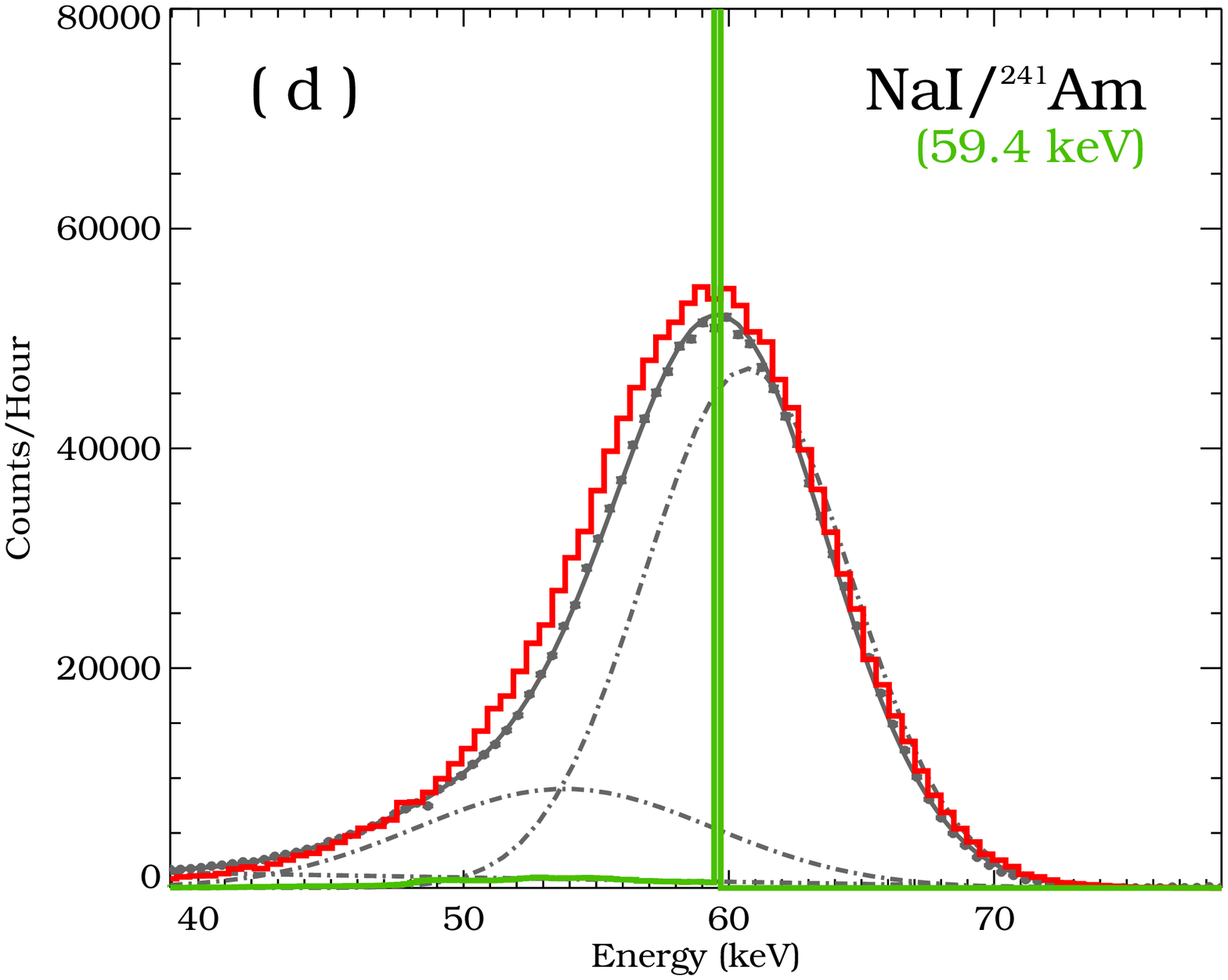}  \\
\includegraphics[height=42mm,bb=0 10 590 458,clip]{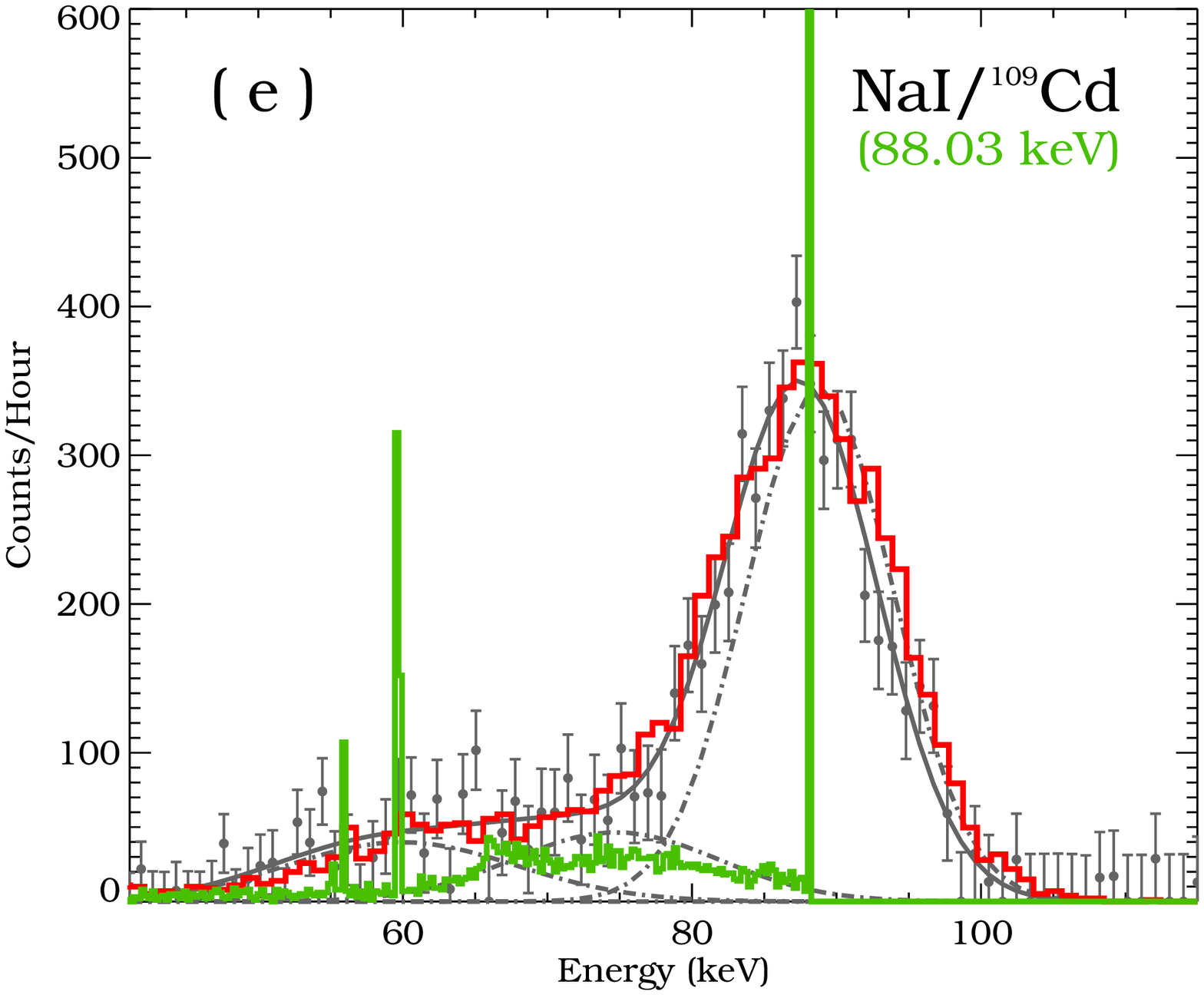}   &                
\includegraphics[height=42mm,bb=0 10 590 458,clip]{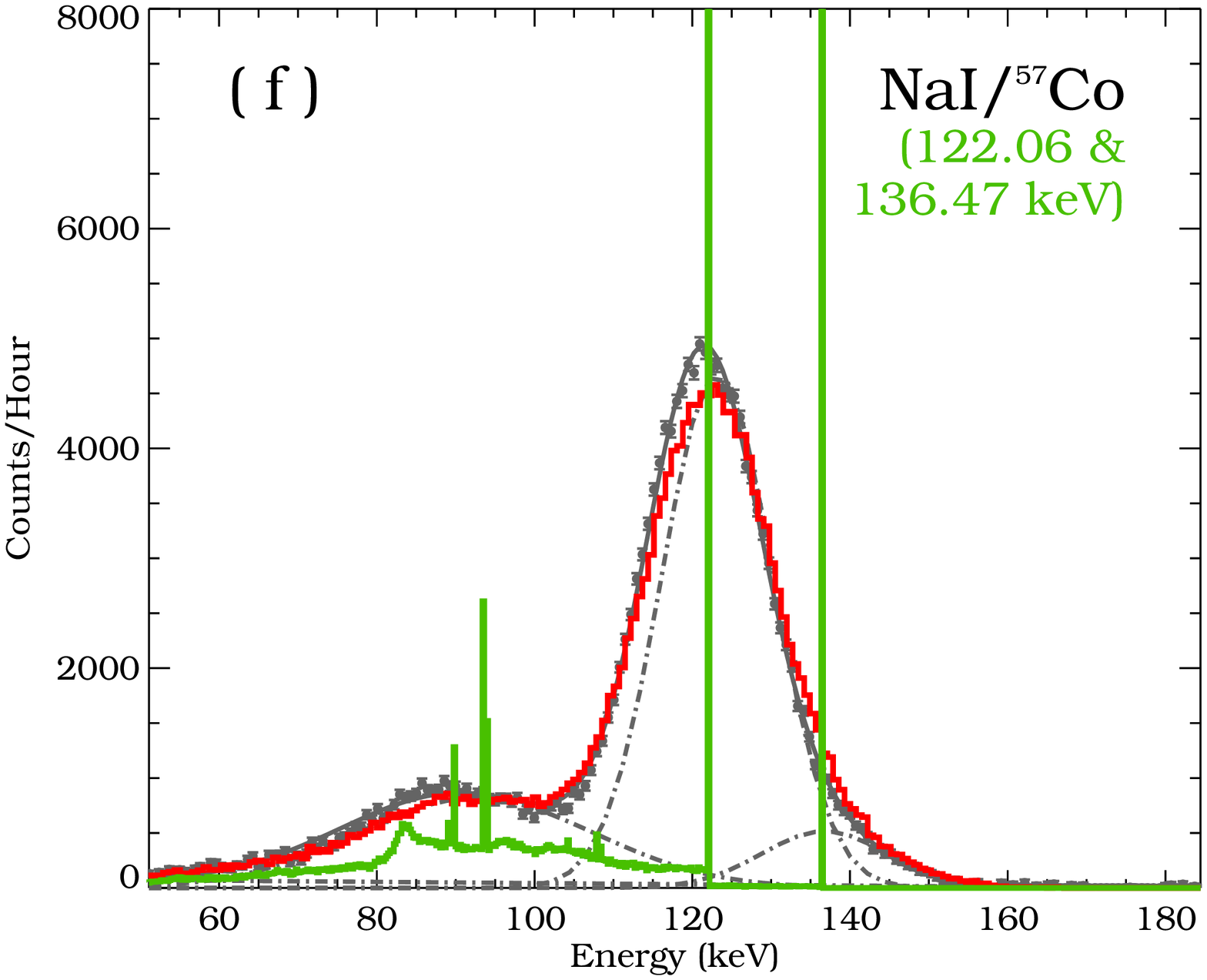}
\end{tabular}
\caption{
Comparison between spectra collected with NaI FM 12, after channel-energy conversion, 
and simulated spectra for 4 radioactive source lines.
Here, line fits and various components, previously described in Figures
\ref{NaI_all_1} and \ref{NaI_all_2}, are shown as {\it solid}
and {\it dashed grey curves}, respectively.
Simulated broadened spectra are overplotted in {\it red}. 
Unbroadened spectra ({\it green histograms}) can be considered as guidelines
for the exact energy positions of lines and secondary escape peaks (see text).
The height of the unbroadened peaks is truncated in most plots to
better show the comparison between real and broadened spectra
}
\label{NaI-SIM_1}
\end{figure}
%
%
\begin{figure}[t!]
\centering
\begin{tabular}{cc} 
\includegraphics[height=42mm,bb=0 10 590 458,clip]{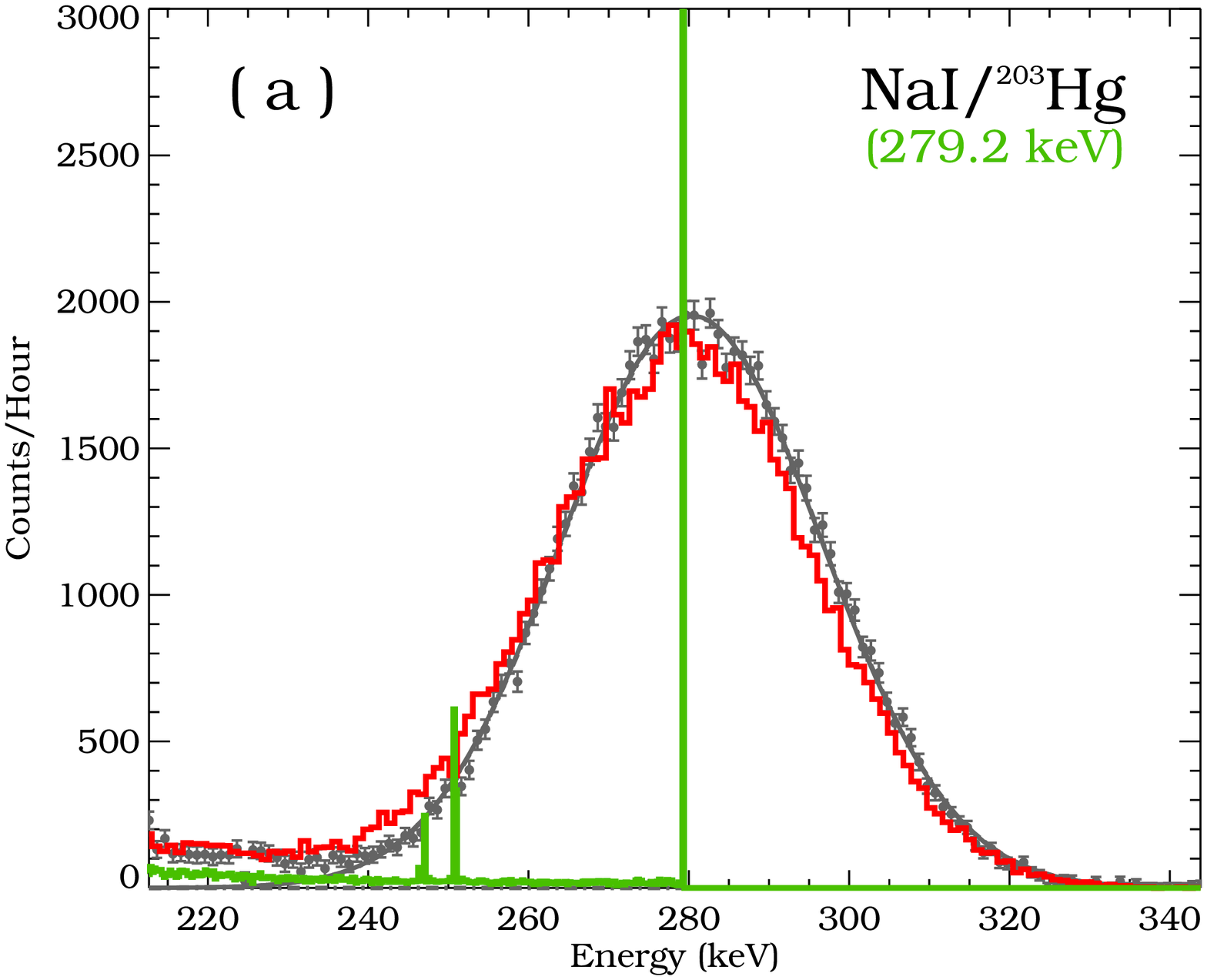}   &    
\includegraphics[height=42mm,bb=0 10 590 458,clip]{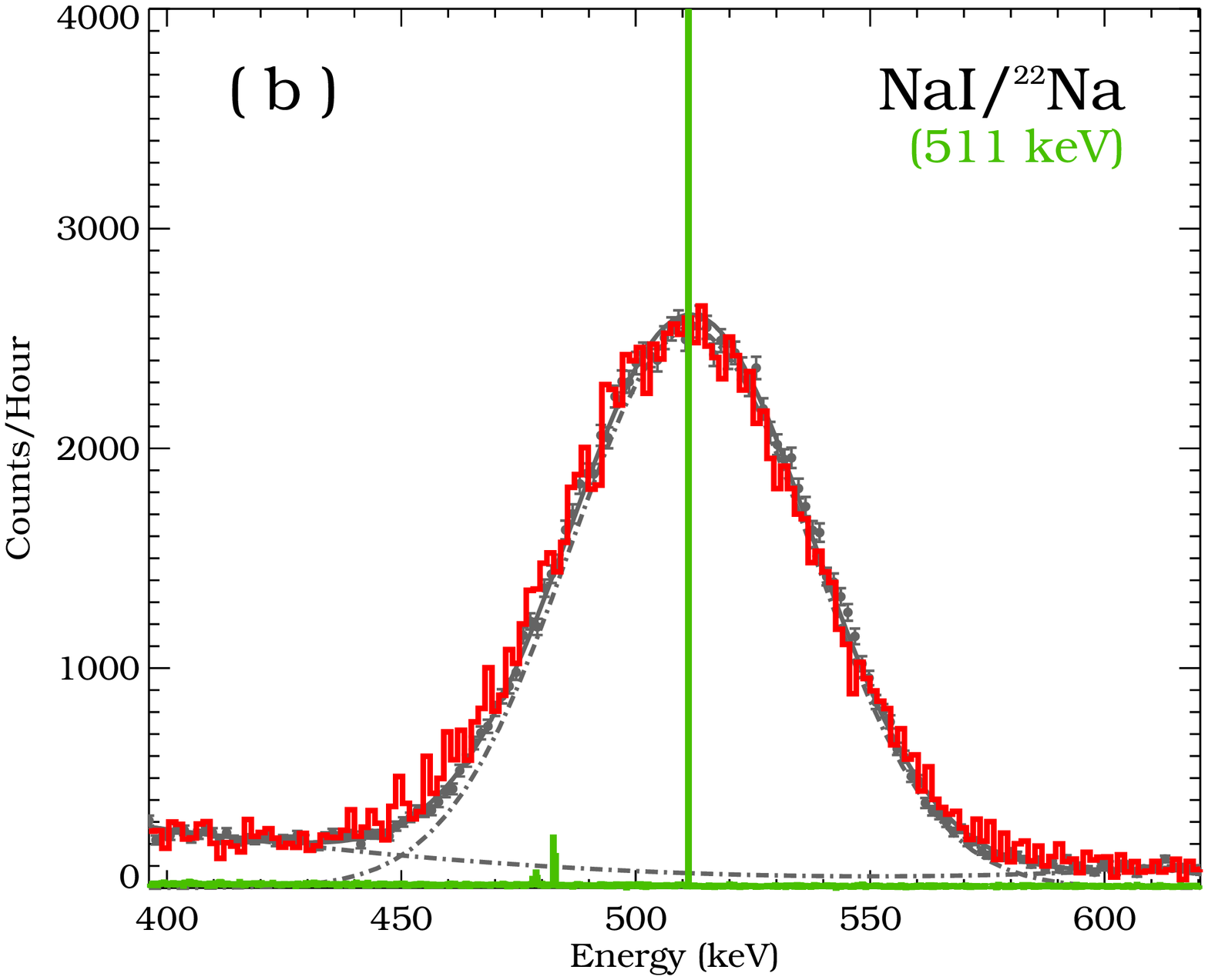}   \\
\includegraphics[height=42mm,bb=0 10 590 458,clip]{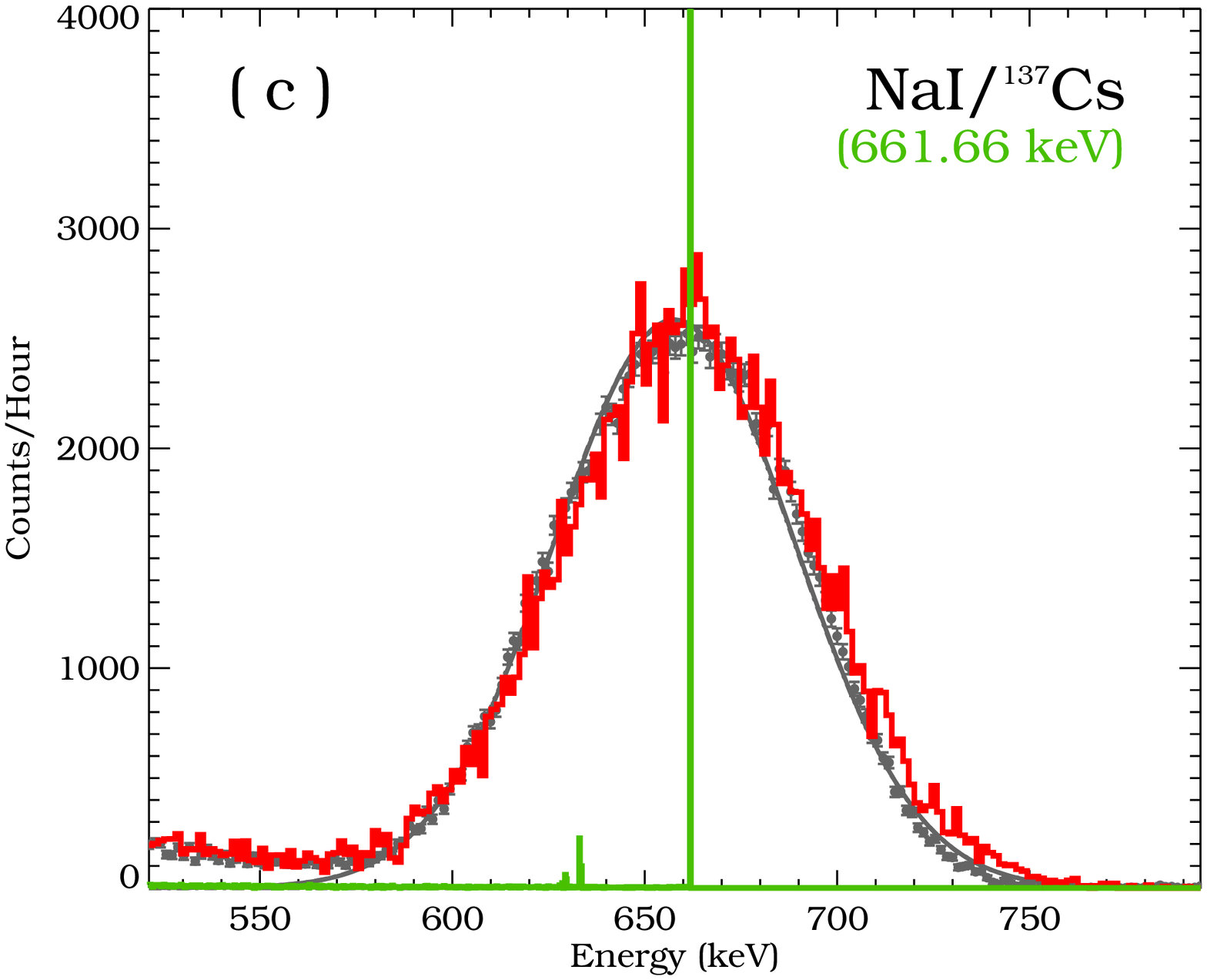}   &   
\includegraphics[height=42mm,bb=0 10 590 458,clip]{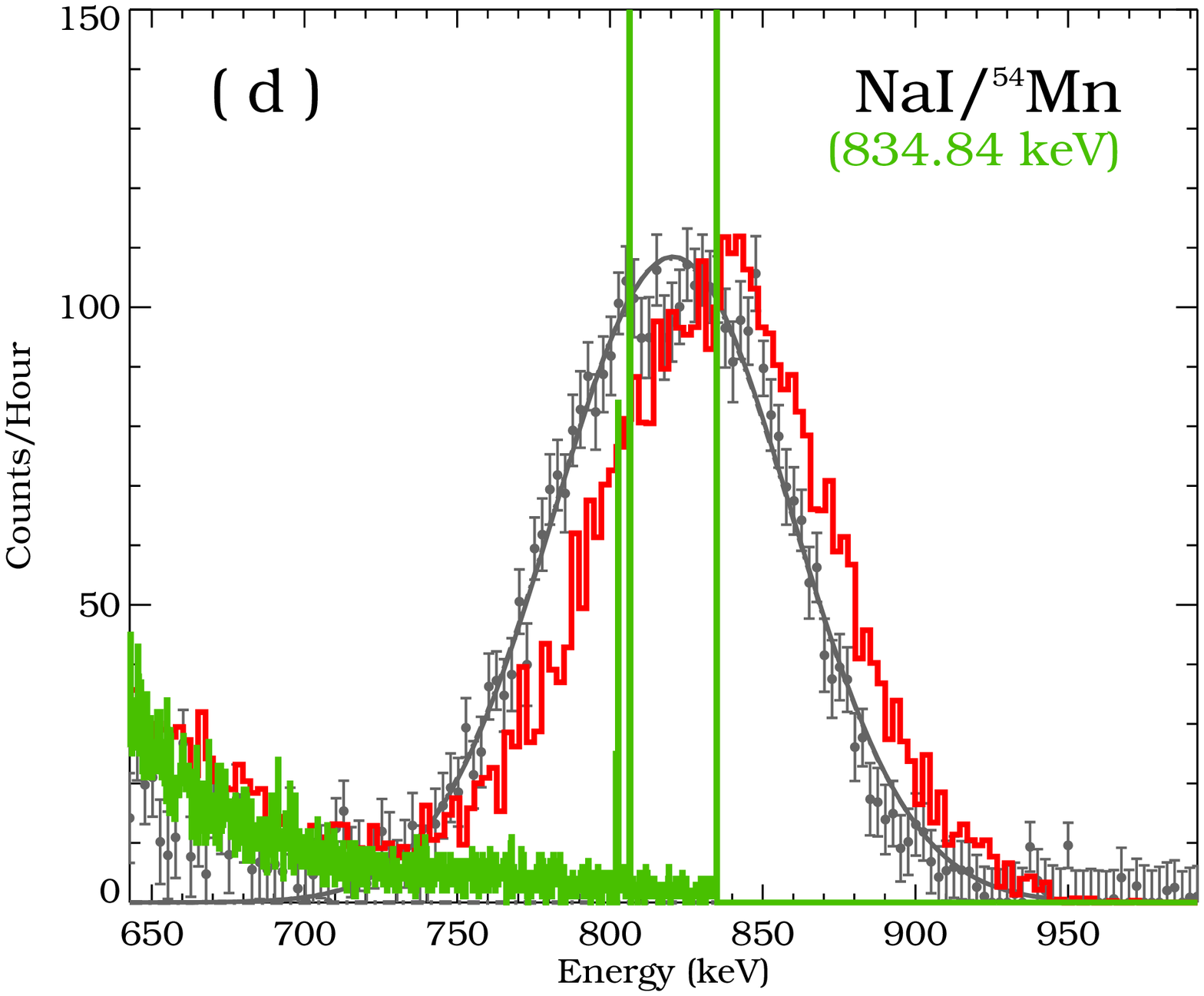}   \\
\multicolumn{2}{c}{\includegraphics[height=42mm,bb=0 10 590 458,clip]{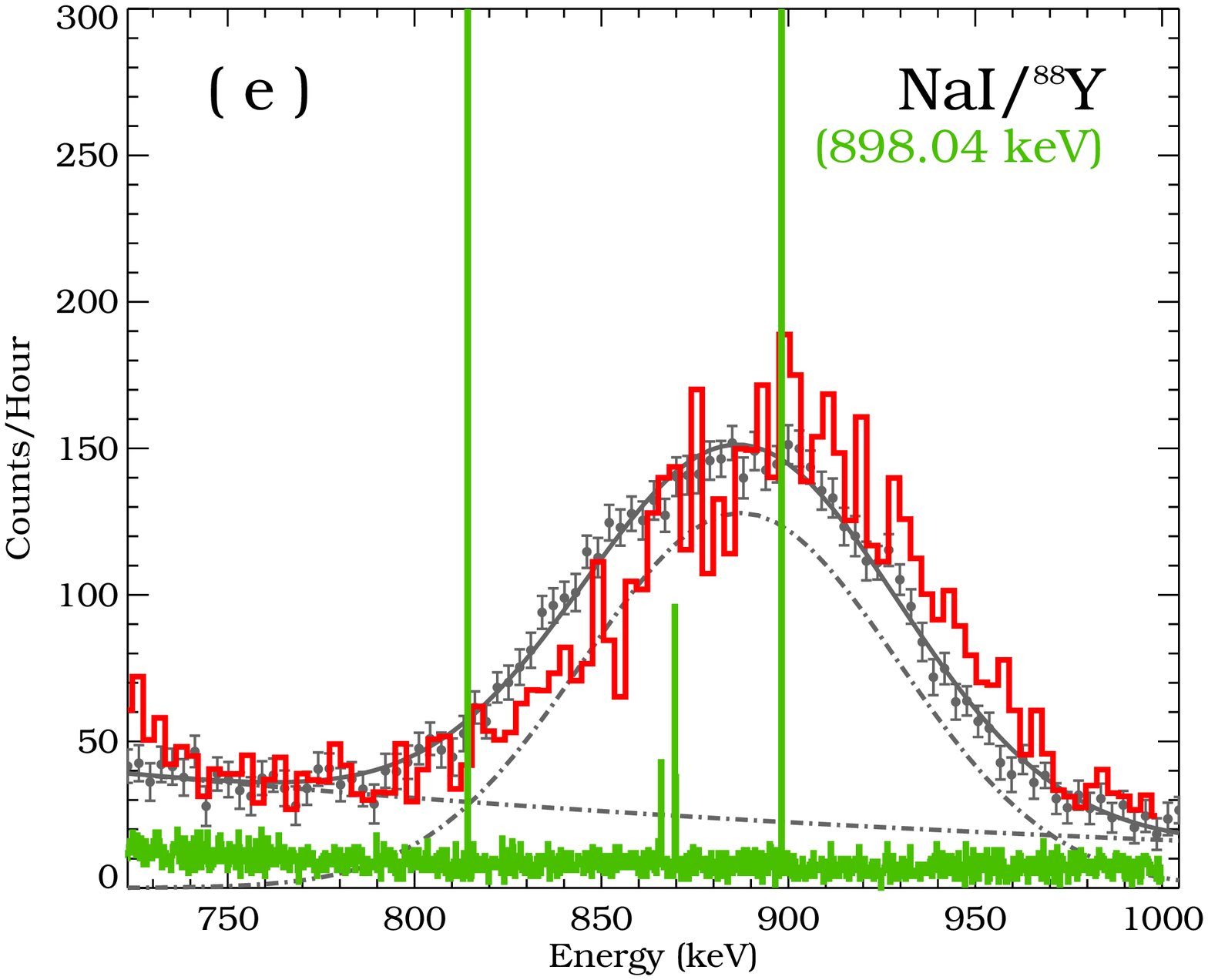}}
\end{tabular}
\caption{
Comparison between spectra collected with NaI FM 12, after channel-energy conversion, 
and simulated spectra for 7 radioactive source lines.
Here, line fits and various components, previously described in Figures
\ref{NaI_all_1} and \ref{NaI_all_2}, are shown as {\it solid}
and {\it dashed grey curves}, respectively.
Simulated broadened spectra are overplotted in {\it red}. 
Unbroadened spectra ({\it green histograms}) can be considered as guidelines
for the exact energy positions of lines and secondary escape peaks (see text).
The height of the unbroadened peaks is truncated in most plots to
better show the comparison between real and broadened spectra
}
\label{NaI-SIM_2}
\end{figure}
%
\subsubsection{Simulation validation}\label{Sim_Val}
In order to check the accuracy of the obtained channel-energy relation 
and to validate simulations presented in Section \ref{Sec_Lab_Sim}, 
radioactive source spectra were compared with simulated data. 
Figures \ref{NaI-SIM_1} and \ref{NaI-SIM_2} show the 14 previously analysed NaI lines (measured 
at normal incidence with detector FM 12\footnote{In this case, simulations were not 
based on measurements performed with detector FM 04, because at that time detector 
FM 12 was the first to have a complete set of collected spectra.})
as a function of energy, that is after applying the channel-energy conversion.
Simulated unbroadened and broadened spectra are overplotted as
{\it green} and {\it red histograms}, respectively.
Sample spectra over the full NaI energy range comparing simulation and measurements 
can be found in \cite{HOO08b}.

Unbroadened lines represent good guidelines to check the exact position of the 
full-energy and escape peaks. A good example is the high-energy double peak of
$^{57}$Co (Fig. \ref{NaI-SIM_1}, {\it panel f}), where simulations confirm the position
of both radioactive lines at 122.06 and 136.47 keV, and the presence of the
Iodine escape peak around $\sim$90 keV.
Still, some discrepancies are evident, e.g. at lower energies.
One likely cause of the discrepancies below 60 keV, mostly resulting
in a higher number of counts of the simulated data compared to the real data,
and which is particularly visible for
the $^{57}$Co line at 14.41 keV  (Fig. \ref{NaI-SIM_1}, {\it panel a}),
is the uncertainty about the detailed composition of the radioactive source. Indeed, 
radioactive isotopes are contained in a small (1 mm) sphere of ``salt''.
Simulations including this salt sphere were performed and a factor of 3.8  
difference in the perceived 14.41 keV line strength was found. The true answer lies 
somewhere between this and the simulation with no source material, as salt
and radioactive isotope are mixed. For the general calibration simulation, a pointsphere of
radioactive material not surrounded by salt had been used. 
Another possible explanation could  probably be the leakage of secondary electrons from 
the surface of the detector leading to a less-absorbed energy.
Further discrepancies at higher energies, which are visible e.g.
in {\it panels d} and {\it e} of Fig. \ref{NaI-SIM_2} are smaller than 1~\%.
%
\begin{figure}[t!]
\centering
\includegraphics[width=70mm,bb=20 10 565 491,clip]{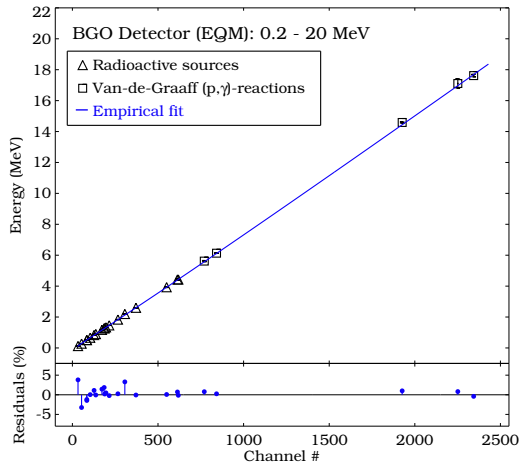}
\caption{
BGO channel-energy relation and corresponding residuals calculated for EQM detector
applying the empirical fit ({\it blue curve}) of Eq. \ref{fitfunc} over the full energy domain between 
0.2 and 20 MeV. Data taken at MPE and at SLAC are marked with \textit{triangles} and \textit{squares}, respectively.
Fit parameters for this fit are
a = -428 $\pm$ 23, b = -54 $\pm$ 3, c = 8.29 $\pm$ 0.06 and
d = 167 $\pm$ 11, with a reduced $\chi^{\,2}$ of 7
}
\label{Fig_BGO_CE}
\end{figure}
%
%
\subsubsection{BGO response}\label{BGO_CE}
The determination of a channel-energy relation for the BGO detectors required 
the additional analysis of the high-energy data taken with the EQM module at SLAC 
to cover the energy domain between 5 MeV and 20 MeV. In order to combine the radioactive
source measurements made with the BGO FMs at MPE and the proton beam induced radiation 
measurements made with the BGO EQM at SLAC (see Section \ref{Sec_SLAC}), it was necessary to 
take into account the different gain settings by application of a scaling factor.
The scaling factor was derived by comparing $^{22}$Na and Am/Be measurements 
(at 511, 1274, and 4430 keV) which were done at both sites.
Due to the very low statistics in the measurements from the
high-energy reaction of the Van-de-Graaf beam on the LiF target (Eq. \ref{eq_SLAC_2}),
first and second electron escape peaks from pair annihilation 
of the 14.6 MeV line could not be considered in this analysis
(see also Fig. \ref{BGO_all_2}, {\it panel f}).
They were mainly used as background reference points in order to help finding 
the exact position of the 17.5 MeV line. In this way, a dataset of 23 detected lines 
was available for determining the BGO EQM channel-energy relation.

For the BGO flight modules analysis (FM 01 and FM 02), a smaller line sample between 125 keV
and 4.4 MeV was adopted, still leading to similar results.
Differences are due to different gains,
caused by the setup constraints, which are described
in more detail in Section \ref{Energy Resolution}.
Similarly to the NaI analysis, BGO datasets were fitted with an
empirical function (Eq. \ref{fitfunc}). Fig. \ref{Fig_BGO_CE} 
shows the BGO EQM channel-energy relation between 0.2 and 20 MeV.
Fit parameters are given in the caption. In this case, fit residuals
are smaller than 4\%.
\subsection{Energy Resolution}\label{Energy Resolution}
The energy resolution $R$ of a detector is conventionally defined as the full width at half
maximum ($w$) of the differential pulse height distribution divided by the location of the 
peak centroid $H_0$\cite{KNO00}. This quantity mainly reflects the statistical fluctuations recorded from pulse to pulse.
In the case of an approximately linear response, the average pulse amplitude is given by $H_0$~=~$K\,N$,
where $K$ is a proportionality constant, and the limiting resolution of a detector can be calculated as
\begin{equation}\label{eq_R}
R \equiv \frac{w}{H_0} \; = \; \frac{2.35\,K\,\sqrt{N}}{K\,N} \; = \; \frac{2.35}{\sqrt{N}} 
\end{equation}
where $N$ represents the average number of charge carriers (in our case, it represents the number of
photoelectrons emitted from the PMT photocathode),
and the standard deviation of the peak in the pulse height spectrum is given by $\sigma$~=~$K\,\sqrt{N}$.
%
%
\begin{figure}[t!]
\centering
\begin{tabular}{cc}
\includegraphics[width=55mm,bb=10 10 556 491,clip]{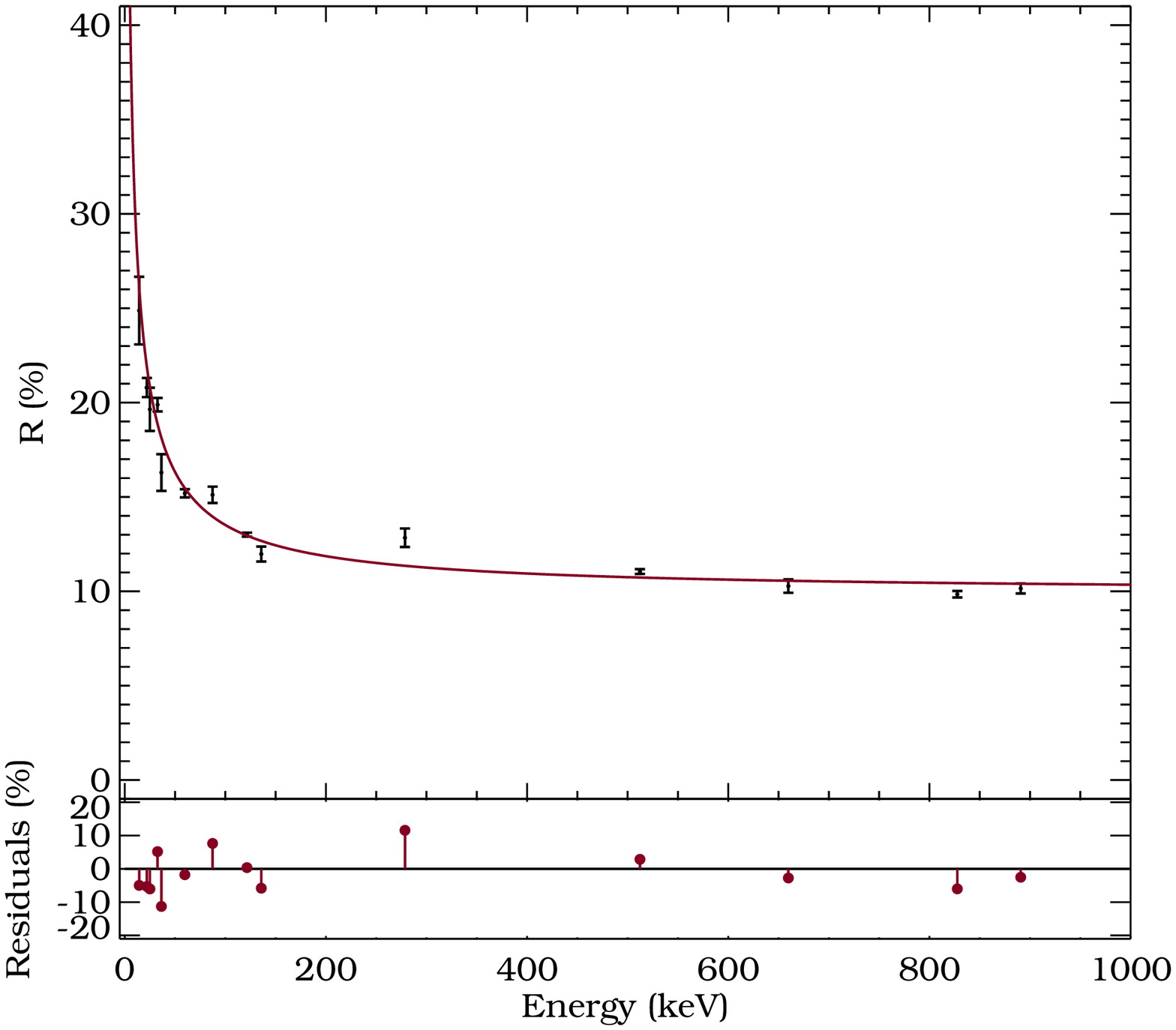} &   
\includegraphics[width=55mm,bb=20 10 575 490,clip]{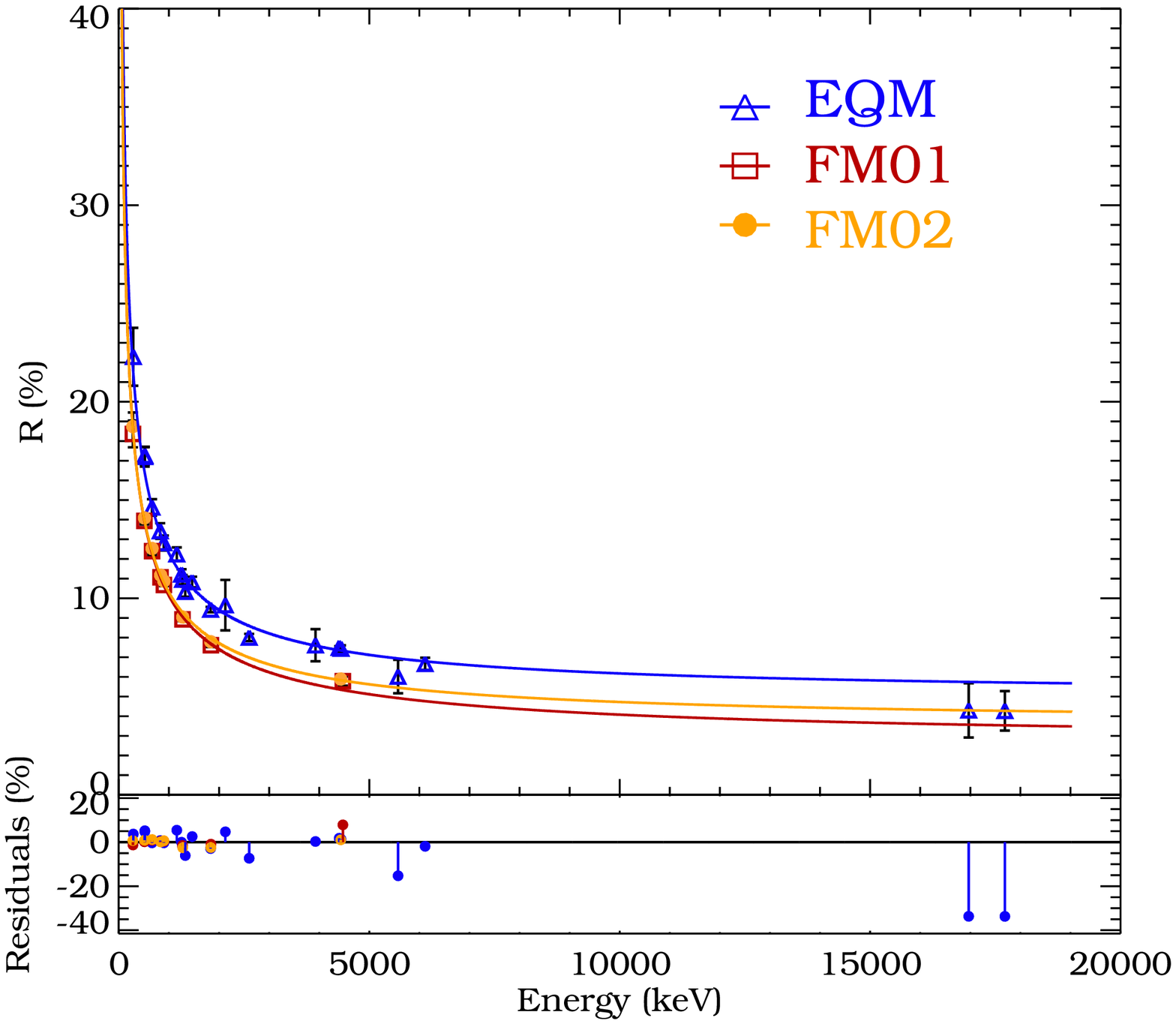}
\end{tabular}
\caption{Energy resolution $R$ in percentage calculated for detector NaI FM 04 ({\it left panel})
and for all BGO detectors ({\it right panel}), i.e. BGO EQM ({\it blue curve}), BGO FM 01 ({\it red curve})
and BGO FM 02 ({\it yellow curve}). Residuals to Eq. \ref{eq_w_2} are given in the panel under the plot.
NaI FM 04 fit parameters to Eq. \ref{eq_w_2} are b = 0.916 $\pm$ 0.014 and c = 0.0994 $\pm$ 0.0010, 
with a reduced $\chi^{\,2}$ of 5. 
In the case of the BGO detectors, the fit parameters for EQM are,
b = 3.54 $\pm$ 0.05 and c = 0.0506 $\pm$ 0.0013, with a reduced 
$\chi^{\,2}$ of 2.2; 
for FM 01, b = 3.09 $\pm$ 0.03 and c = 0.027 $\pm$ 0.003, with a reduced
$\chi^{\,2}$ of 1.3; 
for FM 02, b = 3.050 $\pm$ 0.021 and c = 0.0361 $\pm$ 0.0012, 
with a reduced $\chi^{\,2}$ of 2.4
}
\label{fig_R}
\end{figure}
%
%
However, in real detectors the resolution is not only determined by photoelectron
statistics, but can be affected by other effects, such as
(1) local fluctuations in the scintillation efficiency; (2) nonuniform light collection;
(3) variance of the photoelectron collection over the photocathode;
(4) contribution from the nonlinearity of the NaI scintillation response;
(5) contributions from PMT gain drifts; and (6) temperature drift
(e.g. see \cite{KNO00}).
In order to take all these effects into account, a nonlinear dependence of the energy
resolution was assumed:
\begin{equation}\label{eq_w_2}
w \; = \; \sqrt{ a^{\,2} \; + \; b^{\,2} \, E \; + \; c^{\,2} \, E^{\,2} } 
\end{equation}
This formula is mainly based on traditional physical understanding of scintillation detectors
and produces a physically motivated behavior outside the range of measurements. It consists of
(1) a constant term, $a$, which describes limiting electronic resolution  
(typically not a noticeable effect in scintillators);
(2) a term proportional to the square root of the energy, explaining statistical fluctuations  
in the numbers of scintillation photons and photo electrons; and
(3) a term proportional to the energy, which accounts for the non-ideal ``transfer efficiency'' 
of transporting scintillation photons from their creation sites to the PMT photocathode.
For the actual fits the first parameter $a$ was set to zero, since no significant
electronic broadening was observed.
Spatial non-uniformity of the energy resolution will be discussed in Section \ref{QDE_Uni}.

\begin{figure}[t!]
\centering
\includegraphics[width=70mm,bb=10 10 570 470,clip]{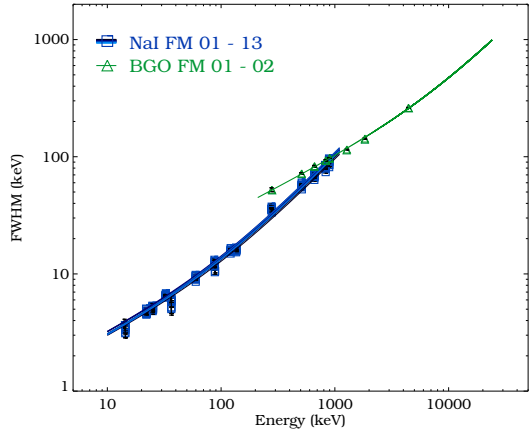}
\caption{FWHM (in keV) as a function of Energy for the 12 NaI FM detectors ({\it blue squares}) 
and for the two BGO FM detectors ({\it green triangles}). For both detector types, the standard
fit to Eq. \ref{eq_w_2} is plotted
}
\label{fig_FWHM-NaI_BGO}
\end{figure}

In order to fit the energy resolution, it was necessary to convert the measured widths
(in channels) to energies in keV by applying to each detector the 
corresponding channel-energy relation previously obtained.
Fit results for detector NaI FM 04 and for the three BGO detectors (EQM, FM 01 and FM 02) 
are displayed in Fig. \ref{fig_R}.
In the case of NaI ({\it left panel}), we noticed that similar results could also be obtained by excluding 
those calibration lines which are affected by greater uncertainties of $w$, 
namely the 14.4 keV line from $^{57}$Co and all secondary lines, i.e. the 25 keV
line from $^{109}$Cd, the 36.6 keV from $^{137}$Cs and the 136.6 keV line from $^{57}$Co.
As regards the BGO energy resolution ({\it right panel}), EQM results (\textit{blue triangles}) 
show poorer energy resolution compared to FM 01 (\textit{red squares}) 
and FM 02 (\textit{yellow dots}), which could be explained by minor differences in
the detector design (optical coupling).
Finally, a common plot of all NaI and BGO flight module detectors 
is shown in Fig. \ref{fig_FWHM-NaI_BGO}.
\subsection{Full-energy Peak Effective Area and Angular Response} \label{Effective_Area}
The full-energy peak effective area in $\rm{cm^2}$ for both NaI and BGO detectors was computed as:
\begin{equation}
A_{\,Eff} \; = \; \frac{A}{a_{\,c}\,\cdot\,P_{\,T}} \,\cdot\,4 \, \pi \,d_{\,s}^{\,2}  \; , 
\end{equation}
where (1) $A$ = Line area (count/s);
(2) $a_{\,c}$ = Current source activity (1/s);
(3) $P_{\,T}$ = Line transition probability;
(4) $d_{\,s}$ = Distance between source and detector (cm).
No additional factor to account for flux attenuation between the source
and the detector was needed, since its effect above 20 keV is less than 1\%.
The different line-transition probabilities for each radioactive nuclide which
were applied for this analysis can be found in Table \ref{Tab_Nuclides}
(column 4). The reference activities were provided in a calibration certificate 
by the supplier of the radioactive sources\footnote{Calibrated radioactive sources were delivered by
AEA Technology QSA GmbH (Braunschweig, Germany) together with a calibration
certificate from the Deutscher Kalibrierdienst (DKD, Calibration laboratory
for measurements of radioactivity, Germany)}.
The radioactive source activities at the day of measurement were calculated by 
taking into account the time elapsed since the calibration reference day.
The relative measurement uncertainty of the given activities for all sources is 3~\%, 
with the exception of the Mercury source ($^{203}$Hg) and the Cadmium source 
($^{109}$Cd), which have an uncertainty of 4~\%.

\begin{figure}[t!]
\centering
\includegraphics[width=70mm,bb=20 10 570 461,clip]{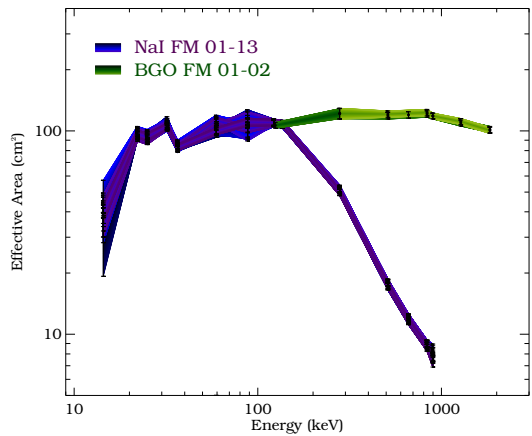}  
\caption{
On-axis effective area of the 12 NaI FM detectors and 
of the two BGO FM detectors
}
\label{figAeff_NaI}
\end{figure}

Results for the on-axis effective area as a function of the energy for all NaI and BGO FM detectors are shown in
Fig. \ref{figAeff_NaI}. The initial drop below 20 keV is due to a Silicone rubber layer placed between the NaI crystal 
and the entrance window, which absorbs low-energy gamma-rays. At energies higher than 300 keV,
NaI detectors become more transparent to radiation and a decrease in the response is observed. 
The BGO on-axis effective area is constant over the energy range 150 keV $-$ 2 MeV, with a mean value of 
120~$\pm$~6 cm$^{\,2}$. Unfortunately, the effective area at 4.4 MeV could not be determined,
since the activity of the Am/Be source was not known. SLAC measurements could not be used for this purpose either.
At 33.17 keV, the effect of the Iodine K-edge is clearly visible as a drop. This energy region was extensively investigated 
during the PTB/BESSY calibration campaign and is further described in Section \ref{QDE_Uni}.
%
%

%
%
\begin{figure}[b!]
\centering
\begin{tabular}{cc}
\includegraphics[width=56mm,bb=20 0 590 463,clip]{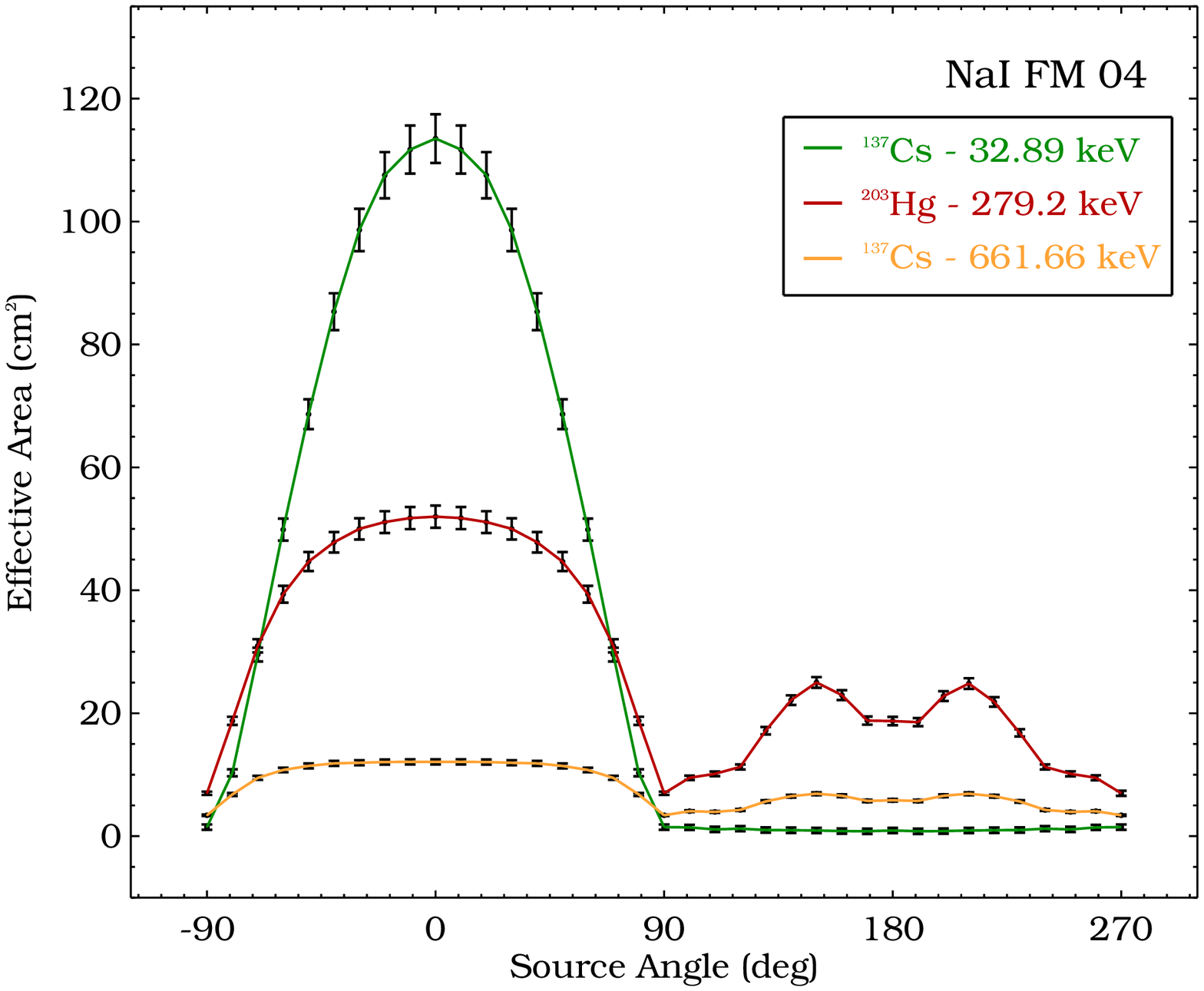} &              
\includegraphics[width=56mm,bb=20 0 590 463,clip]{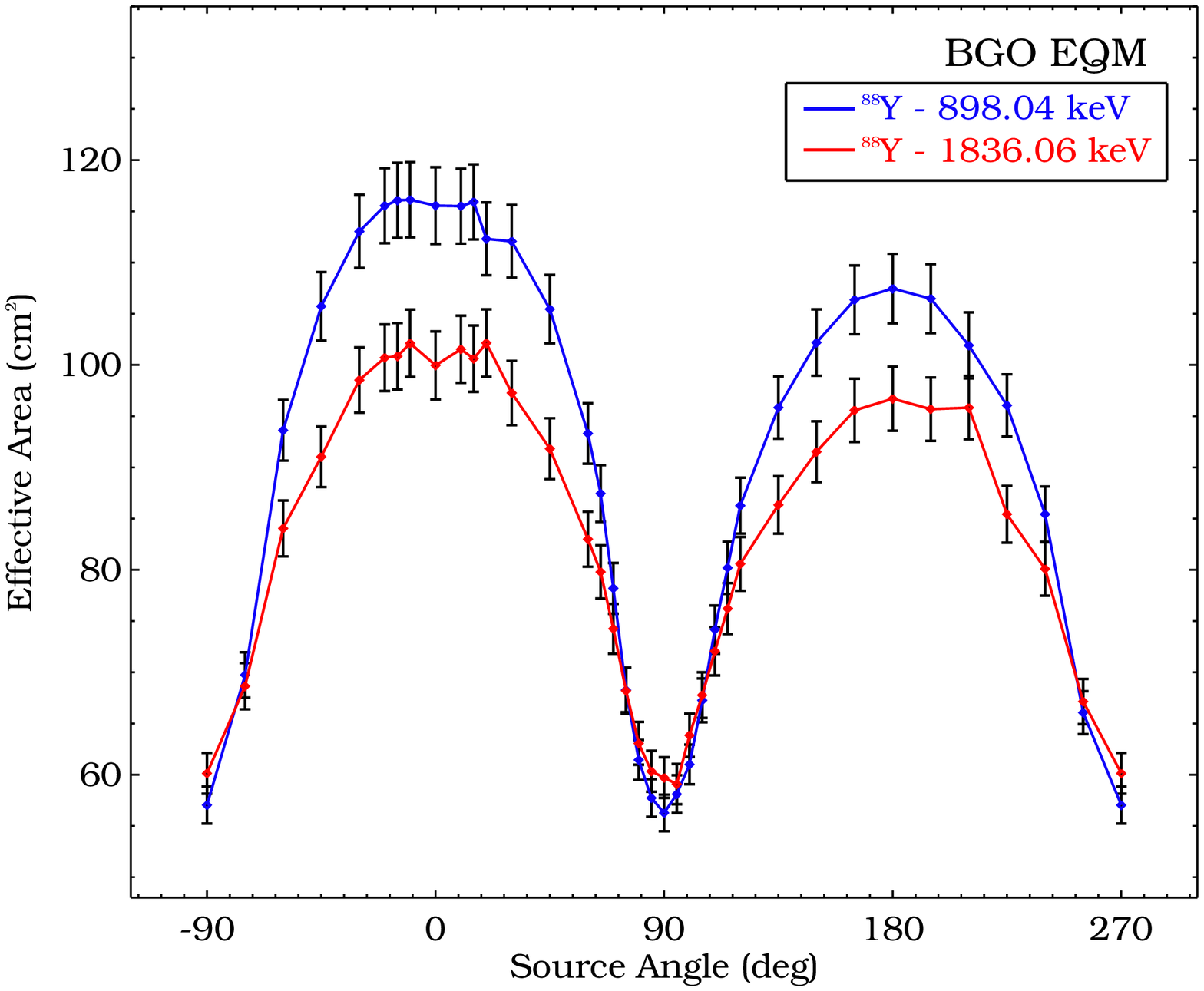}
\end{tabular}
\caption{Off-axis effective area as a function of the irradiation angle (from -90$^o$ to 270 $^o$)
for NaI FM 04 ({\it left panel}) and BGO EQM ({\it right panel}). 
Different curves represent different line-energies.
In the case of NaI, results for
three radioactive lines are shown, namely: 
32.06 keV from $^{137}$Cs ({\it top green curve}),
279.2 keV from $^{203}$Hg ({\it middle red curve}), and 
661.66 keV from $^{137}$Cs ({\it bottom yellow curve}).  
For BGO, two lines from $^{88}$Y are shown:
898.04 keV ({\it blue curve}) and 1836.06 keV ({\it red curve})
}
\label{figAeff_offAxis_all_1}
\end{figure}

For several radioactive sources, off-axis measurements of the NaI and BGO response have been performed. 
These are extremely important for the interpretation of scattered photon flux
both from the spacecraft and from the atmosphere.
Fig. \ref{figAeff_offAxis_all_1} shows results for the NaI and the BGO effective area 
as a function of the irradiation angle over the full 360$^o$.
The {\it left panel} presents NaI FM 04 measurements from 
(i) the 32.89 keV line\footnote{In this case, the double line was fitted with
a single Gaussian, since the response dramatically drops above 90$^o$ and the fit algorithm is not capable of
identifying two separate components.} from $^{137}$Cs ({\it top green curve});
(ii) the 279.2 keV line from $^{203}$Hg ({\it middle red curve}); 
and (iii) the 661.66 keV line from $^{137}$Cs ({\it bottom yellow curve}). It's worth noting
that all curves, especially the middle one, trace the detector's structure
(crystal, housing, and PMT).
Furthermore, the bottom curve (661.66 keV) varies very little with the inclination angle
because of the high penetration capability of gamma-rays at those energies.
In the case of BGO, measurements performed with $^{88}$Y at 898.04 and 1836.06 keV
highlight the two drops in the response due to the presence of the PMTs on 
both sides of the crystal. Although the BGO detectors are symmetrical, an asymmetry in the curves is caused by the
Titanium bracket on one side of the crystal housing, which is necessary for mounting
the detectors onto the spacecraft (Figure \ref{Fig_MPE_Lab}, {\it right panel}). 
Comparisons of the effective area for NaI and BGO with simulations can be found in \cite{HOO08b}.
%
%
\begin{figure}[b!]
\centering
\includegraphics[width=70mm,bb=20 10 570 461,clip]{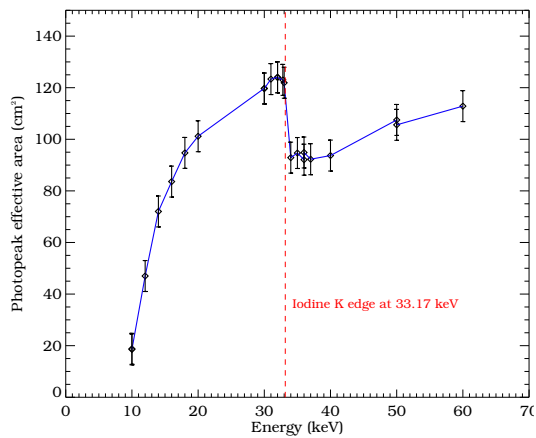}
\caption{NaI FM 04 effective area calculated from QDE determination of 
PTB/BESSY data collected between 10 and 60 keV.
A relative uncertainty of 5\%\ has been estimated
}
\label{fig_A_eff_33}
\end{figure}
%
\subsection{Quantum Detection Efficiency (QDE) and Spatial Uniformity of NaI Detectors}\label{QDE_Uni}
\begin{figure}[t!]
\centering
\begin{tabular}{|c|c|}
\hline\noalign{\smallskip}
\includegraphics[width=55mm,bb=0 81 630 693,clip]{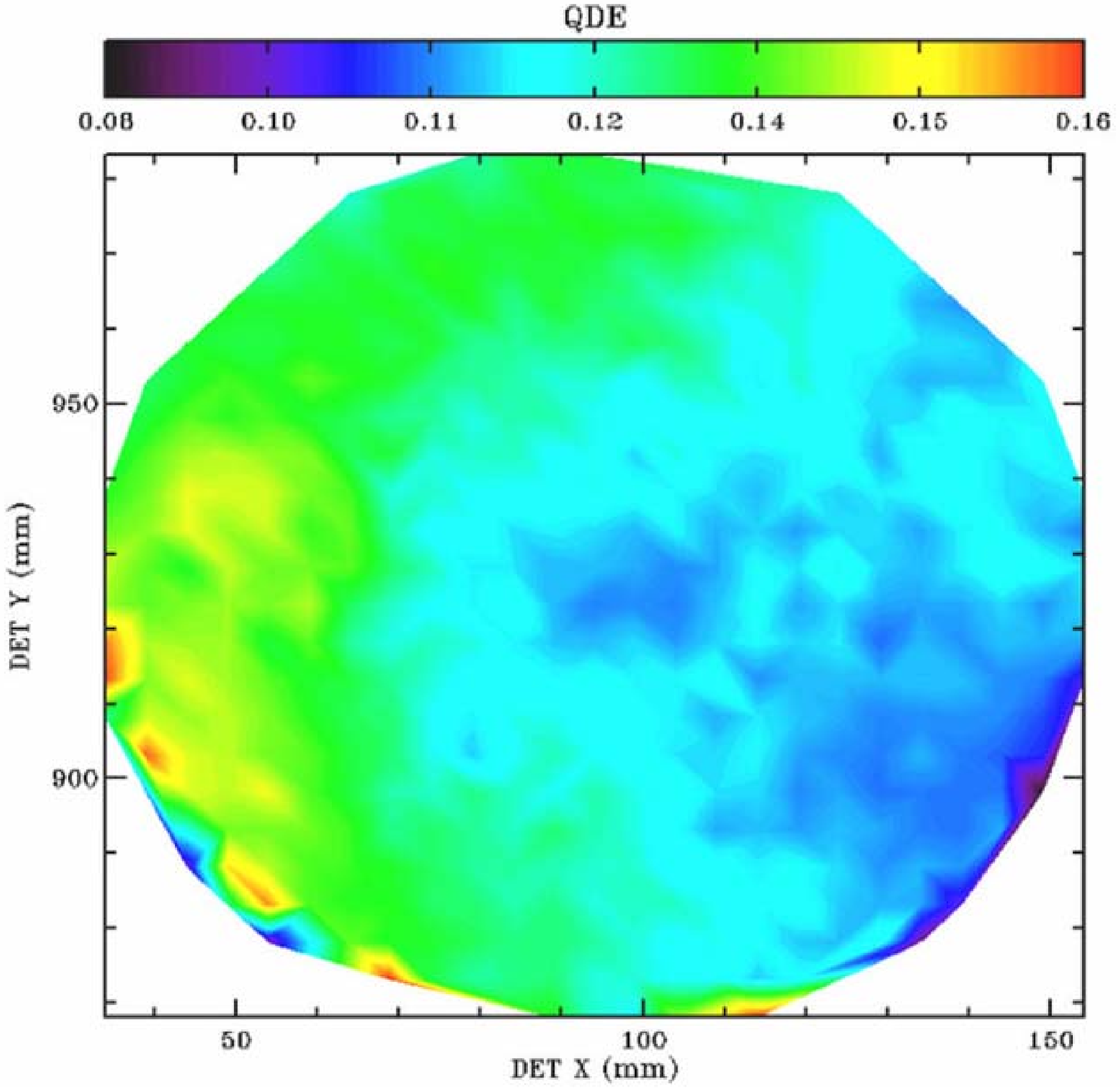}  & 
\includegraphics[width=55mm,bb=0 90 621 702,clip]{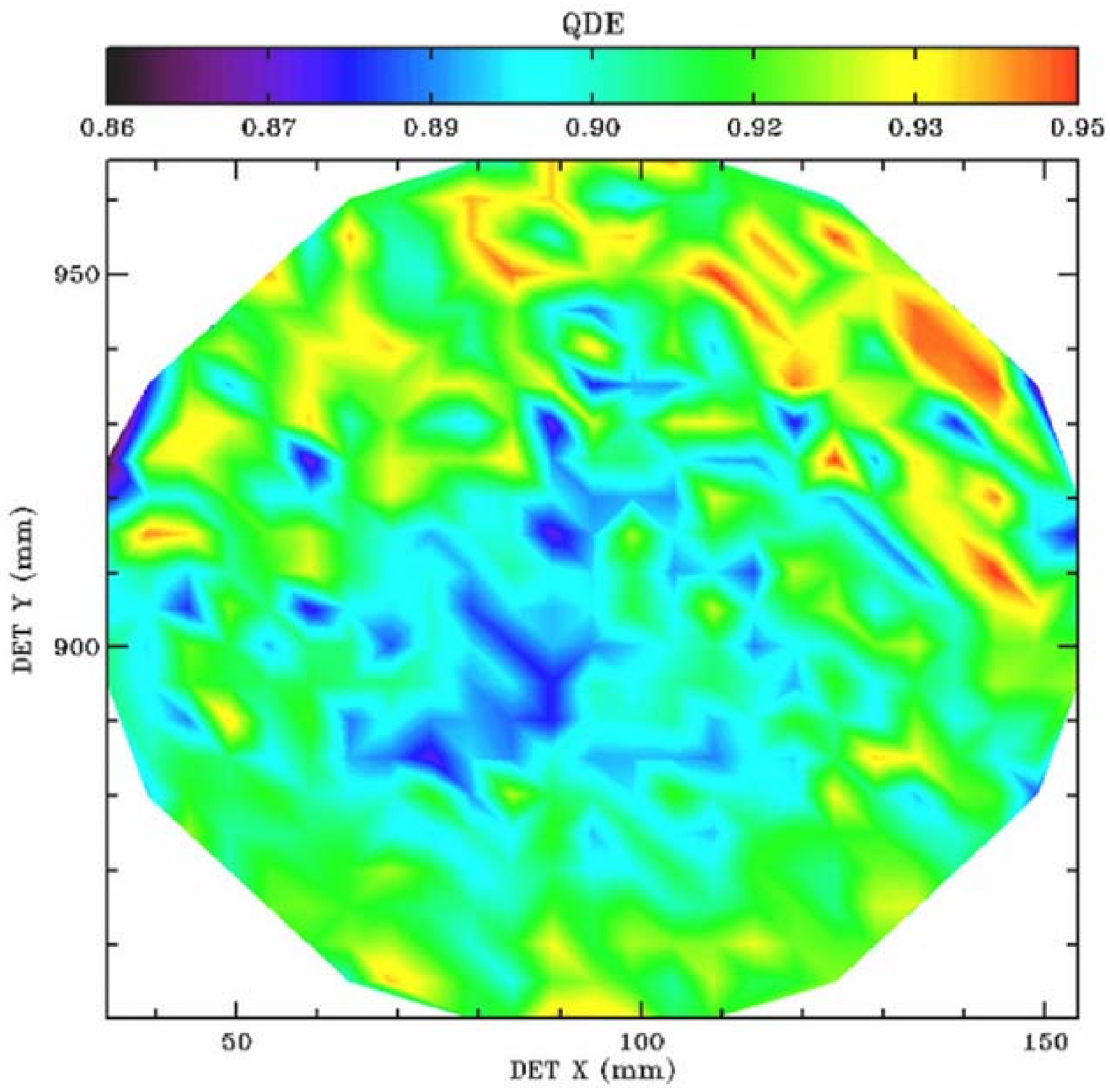}  \\
\noalign{\smallskip}\hline\noalign{\smallskip}
\includegraphics[width=55mm,bb=0 81 630 693,clip]{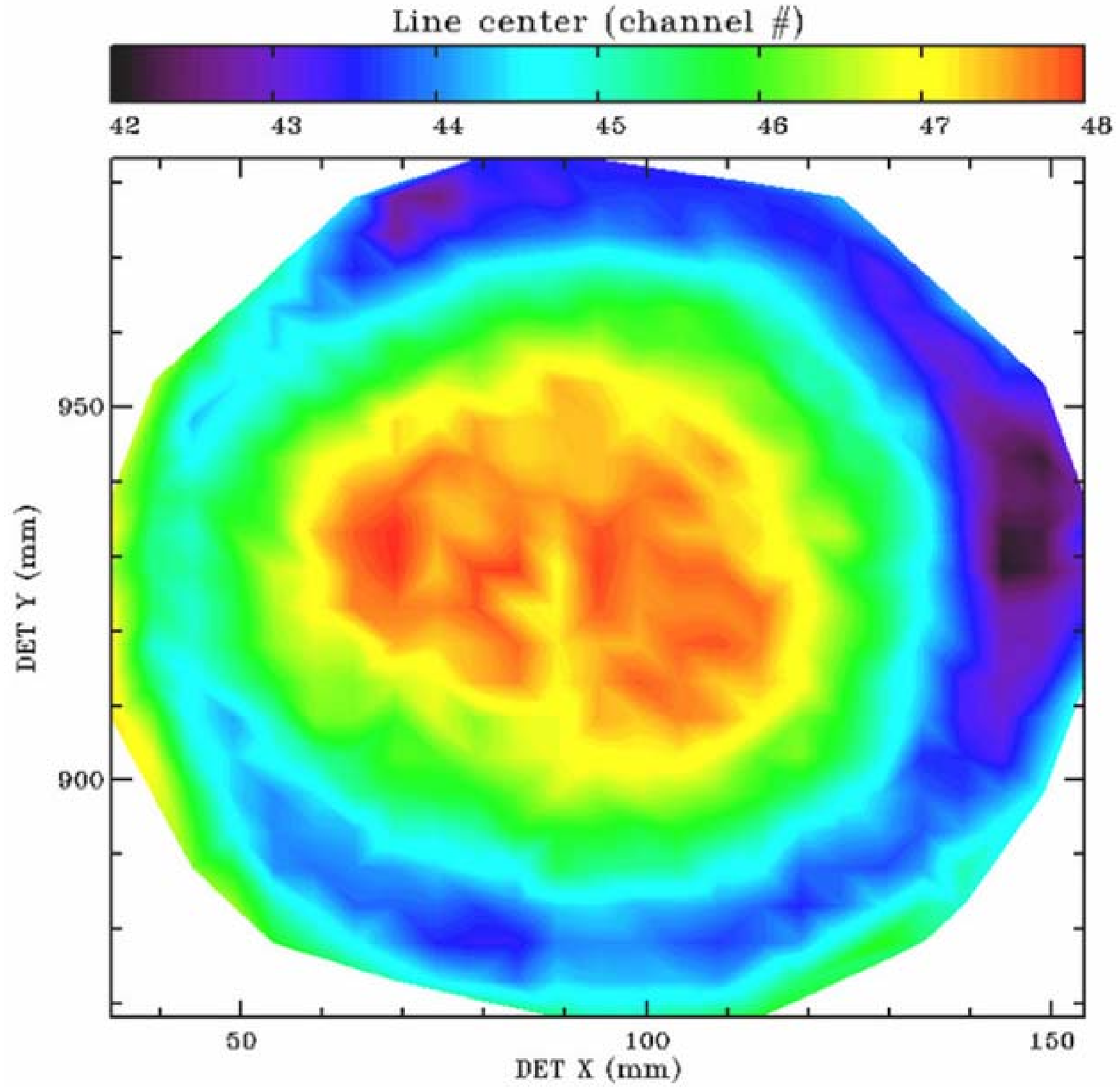}  & 
\includegraphics[width=55mm,bb=0 90 612 702,clip]{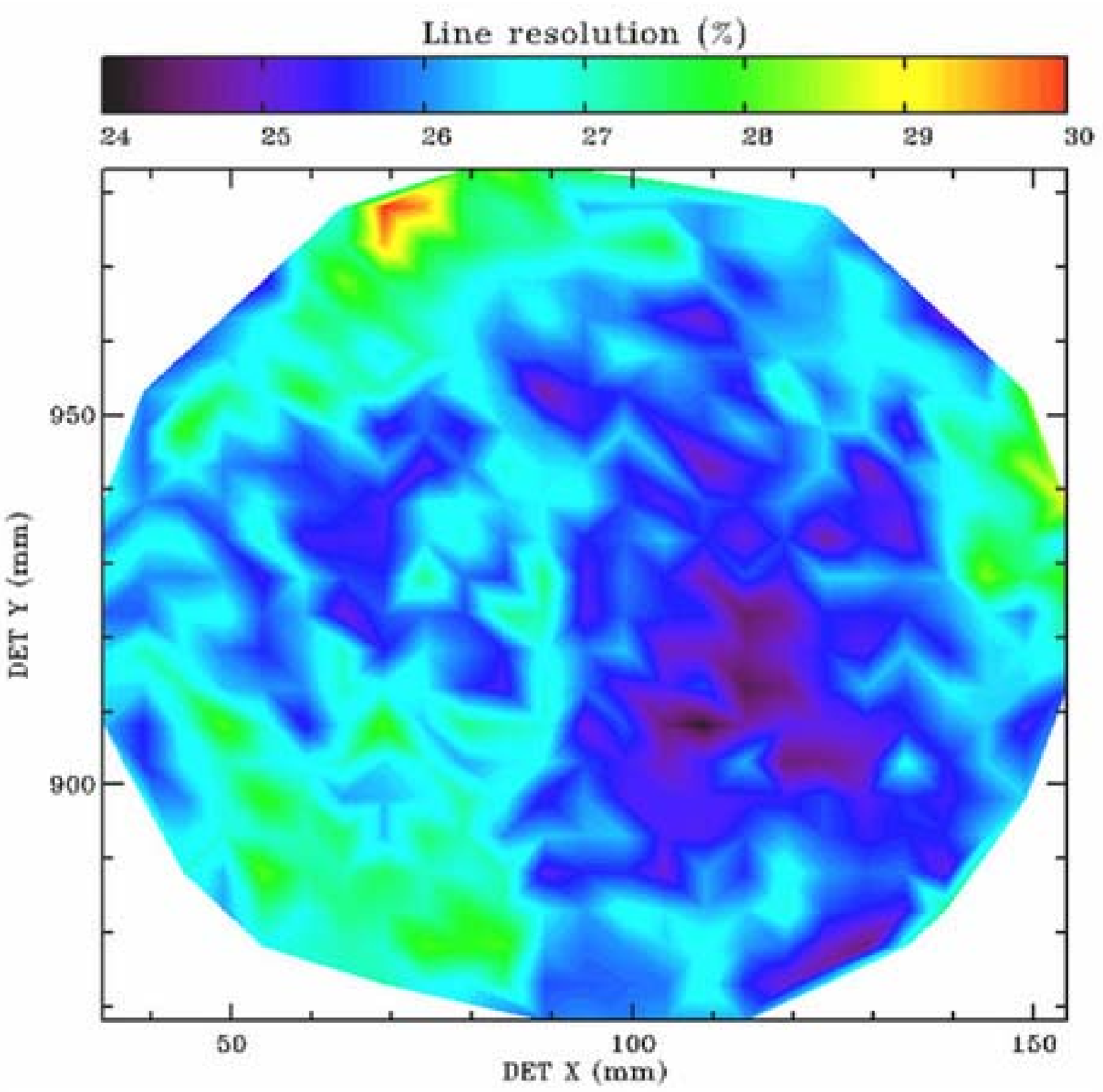} \\
\noalign{\smallskip}\hline
\end{tabular}
\caption{Contour plots showing results for the PTB/BESSY rasterscans
of detector NaI FM04. {\it At the top}: 
QDE calculated over the active area (126 cm$^{\,2}$) 
at 10 and 60 keV ({\it left} and {\it right panel}, respectively).
{\it At the bottom}:
Calculated fit parameters of the full-energy peak 
as a function of the beam position (in mm)
for the 10 keV rasterscan: 
(1) line-center (in channel \#, \textit{left panel});
(2) line resolution (in \%, \textit{right panel}).
The color code is indicated in the colorbars above each plot
}
\label{fig46}
\end{figure}

As already mentioned in Section \ref{Sec_BESSY}, the QDE for detector NaI FM 04 could be 
determined through detailed measurements performed at PTB/BESSY by firstly relating the fitted full-energy peak area 
of the NaI spectrum with the sum of counts detected by the HPGe detector (for which QDE~=~1)
in the same region of interest around the full-energy peak 
(which includes the small Ge escape peak appearing above 11 keV), and subsequently
accounting for the different beam fluxes.
The QDE was determined at all energies by analyzing a spectrum taken in the 
center of the detector's surface and at three energies (rasterscans at 10, 36 and 60 keV) by analyzing spectra taken
over the whole surface area (see Fig. \ref{fig46}, {\it top panels}, for results obtained at 10 and 60 keV).
The effective area (in cm$^{\,2}$) can then be calculated as the integral
of the QDE over the detector's active area (126.7 cm$^{\,2}$). Results for 19 lines measured
between 10 and 60 keV are shown in Fig. \ref{fig_A_eff_33}. This Figure can
be considered as a zoom at low energies of the NaI effective area in Fig. \ref{figAeff_NaI}.
%

The detector's spatial homogeneity was investigated at PTB/BESSY by means of rasterscans of detector NaI FM 04 
at three distinct energies, namely 10, 36 and 60 keV. During each rasterscan, $\sim$~700 runs per detector
were recorded with a spacing of 5~mm, and for each one the full-energy peak was 
analysed as previously described in Section \ref{Ana_Photopeak}. 
For the rastercsan at 10 keV, the dependence of line-center (in channel \#) and line resolution (in \%) 
on the rasterscan position in mm (DET\_X and DET\_Y)
is shown in Fig. \ref{fig46}, {\it bottom panels}. Results for the line area are not further shown, since the QDE plots
previously discussed in Fig. \ref{fig46}, {\it top panels}, already reflect the surface behavior for this parameter).

From the line center spatial dependence one can notice that some border effects appear toward
the edge of the NaI crystal, shifting the full-energy peak to lower channel numbers i.e. energies.
This effect is of the order of 12 \%\ at 10 and 60 keV and of 7 \%\ at 36 keV.
The line resolution is homogeneous over the whole detector's area, with a mean value
of 25\%, 15\%\ and 10\%\ at 10, 36 and 60 keV, respectively. While the first two resolutions
are comparable to the results obtained with radioactive sources at 14.4 and 36.6 keV,
the 60 keV rasterscan gives an improved resolution when compared to the result of 15\%\ obtained 
with the $^{241}$Am source at 59.4 keV (see Fig. \ref{NaI_all_1}, {\it panel f}).
\section{Conclusions}\label{Concl}
After the successful launch of the Fermi mission and the proper activation of its
instruments, the 14 GBM detectors started to collect scientific data.
The spectral overlap of the two BGO detectors (0.2 to 40 MeV) with the LAT lower limit of $\sim$~20 MeV
opens a promising epoch of investigation of the high-energy prompt and afterglow GRB emission
in the yet poorly explored MeV-GeV energy region.

On ground, the angular and energy response of each GBM detector was calibrated
using various radioactive sources between 14.4 keV and 4.4 MeV. 
The channel-energy relations, energy resolutions,
on- and off-axis effective areas of the single detectors were determined.
Additional calibration measurements were performed for NaI detectors at PTB/BESSY below 60 keV and 
for BGO detectors at SLAC above 5 MeV, thus covering the whole GBM energy domain.
As already mentioned in the introduction, further calibration measurements 
at system level and after integration onto the spacecraft were carried out.
All those measurements crucially contribute to the validation of Monte Carlo simulations of the 
direct GBM detector response. These incorporate detailed models of the Fermi observatory, 
including the GBM detectors, instruments, and all in-flight spacecraft components, plus 
the scattering of gamma-rays from a burst in the spacecraft and in the Earth's atmosphere \cite{HOO08b}.
The response as a function of photon energy and direction is finally captured in a 
Direct Response Matrix (DRM) database, allowing the determination of the 
true gamma-ray spectrum from the measured data.

The results reported in this paper directly contribute to the DRM final determination,
and they fully follow physical expectations \cite{KIE04}.
Measurements and fit results for two sample detectors (NaI FM 04 and BGO FM 02)
are given in Tables \ref{Tab_Results_NaI_FM04} and \ref{Tab_Results_BGO_FM02},
respectively (see Appendix). Fit parameters for the energy-channel relations and
energy resolutions of those detectors are always reported in the captions
of the corresponding plots (see Figures \ref{figNaI_FM04_low-high},
\ref{Fig_BGO_CE} and \ref{fig_R}). It is worth noting that these parameters 
reflect the characteristics of all other NaI and BGO detectors not reported in this
paper, thus showing that all detectors behave the same within statistics.

The channel-energy relation parameters obtained throughout this analysis
are not directly used for in-flight calculations because of the different electronic setup 
which was used during detector level calibration and in-flight (see Section \ref{MPE_calib_setup}).
The same analysis method described in Section \ref{Ana_Photopeak}
and the same fitting procedures of Section \ref{CE Relation}
were adopted to analyze data collected during system level calibration.
These results confirm that the systematic uncertainties in the channel-energy conversions 
arise from the discussed sources, i.e. fitting procedure, limited statistics in the case of high-energy 
lines of BGOs, electronics and non-uniform responses of detectors,
and are fully consistent with measurements presented in this paper.

The GBM detectors will play an important role in the GRB field in the next decade.
The unprecedented synergy between the GBM and the LAT will allow to observe burst spectra
covering $\sim$~7 decades in energy. Moreover, simultaneous observations by the 
large number of gamma-ray burst detectors operating in the Fermi era 
will complement each other.
A nice overview of the currently operating space missions can be found 
in Table 1 of \cite{BAN08}. Here, instrument characteristics such as FOVs, effective
areas, localization uncertainties and energy bands are compared.
The GBM detectors fit in this overall picture by providing a higher trigger energy range (50-300 keV)
than e.g. Swift-BAT \cite{GEH04} (15-150 keV) and a spectral coverage up to 30 MeV,
an energy limit which can only be investigated with the LAT and the Mini-Calorimeter on-board AGILE \cite{TAV06}.
New insights into the GRB properties are therefore expected from GBM, thus advancing
the study of GRB physics.
\section*{Appendix}\label{appendix}
\renewcommand{\tabcolsep}{1mm}
\begin{table}[ht!]
\centering
\caption{
List of all calibrated GBM detectors. Columns 2 and 3 indicate the 
detector numbering schemes adopted during the presented calibration analysis 
and in-flight, respectively. A checklist showing which detectors were
employed at each detector-level calibration campaign is given in columns
4 to 6
}
\begin{tabular}{cccccc}
\hline\noalign{\smallskip}
  & \multicolumn{2}{c}{\centering \bf Detector \#} & \multicolumn{3}{c}{\bf Calibration campaign} \\
  & \multirow{2}{15mm}{\centering \bf This work} & \multirow{2}{12mm}{\centering \bf In-flight} & \bf MPE  & \bf PTB/BESSY  &  \bf SLAC \\
  &    &   & \bf (14.4 keV $\mathbf{-}$ 4.4 MeV)  & \bf (10 $\mathbf{-}$ 60 keV)  & \bf (4.4 $\mathbf{-}$ 17.6 MeV) \\
\noalign{\smallskip}\hline\noalign{\smallskip}
\bf NaI  &  EQM    &  $-$  &  $\surd$  &  $-$      &  $-$  \\
         &  FM 01  &  n0   &  $\surd$  &  $\surd$  &  $-$  \\
         &  FM 02  &  n1   &  $\surd$  &  $\surd$  &  $-$  \\
         &  FM 03  &  n2   &  $\surd$  &  $\surd$  &  $-$  \\
         &  FM 04  &  n3   &  $\surd$  &  $\surd$  &  $-$  \\
         &  FM 05  &  n4   &  $\surd$  &  $-$      &  $-$  \\
         &  FM 07  &  n6   &  $\surd$  &  $-$      &  $-$  \\
         &  FM 08  &  n7   &  $\surd$  &  $-$      &  $-$  \\
		     &  FM 09  &  n8   &  $\surd$  &  $-$      &  $-$  \\
		     &  FM 10  &  n9   &  $\surd$  &  $-$      &  $-$  \\
		     &  FM 11  &  na   &  $\surd$  &  $-$      &  $-$  \\
		     &  FM 12  &  nb   &  $\surd$  &  $-$      &  $-$  \\
		     &  FM 13  &  n5   &  $\surd$  &  $-$      &  $-$  \\
\bf BGO  &  EQM    &  $-$  &  $\surd$  &  $-$      &  $\surd$ \\
		     &  FM 01  &  b0   &  $\surd$  &  $-$      &  $-$  \\
		     &  FM 02  &  b1   &  $\surd$  &  $-$      &  $-$  \\
\noalign{\smallskip}\hline
\end{tabular}
\label{Tab_Num_Dec}
\end{table}
\begin{table}[ht!]
\centering
\caption{
Measured and fitted quantities with fit residuals (Res.) 
for 14 lines from radioactive sources collected with detector NaI FM 04
}
\begin{tabular}{c|ccc|cccc|c}
\hline\noalign{\smallskip}
  &  \multicolumn{8}{c}{\bf Full-energy peak parameters} \\
\noalign{\smallskip}\cline{2-9}\noalign{\smallskip}
\multirow{3}{15mm}{\centering{\bf Tabulated \\ \bf line energy \\ \bf (keV) }} & \multicolumn{3}{c|}{\bf Center} & \multicolumn{4}{c|}{\bf FWHM}  &  \bf Eff. area \\ 
\noalign{\smallskip}\cline{2-9}\noalign{\smallskip}
 & \multicolumn{1}{p{20mm}}{\centering \bf Measured}  & \bf Fitted & \bf Res. & \multicolumn{1}{p{20mm}}{\centering \bf Measured} & \multicolumn{2}{c}{\bf Fitted} & \bf Res. & \multirow{2}{13mm}{\centering \bf cm$\mathbf{^{\,2}}$} \\
 & \bf (channels)& \bf (keV)  & \bf (\%)      & \bf (channels) & \bf (keV)   & \bf (\%)       & \bf (\%)      &     \\
\noalign{\smallskip}\hline\noalign{\smallskip}
   14.41   &     78.3   $\pm$   0.8    &    14.39    &     0.15		&   18.8    $\pm$   1.4   &   4    &     26   &    -4     &   46    $\pm$   4  \\
   22.1    &    116.35  $\pm$   0.19   &    22.04    &     0.29   &   21.7    $\pm$   0.5   &   5    &     21   &    -5     &   98    $\pm$   4  \\
   25      &    130.4   $\pm$   0.5    &    25.05    &    -0.20   &   22.5    $\pm$   1.3   &   5    &     21   &    -6     &   95    $\pm$   4  \\
   32.06   &    161.72  $\pm$   0.22   &    32.14    &    -0.26   &   25.9    $\pm$   0.5   &   6    &     19   &     5     &  114    $\pm$   4  \\
   36.6    &    179.1   $\pm$   0.6		 &     $-$     &     $-$		&   28.6    $\pm$   1.7   &   7    &     18   &    -11    &   85    $\pm$   3  \\
  59.4     &    291.9   $\pm$   0.3		 &    59.8     &    -0.6    &   41.2    $\pm$   0.6   &   9    &     15   &    -1.8   &  101    $\pm$   4  \\
  88.03    &    412.8   $\pm$   0.6		 &    87.4     &     0.8    &   56.4    $\pm$   1.6   &   12   &     14   &     8     &  122    $\pm$   5  \\
  122.06   &    556.25  $\pm$   0.23 	 &   121.5     &     0.4    &   65.5    $\pm$   0.5   &   16   &     13   &     0.4   &  111    $\pm$   4  \\
  136.47   &    615.1   $\pm$   0.8 	 &    $-$      &     $-$    &   67.0    $\pm$   2.2   &   17   &     13   &    -6     &  109    $\pm$   4  \\  
  279.2    &   1195     $\pm$   6 		 &   278.60    &     0.21   &  144      $\pm$   5     &   32   &     11   &    12     &   52.0  $\pm$   1.8  \\  
  511      &   2136.1   $\pm$   0.5		 &   512.23    &    -0.24   &  228.4    $\pm$   2.6   &   55   &     11   &     2.9   &   18.0  $\pm$   0.7  \\
  661.66   &   2732     $\pm$   9		   &   659.76    &     0.29   &  274      $\pm$   9     &   70   &     10   &    -2.7   &   12.1  $\pm$   0.4  \\
  834.84   &   3413.3   $\pm$   2.2		 &   827.7     &     0.9    &  331      $\pm$   6     &   87   &     10   &    -6     &    9.0  $\pm$   0.3  \\
  898.04   &   3669.6   $\pm$   2.8		 &   890.8     &     0.8    &  367      $\pm$  10     &   93   &     10   &    -2.5   &    8.2  $\pm$   0.4  \\
\noalign{\smallskip}\hline
\end{tabular}

\label{Tab_Results_NaI_FM04}
\end{table}
\begin{table}[ht!]
\centering
\caption{
Measured and fitted quantities with fit residuals (Res.) 
for 8 lines from radioactive sources collected with detector BGO FM 02
}
\begin{tabular}{c|ccc|cccc|c}
\hline\noalign{\smallskip}
  &  \multicolumn{8}{c}{\bf Full-energy peak parameters} \\
\noalign{\smallskip}\cline{2-9}\noalign{\smallskip}
\multirow{3}{15mm}{\centering{\bf Tabulated \\ \bf line energy \\ \bf (keV) }} & \multicolumn{3}{c|}{\bf Center} & \multicolumn{4}{c|}{\bf FWHM}  &  \bf Eff. area \\ 
\noalign{\smallskip}\cline{2-9}\noalign{\smallskip}
 & \multicolumn{1}{p{20mm}}{\centering \bf Measured}  & \bf Fitted & \bf Res. & \multicolumn{1}{p{20mm}}{\centering \bf Measured} & \multicolumn{2}{c}{\bf Fitted} & \bf Res. & \multirow{2}{13mm}{\centering \bf cm$\mathbf{^{\,2}}$} \\
 & \bf (channels)& \bf (keV)  & \bf (\%)      & \bf (channels) & \bf (keV)   & \bf (\%)       & \bf (\%)      &     \\
\noalign{\smallskip}\hline\noalign{\smallskip}
124.59   &    47.68    $\pm$   0.03   &    124.59   &     0.002   &   14.2   $\pm$    0.5   &    51.96   &    18.61   &      0.7    &  107  $\pm$  3   \\ 
279.2    &    90.13    $\pm$   0.19   &    279.27   &    -0.023   &   19.46   $\pm$    0.29  &    71.36   &    13.97   &      0.8    &  122  $\pm$  7   \\
511      &   152.87    $\pm$   0.06   &    510.82   &     0.03    &   22.4   $\pm$    0.4   &    82.07   &    12.39   &      1.2    &  121  $\pm$  4   \\
661.66   &   193.90    $\pm$   0.08   &    662.66   &    -0.15    &   25.2   $\pm$    0.4   &    93.07   &    11.16   &      0.12   &  121  $\pm$  4   \\
834.84   &   240.15    $\pm$   0.12   &    833.88   &     0.12    &   26.37   $\pm$    0.14  &    96.95   &    10.80   &      0.7    &  119  $\pm$  4   \\
898.04   &   257.40    $\pm$   0.06   &    897.73   &     0.03    &   31.2    $\pm$    0.3   &   118.23   &     9.27   &     -2.5    &  118  $\pm$  4   \\
1274.54  &   359.42    $\pm$   0.10   &   1275.18   &    -0.05    &   38.7    $\pm$    0.5   &   146.50   &     7.98   &     -2.5    &  112  $\pm$  4   \\
1836.06  &   511.19    $\pm$   0.12   &   1835.92   &     0.008   &   71.1    $\pm$    0.7   &   258.30   &     5.83   &      1.0    &  101  $\pm$  3   \\
\noalign{\smallskip}\hline
\end{tabular}

\label{Tab_Results_BGO_FM02}
\end{table}
\end{document}